\documentclass[11pt,eadjoint tfnpsf]{report}

\usepackage{amsmath,amsbsy,amssymb,latexsym}

\newcommand{\NN}{\nonumber}

\newcommand{\expect}[1]{\langle{#1}\rangle}

\newcommand{\BE}{\begin{equation}}
\newcommand{\EE}{\end{equation}}
\newcommand{\BEA}{\begin{eqnarray}}
\newcommand{\EEA}{\end{eqnarray}}
\newcommand{\BSE}{\begin{subequations}}
\newcommand{\ESE}{\end{subequations}}
\newcommand{\BA}{\begin{align}}
\newcommand{\EA}{\end{align}}

\newcommand\CD{{\mathcal D}}
\newcommand\CF{{\mathcal F}}
\newcommand\CG{{\mathcal G}}
\newcommand\CH{{\mathcal H}}
\newcommand\CK{{\mathcal K}}
\newcommand\CL{{\mathcal L}}
\newcommand\CM{{\mathcal M}}
\newcommand\CN{{\mathcal N}}
\newcommand\CP{{\mathcal P}}
\newcommand\CQ{{\mathcal Q}}
\newcommand\CR{{\mathcal R}}
\newcommand\CS{{\mathcal S}}

\newcommand\CU{{\mathcal U}}
\newcommand\CV{{\mathcal V}}
\newcommand\CW{{\mathcal W}}

\newcommand\I{{\mathbb I}}

\def\LLangle{\left< \hspace{-2.5mm} \left<}
\def\RRangle{\right> \hspace{-2.5mm} \right>}

\def\llangle{\left< \hspace{-1mm} \left<}
\def\rrangle{\right> \hspace{-1mm} \right>}

\def\a{\alpha}

\def\b{\beta}
\def\g{\gamma}

\def\G{\Gamma}

\def\e{\epsilon}

\def\u1{\underline{ij}}
\def\G2{{\; \rm GeV/}c2}
\def\G{\; \rm GeV}

\def\Im{\mathop{\rm Im}}
\def\Re{\mathop{\rm Re}}

\catcode`\@=11

\def\section{\@startsection {section}{1}{\z@}{-4.0ex plus -0.5ex minus -.2ex}
{1.5ex plus 0.3ex minus .2ex}{\large\bf\raggedright}}
\def\subsection{\@startsection{subsection}{2}{\z@}{-4.0ex plus -0.5ex minus -.2ex}
{1.5ex plus .2ex}{\normalsize\bf\raggedright}}
\def\subsubsection{\@startsection{subsubsection}{3}{\z@}{-4.0ex plus -0.5ex minus -.2ex}
{1.5ex plus .2ex}{\normalsize\bf\raggedright}}
\def\paragraph{\@startsection{paragraph}{4}{\z@}{1.5ex plus
   0.5ex minus .2ex}{-1em}{\normalsize\bf}}
\def\subparagraph{\@startsection{subparagraph}{5}{\parindent}{1.5ex plus
   0.5ex minus .2ex}{-1em}{\normalsize\bf}}

\makeatletter

\@addtoreset{equation}{section}
\makeatother

\begin{document}

\begin{titlepage}

\begin{flushright}
\normalsize
~~~~
July, 2006 \\
OCU-PHYS 250 \\
hep-th/0607047 \\
\end{flushright}

\begin{center}
\vspace*{33pt}

{\LARGE Gauged $\CN=2$ Supergravity and Partial Breaking of}
\vspace*{11pt}

{\LARGE Extended Supersymmetry}

\vspace*{33pt}
{\large Kazunobu Maruyoshi\footnote[2]{E-mail: maruchan@sci.osaka-cu.ac.jp}}

\vspace*{11pt}
\textit{Department of Mathematics and Physics, Graduate School of Science} 

\textit{Osaka City University}

\textit{3-3-138, Sugimoto, Sumiyoshi-ku, Osaka, 558-8585, Japan}

\end{center}

\vspace*{44pt}
\begin{center}
{\large Abstract}
\end{center}

We review the general gauged $\CN=2$ supergravity coupled to an arbitrary number of vector multiplets and hypermultiplets.
We consider two different models where $\CN=2$ supersymmetry is broken to $\CN=1$ spontaneously, 
one has a $U(1)$ vector multiplet and the other has a $U(N)$ vector multiplet.
In both cases, partial breaking of $\CN=2$ supersymmetry is accomplished by the Higgs and the super-Higgs mechanisms.
The mass spectrum can be evaluated and we conclude that the resulting models have $\CN=1$ supersymmetry.
This is based on master thesis submitted to Graduate School of Science, Osaka City University, in March 2006.

\end{titlepage}

\tableofcontents

\chapter{Introduction}
  The standard model is a successful theory describing the high energy physics.
  The remarkable agreement of theory and experiment gives us confidence 
  that it is the correct description of strong, weak and electric interactions of elementary particles.
  However there are several serious problems, 
  such as the gauge hierarchy problem and the inability to include the general relativity.
  These facts suggest that the standard model is not final theory 
  but rather an effective theory describing the electroweak energy scale region.
  
  Supersymmetry has become the dominant framework of formulating physics beyond the standard model.
  Supersymmetric extension of the standard model can solve some problems.
  Since supersymmetry requires that each scalar field have fermionic partner of the same mass,
  the quadratic divergences of the scalar mass terms automatically vanish. 
  Also three gauge coupling constants modify to meet accurately at very high energy.
  But supersymmetry is not observed in nature, so it must be broken at low energy scale.
  If a supersymmetry is broken spontaneously, a massless fermion called Nambu-Goldstone fermion appears.
  Thus, global supersymmetry should not be broken spontaneously.
  Though, in the local supersymmetry, 
  the Nambu-Goldstone fermion is absorbed by gravitino through the super-Higgs mechanism.
  This motivates to study supergravity as a candidate beyond the standard model.
  
  The fact that extended ($\CN>1$) supersymmetric theories cannot have the chiral structure of the standard model 
  make it difficult to deal with phenomenologically.
  But many theorists have been researched extended supersymmetric theories using their rich geometrical structures.
  In $\CN=2$ supersymmetric Yang-Mills theory, by using its symplectic structure, Seiberg and Witten have showed 
  that the prepotential function can be determined exactly including the non-perturbative effects 
  \cite{SeibergWitten1, SeibergWitten2}.
  Similarly, $\CN=2$ supergravity has symplectic structure.
  The most general form of Lagrangian coupled to an arbitrary number of vector multiplets and hypermultiplets 
  in presence of a general gauging of the isometries of both vector and hypermultiplet scalar manifolds 
  has been obtained by \cite{D'Auria, Andrianopoli}.
  This construction uses a coordinate independent and manifestly symplectic covariant formalism 
  which in particular does not require the use of a prepotential function.
  
  Partial supersymmetry breaking plays an important role of relating 
  the extended supersymmetric field theories with the phenomenological models.
  It has been shown that, 
  in $U(1)$ gauged global $\CN=2$ supersymmetric theory with electric and magnetic Fayet-Iliopoulos term, 
  the $\CN=2$ supersymmetry is broken to $\CN=1$ \cite{APT, APT2,APT3}.
  It has been generalized to the case of $U(N)$ gauged theory by \cite{FIS1, FIS2, FIS3}.
  In $\CN=2$ supergravity, partial supersymmetry breaking have been accomplished 
  by simultaneous realization of the Higgs and the super-Higgs mechanisms 
   \cite{Cecotti2, Ferrara1, Ferrara2, Fre, Porrati, Louis, Andrianopoli1, Andrianopoli2}.
  But it is notable that these results have been obtained in the microscopic theories.
  Also, it is known that $\CN=2$ supergravity relates to $\CN=2$ global supersymmetric theory \cite{Ferrara2, David}.
  In this thesis, 
  we will see that in $U(N)$ gauged $\CN=2$ supergravity as low energy effective theory 
  the half of supersymmetry is broken to $\CN=1$ counterpart spontaneously \cite{IM}.
\subsubsection{The Organization of This Thesis}
  The first half of this thesis (chapter 2 $\sim$ chapter 4) 
  is review of the general matter coupled $\CN=2$ supergravity in four dimensions.
  On the other hand, in the latter half of this thesis (chapter 5 and chapter 6), 
  we deal with the partial supersymmetry breaking.
  
  In chapter 2, we define special K\"ahler manifolds and quaternionic K\"ahler manifolds.
  These manifolds describe scalar sector of vector multiplet and hypermultiplet respectively.
  
  In chapter 3, we review the gauging procedure of special and quaternionic K\"ahler manifolds.
  Notion of momentum map which is introduced in section 3.1 plays an important role for the gauging.
  
  Finally, the general matter coupled Lagrangian is given in chapter 4.
  It is consistent with the Bianchi identities which are obtained in appendix B.
  Also, the supersymmetry transformation laws of all the fields are introduced at the end of this chapter.
  
  In chapter 5, 
  we review the simplest $\CN=2$ model where the $\CN=2$ supersymmetry is broken to $\CN=1$ \cite{Ferrara1}.
  This model has $U(1)$ gauged vector multiplet and a hypermultiplet.
  If we choose symplectic section such as no prepotential exists, partial supersymmetry breaking occurs.
  It is accomplished by simultaneous realization of the Higgs and the super-Higgs mechanisms.
  We, also, see that the mass spectrum.
  In this model, the symplectic section has chosen to be the simplest function.
  Therefore, this model is microscopic theory such that higher order coupling terms of the scalar fields are not exist.
  
  Chapter 6 is the main part of this thesis. 
  We extend the model in the chapter 5 to $U(N)$ gauged model.
  We give $\CN=2$ supergravity model coupled to a $U(N)$ gauged vector multiplet and a hypermultiplet.
  In particular, we do not choose the symplectic section as the simplest function.
  It leads to the effective theory which contains higher order coupling terms of the scalar fields.
  We observe that the $\CN=2$ supersymmetry is broken to $\CN=1$ spontaneously.
  Furthermore, 
  by redefining the fluctuations of the scalar fields from the vacuum expectation value as new $\CN=1$ scalar fields,
  we can write down $\CN=1$ supergravity model in terms of the superpotentials.

\subsubsection{Field Contents}
  Before going to the next chapter, we present, here, the field contents of $\CN=2$ supergravity.
  The general matter coupled $\CN=2$ supergravity contains a gravitational multiplet, $m$ vector multiplets 
  and $k$ hypermultiplets.
  All the component fields of these multiplets are massless and describe as follows:
    \begin{itemize}
      \item \textit{a gravitational multiplet}
      \\
      This multiplet is described by the vierbein $ e^i_\mu $ ($i,\mu = 0,1,2,3$), 
      two gravitini $ \psi^A_\mu $ ($A=1,2$) and the graviphoton $ A^0_\mu $.
      (The upper and lower position of the index $A$ represent left and right chirality, respectively.)
      \item \textit{$m$ vector multiplets}
      \\
      Each vector multiplet contain a gauge boson $ A^a_\mu $ ($ a = 1,\dots,m $), 
      two gaugino $ \lambda^{aA} $ and a complex scalar $ z^a $.
      (The chirality notation is opposite to that of gravitinos, that is, 
      the upper and lower position denote right and left chirality, respectively.)
      \item \textit{$k$ hypermultiplets}
      \\
      Each multiplet contain two hyperini $ \zeta^\alpha $ ($ \alpha = 1,\ldots,2k $) 
      and four real scalar $ b^u $ ($ u = 1,\ldots,4k $).
      (The upper and lower position of the index $\a$ represent left and right chirality, respectively.)
    \end{itemize}
  The chirality notations are explained in detail in the appendix A.
\chapter{Special and quaternionic K\"ahler manifolds}
  The vector multiplets and the hypermultiplets contain the different kinds of the scalar fields 
  and they are governed by the different kinds of geometries.
  The scalar field sector of vector multiplets is described by special K\"ahler manifold $\CS\CM$, 
  while the scalar sector of hypermultiplets is described by quaternionic K\"ahler manifold $\CH\CM$.
  Special K\"ahler manifold is Hodge-K\"ahler manifold which has the particular bundle structure.
  Therefore, in section 2.1, we define a Hodge-K\"ahler manifold.
  The definition of the special K\"ahler manifold is given in section 2.2.
  This definition is independent of whether prepotential exists or not.
  We also give a different definition of the special K\"ahler manifold 
  which depend on the existence of the prepotential.
  On the other hand, quaternionic K\"ahler geometry is defined in section 2.3.
  
  There are many works on special and quaternionic K\"ahler geometries.
  Special K\"ahler geometry based on the special coordinates was introduced in \cite{deWitVan, deWitLauwers, Cremmer}.
  The definition of the special K\"ahler geometry in terms of the symplectic bundles is 
  in \cite{Strominger, Castellani, D'Auria, Ceresole, Andrianopoli, Van, Van3, Van4, Van2}.
  On the other hand, the geometric interpretation of the coupling of hypermultiplets
  was introduced in \cite{BaggerWitten} and \cite{deWitLauwers2, Galicki, D'Auria, Andrianopoli}.
  Here we use the notations of \cite{Andrianopoli}.

\section{Hodge-K\"ahler manifolds}
\subsubsection{K\"ahler manifolds}
  First of all, let us see the notations of K\"ahler manifold.
  A K\"ahler manifold is defined as follows.
  Consider a complex manifold $M$ equipped hermitian metric $g$.
  A K\"ahler manifold $\CM$ is a Hermitian manifold $M$ whose K\"ahler 2-form $K$ is closed: $dK=0$.
  In this case, the metric $g$ is called K\"ahler metric of $\CM$.
  The K\"ahler 2-form can be written in terms of the K\"ahler metric as,
    \BE
      K
       =     \frac{i}{2 \pi} g_{ab^*} d z^a \wedge d \bar{z}^{b^*},
    \EE
  where $a = 1, \dots, n$, and $n$ is complex dimension of the K\"ahler manifold.
  The K\"ahler metric is locally given by
    \BE
      g_{ab^*}
       =     \partial_a \partial_{b^*} \CK(z, \bar{z}),
    \EE
  where $\CK$ is real function and is called the K\"ahler potential.
  It is defined up to holomorphic function $f(z)$, that is, the K\"ahler metric is invariant 
  under the K\"ahler transformation:
    \BE
      \CK
       \rightarrow
             \CK + f(z) + \bar{f}(\bar{z}).
             \label{Kahlertr}
    \EE
  
  The only non-vanishing components of the Levi-Civita connection is $\Gamma^a_{bc}$ and $\Gamma^{a^*}_{b^* c^*}$.
  With these components, for example, the covariant derivatives of vector $V_a$ are
    \BEA
      \nabla_a V_b
      &=&    \partial_a V_b + \Gamma_{ab}^c V_c,
             \NN \\
      \nabla_{a^*} V_b
      &=&    \partial_{a^*} V_b.
    \EEA
  We require the metric compatibility which can be written, in components, as
    \BE
      \nabla_a g_{bc^*}
       =     0,
    \EE
  which leads to the following relations:
    \BEA
      \Gamma^a_{bc}
      &=&  - g^{ad^*} \partial_c g_{bd^*},
             \NN \\
      \Gamma^{a^*}_{b^* c^*}
      &=&  - g^{a^* d} \partial_{c^*} g_{b^* c}.
             \label{LeviCivita}
    \EEA
  Furthermore, the non-vanishing components of curvature 2-form are written as
    \BEA
      R^a_{~b}
      &=&    R^a_{~bc^*d} d \bar{z}^{c^*} \wedge dz^d,
             ~~~~~
      R^a_{~bc^*d}
       =     \partial_{c^*} \Gamma^a_{bd},
             \NN \\
      R^{a^*}_{~b^*}
      &=&    R^{a^*}_{~b^* c d^*} d z^c \wedge d \bar{z}^{d^*},
             ~~~~~
      R^{a^*}_{~b^* c d^*}
       =     \partial_{c} \Gamma^{a^*}_{b^* d^*}.
    \EEA
\subsubsection{Hodge-K\"ahler manifolds}
  Let us consider a line bundle $ \CL \xrightarrow{\pi} \CM $ over a K\"ahler manifold.
  This is the holomorphic vector bundle whose fibre is $\mathbb{C}^1$ 
  and its structure group is subgroup of $GL(1, \mathbb{C})$.
  Thus, the Chern class is written as,
    \BE
      c(\CL)
       =     1 + \frac{i}{2\pi} F,
    \EE
  where $F$ is the field strength of the $GL(1, \mathbb{C})$ connection.
  Since $c_j=0$ for $j>1$, the only available Chern class is $c_1$:
    \BE
      c_1(\CL)
       =     \frac{i}{2\pi} F
       =     \frac{i}{2\pi} \bar{\partial} \theta
       =     \frac{i}{2\pi} \bar{\partial} (h^{-1} \partial h)
       =     \frac{i}{2\pi} \bar{\partial} \partial \log h,
             \label{c1}
    \EE
  where $h(z, \bar{z})$ and $\theta$ are the hermitian metric and the canonical hermitian connection 
  on $\CL$ respectively. 
  In (\ref{c1}), we have used the relation between $h(z, \bar{z})$ and $\theta$:
    \BEA
      \theta
      &=&    h^{-1} \partial h 
       =     h^{-1} \partial_a h dz^a,
             \nonumber \\
      \bar{\theta}
      &=&    h^{-1} \bar{\partial} h
       =     h^{-1} \partial_{a^*} h d\bar{z}^{a^*}.
             \label{metricconnection}
    \EEA
  Let $f(z)$ be a holomorphic section of $\CL$. 
  The norm of $f(z)$ is given as $||f(z)||=h(z, \bar{z}) \bar{f}(\bar{z}) f(z)$.
  Since under the action of the operator $\bar{\partial} \partial$ 
  the term $\log (\bar{f}(\bar{z}) f(z))$ yields a vanishing contribution, 
  we can rewrite (\ref{c1}) as
    \BE
      c_1(\CL)
       =     \frac{i}{2\pi} \bar{\partial} \partial \log || ~f(z)~ ||^2.
             \label{C1}
    \EE
  \\
  \textit{Definition 2.1.}
  A K\"ahler manifold $\CM$ is a Hodge-K\"ahler manifold $\CM_{\rm{Hodge}}$
  if there exists a line bundle $ \CL \xrightarrow{\pi} \CM $ 
  whose first Chern class is equivalent to the cohomology class of the K\"ahler 2-form $K$, 
    \BE
      c_1(\CL) 
       =     [K].
             \label{K}
    \EE
  Eq. (\ref{K}) implies that there is a holomorphic section $W(z)$ such that:
    \BEA
      K
      &=&    \frac{i}{2\pi} g_{ab^*} dz^a \wedge  d\bar{z}^{b^*}
             \NN \\
      &=&    \frac{i}{2\pi} \bar{\partial} \partial \log || ~W(z)~ ||^2
             \NN \\
      &\equiv&
             \frac{i}{2\pi} \partial_a \partial_{b^*} \log || ~W(z)~ ||^2 d z^a \wedge  d \bar{z}^{b^*}.
             \label{Kahlerform}
    \EEA
  It follows from (\ref{Kahlerform}) that if the manifold $\CM$ is a Hodge-K\"ahler manifold, 
  then the exponential of the K\"ahler potential can be interpreted as the metric on a line bundle $\CL$:
    \BE
      h(z, \bar{z}) 
       =     \exp(\CK(z, \bar{z})).
             \label{Hodgecondition}
    \EE
  By substituting (\ref{Hodgecondition}) into (\ref{metricconnection}), we obtain
    \BEA
      \theta
      &=&    \partial \CK
       =     \partial_a \CK dz^a,
             \NN \\
      \bar{\theta}
      &=&    \bar{\partial} \CK
       =     \partial_{a^*} \CK d\bar{z}^{a^*}.
    \EEA
  In the $\CN=1$ supergravity, $c_1(\CK)=[K]$ and this restricts $\CM$ to be a Hodge-K\"ahler manifold.
\subsubsection{Principal $U(1)$ bundle}
  It is known that there exists a correspondence between line bundles and principal $U(1)$ bundles. 
  Thus, the covariant derivatives with respect to the canonical connection of the line bundle are related 
  to those of the $U(1)$ bundle.
  We can express this correspondence, in terms of the canonical connection, as 
    \BE
      \CQ
       =     \Im \theta
       =   - \frac{i}{2} (\theta - \bar{\theta}),
    \EE
  where $\CQ$ is the canonical connection on $U(1)$ bundle $\CU \rightarrow \CM$.
  If we apply the above formula to the case of the $U(1)$ bundle $\CU \rightarrow \CM$ 
  associated with Hodge-K\"ahler manifold, that is, 
  the line bundle $\CL$ whose first Chern class equals the K\"ahler 2-form, we get
    \BE
      \CQ
       =   - \frac{i}{2} 
             \left(
             \partial_a \CK dz^a - \partial_{a^*} \CK d\bar{z}^{a^*}
             \right).
             \label{Q}
    \EE
  
  Let us consider a principal $U(1)^p$ bundle.
  Let $\Phi(z, \bar{z})$ be a section of $\CU^p$. 
  Its covariant derivative is
    \BE
      \nabla \Phi
       =     (d + i p \CQ) \Phi,
             \label{U1bundlesection}
    \EE
  or, in components,
    \BEA
      \nabla_a \Phi
      &=&    (\partial_a + \frac{1}{2} p \partial_a \CK ) \Phi,
             \nonumber \\
      \nabla_{a^*} \Phi
      &=&    (\partial_{a^*} - \frac{1}{2} p \partial_{a^*} \CK) \Phi.
             \label{U1bundleCD}
    \EEA
  A covariantly holomorphic section of $\CU$ is defined by the equation: $\nabla_{a^*} \Phi =0$.
  We can easily map each section $\Phi(z, \bar{z})$ of $\CU^p$ 
  into a section $\tilde{\Phi}$ of the line bundle $\CL$ by setting,
    \BE
      \tilde{\Phi}
       =     e^{- p \CK/2} \Phi
    \EE
  Under the map covariantly holomorphic sections of $\CU$ flow into holomorphic section of $\CL$ and viceversa, 
  that is,
    \BE
      \partial_{a^*} \tilde{\Phi}
       =     e^{- p \CK/2} \nabla_{a^*} \Phi
    \EE
\section{Special K\"ahler manifolds}
  The scalar field sector of vector multiplets in $\CN=2$ supergravity is described 
  by special K\"ahler geometry of local type.
  Suppose that there are $m$ vector multiplets, 
  so there exist $m$ complex scalar fields $z^a$, $a=1,\ldots,m$.
  It is known that these scalar fields span a special K\"ahler manifold $\CS \CM$.
  The number of the complex scalar fields, $m$, corresponds to the complex dimension of the special K\"ahler manifold.
  
  Here we only give definition of the special K\"ahler manifold.
  Thus, there is no guarantee that the vector multiplet scalar sector of $\CN=2$ supergravity is described 
  by the special K\"ahler geometry.
  However, in the appendix B, we will confirm that it is true.
\subsection{Definitions}
  We consider, in addition to the line bundle $\CL$ which has been introduced in the Hodge-K\"ahler manifold, 
  the bundle $\CS\CV$ as follows:
  $ \CS\CV \rightarrow \CS\CM $ denotes a holomorphic flat vector bundle 
  with structure group $Sp(2m + 2, \mathbb{R})$.
  Consider tensor bundle of the type $\CH = \CS\CV \otimes \CL$.
  A holomorphic section $\Omega$ of such a bundle will have the following structure,
    \BE
      \Omega(z)
       =     \left(
             \begin{array}{c}
             X^{\Lambda} (z) 
             \\
             F_{\Sigma} (z)
             \end{array}
             \right),
             ~~~~~\Lambda,\Sigma = 0,1,\ldots,m.
    \EE
  \\
  \textit{Definition 2.2.}
  A special K\"ahler manifold $\CS\CM$ of the local type is an $m$-dimensional Hodge-K\"ahler manifold $\CM_{\rm{Hodge}}$ 
  if there exists a bundle $\CH$ with the following properties.
    \begin{enumerate}
      \item For some holomorphic section $\Omega$, the K\"ahler 2-form is given by
        \BE
         K 
          =   - \frac{i}{2\pi}
                \bar{\partial} \partial \log (i \langle \Omega | \bar{\Omega} \rangle).
                \label{K}
        \EE
      \item On overlaps of charts $ i $ and $ j $, 
            the symplectic section $\Omega$ are connected by transition functions of the following form
              \BE
                \Omega_{(i)}
                 =    \left(
                      \begin{array}{c}
                      X^\Lambda \\ 
                      F_\Sigma
                      \end{array}
                      \right)_{(i)}
                 =    e^{f_{ij}(z)} M_{ij} 
                      \left(
                      \begin{array}{c}
                      X^\Lambda \\
                      F_\Sigma
                      \end{array}
                      \right)_{(j)},
                      \label{symplecticvector}
              \EE
            with $ f_{ij} $ holomorphic and $ M_{ij} \in Sp(2m + 2, \mathbb{R})$ ,
      \item The transition functions satisfy the following cocycle conditions on overlap regions of three charts:
              \BEA
                e^{f_{ij} f_{jk} f_{ki}}
                &=&    1, \NN \\
                M_{ij} M_{jk} M_{ki}
                &=&    1.
                       \label{cocyclecondition}
              \EEA
    \end{enumerate}
  Eq. (\ref{K}) implies 
  that we have a local expression for the K\"ahler potential in terms of the holomorphic section $\Omega$:
    \BE
      \CK
       =   - \log (i\langle \Omega | \bar{\Omega} \rangle)
       =   - \log [i(\bar{X}^\Lambda F_\Lambda - \bar{F}_\Lambda X^\Lambda)].
             \label{CK}
    \EE

  We introduce a non-holomorphic section $V$ of the bundle $\CH$ according to,
    \BE
      V
       =     \left(
             \begin{array}{c}
             L^\Lambda   
             \\
             M_\Sigma  
             \end{array}
             \right)
       \equiv
             e^{\CK/2} \Omega
       =     e^{\CK/2}
             \left(
             \begin{array}{c}
             X^\Lambda   
             \\
             F_\Sigma  
             \end{array}
             \right),
             \label{sectionV}
    \EE
  so that (\ref{CK}) becomes
    \BE
      1
       =     i \langle V | \bar{V} \rangle
       =     i (\bar{L}^\Lambda M_\Lambda - \bar{M}_\Lambda L^\Lambda).
             \label{sectionV2}
    \EE
  It can be seen as the section on the associated $U(1)$ bundle (\ref{U1bundlesection}).
  Therefore, by using (\ref{U1bundleCD}), its covariant derivatives are written as, 
    \BEA
      U_a
      &\equiv&
             \nabla_a V
       =     \left(
             \partial_a + \frac{1}{2} \partial_a \CK
             \right) V
       =     \left(
             \begin{array}{c}
             f^\Lambda_a
             \\
             h_{\Sigma | a}
             \end{array}
             \right),
             \label{Ua}
             \\
      U_{a^*}
      &\equiv&
             \nabla_{a^*} V
       =     \left(
             \partial_{a^*} - \frac{1}{2} \partial_{a^*} \CK
             \right) V
       =     0,
             \\
      \bar{U}_{a^*}
      &\equiv&
             \nabla_{a^*} \bar{V}
       =     \left(
             \partial_{a^*} + \frac{1}{2} \partial_{a^*} \CK
             \right) \bar{V}
       =     \left(
             \begin{array}{c}
             \bar{f}^\Lambda_{a^*}
             \\
             \bar{h}_{\Sigma | a^*}
             \end{array}
             \right),
    \EEA
  where $\bar{U}_{a^*}$ has been defined to be complex conjugate of $U_a$.
  At this stage, we can derive the following equations
    \BEA
      \langle V | U_a \rangle
      &=&    \langle V | U_{a^*} \rangle
       =     \langle V | \bar{U}_{a^*} \rangle
       =     0,
             \label{VU}
             \\
      g_{ab^*}
      &=&  - i \langle U_a | \bar{U}_{b^*} \rangle,
             \\
      \nabla_{[a} U_{b]}
      &=&    0,
    \EEA
  where $\nabla_a$ denotes the covariant derivative containing both the canonical connection $\theta$ 
  on the line bundle $\CL$ and the Levi-Civita connection:
    \BE
      \nabla_a U_b
       =     \partial_a U_b
           + \frac{1}{2} \partial_a \CK U_b
           + \Gamma^c_{ab} U_c.
             \label{nablaU}
    \EE
  Defining
    \BE
      C_{abc}
       =     \langle \nabla_a U_b | U_c \rangle,
    \EE
  we can see that it is completely symmetric in its indices.
  With this, (\ref{nablaU}) can be written as,
    \BE
      \nabla_a U_b
       =     i C_{abc} g^{cd^*} \bar{U}_{d^*}.
             \label{Cabc}
    \EE
  \\
  \textit{Lemma 2.1.}
  The completely symmetric tensor $C_{abc}$ is covariantly holomorphic, namely, 
    \BE
      \nabla_{d^*} C_{abc}
       =     0.
    \EE
  \\
  \textit{Proof.}
  First of all, we have to observe that $U_a$ satisfies the following condition:
    \BEA
      \langle U_a | U_b \rangle
      &=&    \nabla_b \langle U_a | V \rangle
           - \langle \nabla_b U_a | V \rangle
             \NN \\
      &=&  - i C_{abc} g^{cd^*} \langle \bar{U}_{d^*} | V \rangle
             \NN \\
      &=&    0,
             \label{Cabc2}
    \EEA
  where we have used (\ref{VU}) and (\ref{Cabc}).
  We can also evaluate the covariant derivatives, $\nabla_a \nabla_{b^*} U_c$ and $\nabla_{b^*} \nabla_a U_c$, 
  as follows:
    \BEA
      \nabla_a \nabla_{b^*} U_c
      &=&    \partial_a (\nabla_{b^*} U_c) 
           + \frac{1}{2} \partial_a \CK (\nabla_{b^*} U_c)
           + \Gamma^d_{ac} (\nabla_{b^*} U_d)
             \NN \\
      &=&    \partial_a \partial_{b^*} U_c 
           - \frac{1}{2} \partial_a (\partial_{b^*} \CK U_c)
           + \frac{1}{2} \partial_a \CK \nabla_{b^*} U_c
           + \Gamma^d_{ac} \nabla_{b^*} U_d,
             \NN \\
      \nabla_{b^*} \nabla_a U_c
      &=&    \partial_{b^*} (\nabla_a U_c)
           - \frac{1}{2} \partial_{b^*} \CK (\nabla_a U_c)
             \NN \\
      &=&    \partial_a \partial_{b^*} U_c 
           + \frac{1}{2} g_{ab^*} U_c
           - \frac{1}{2} \partial_{b^*} \partial_a U_c
           + \frac{1}{2} \partial_a \nabla_{b^*} U_c
           + \Gamma^d_{ac} \nabla_{b^*} U_d
           + (\partial_b^* \Gamma^d_{ac}) U_d,
             \NN
    \EEA
  which lead to
    \BE
      [ \nabla_a \nabla_{b^*} - \nabla_{b^*} \nabla_a ] U_c
       =  - g_{ab^*} U_c - (\partial_b^* \Gamma^d_{ac}) U_d.
    \EE
  Thus, using (\ref{Cabc2}), we derive
    \BEA
      \nabla_{d^*} C_{abc}
      &=&    \langle \nabla_{d^*} \nabla_a U_b | U_c \rangle
           + \langle \nabla_a U_b | \nabla_{d^*} U_c \rangle
             \NN \\
      &=&    \langle (\nabla_a \nabla_{d^*} U_b | U_c \rangle
           + i C_{abe} g^{ef^*} g_{cd^*} \langle \bar{U}_{f^*} | V \rangle
             \NN \\
      &=&    0,
    \EEA
  in the last equality, we have used $\nabla_a \nabla_{d^*} U_b = \nabla_a (g_{bd^*} V) = g_{bd^*} U_a$.
  Therefore, the lemma 2.1 has been proved.
  \hspace{330pt}
   
  \vspace{11pt}
  
  We have to introduce one more important quantity, the generalized gauge coupling matrix $\CN$ 
  which will appear in the Lagrangian of the $\CN=2$ supergravity
    \footnote[2]{We will see it in the chapter 4.}.
  It is introduced by the following relations,
    \begin{equation}
      \bar{M}_\Lambda 
       =    \bar{\mathcal{N}}_{\Lambda \Sigma} \bar{L}^\Sigma,~~~~
      h_{\Lambda \vert a}
       =    \bar{\mathcal{N}}_{\Lambda \Sigma} f^\Sigma_a,
    \end{equation}
  which can be solved constructing the two $ (N +1) \times (N + 1) $ matrices
    \begin{equation}
      f^\Lambda_I
       =    \left(\begin{array}{cc}
            f^\Lambda_a \\
            \bar{L}^\Lambda \\
            \end{array}
            \right),~~~~
     h_{\Lambda \vert I}
      =     \left(\begin{array}{cc}
            h_{\Lambda \vert a} \\
            \bar{M}_\Lambda \\
            \end{array}
            \right),
    \end{equation}
  and setting:
    \begin{equation}
      \bar{\mathcal{N}}_{\Lambda \Sigma}
       =    h_{\Lambda \vert I} \circ  (f^{-1})^I_\Sigma.
            \label{CouplingMatrix}
    \end{equation}
    
  From the previous formulae it is easy to derive a set of useful relations:
    \BEA
      (\Im \CN_{\Lambda \Sigma}) L^\Lambda \bar{L}^\Sigma
      &=&  - \frac{1}{2},
             \\
      U^{\Lambda \Sigma}
       \equiv
             f^\Lambda_a \bar{f}^\Sigma_{b^*} g^{ab^*}
      &=&  - \frac{1}{2} (\Im \CN)^{-1 | \Lambda \Sigma}
           - \bar{L}^\Lambda L^\Sigma,
             \\
      C_{abc}
      &=&    f^\Lambda_a \partial_b \bar{\CN}_{\Lambda \Sigma} f^\Sigma_c
       =     (\CN - \bar{\CN})_{\Lambda \Sigma} f^\Lambda_a \partial_b f^\Sigma_c,
    \EEA
\subsection{The holomorphic prepotential}
  So far we have not mentioned about the holomorphic prepotential $F(X)$.
  Indeed, when the definition of special K\"ahler manifolds is given in intrinsic terms, 
  as we did in the previous section, the holomorphic prepotential $F$ can be dispense of.
  Actually, it appears that some physically interesting cases (for example, partial supersymmetry breaking) 
  are precisely instances where $F(X)$ does not exist.
  
  However, we give the definition of special K\"ahler manifold,
  which depends on the existence of the prepotential here.
  \vspace{12pt}
  \\
  \textit{Definition 2.3.}
  A special K\"ahler manifold is an $ m $-dimensional Hodge-K\"ahler manifold with the following properties.
    \begin{enumerate}
      \item On every chart there exist complex projective coordinate functions $ X^{\Lambda}(z) $, 
            where $ \Lambda = 0,\ldots,m $ 
            and a holomorphic function (prepotential) $ F(X^{\Lambda}) $ which is homogeneous of second degree, 
            such that the K\"ahler potential is
              \BE
                \CK 
                 =  - \log i
                      \left[
                      \bar{X}^{\Lambda} \frac{\partial }{\partial X^{\Lambda}} F(X) 
                    - X^{\Lambda} \frac{\partial }{\partial \bar{X}^{\Lambda}} \bar{F}(\bar{X})
                      \right],
              \EE
      \item On overlaps of charts $ i $ and $ j $, 
            the symplectic vector $\Omega$ which is constructed from $ X $ and $ F $ in property 1 are connected 
            by transition functions of the following form
              \BE
                \Omega_{(i)}
                 =    \left(
                      \begin{array}{c}
                      X \\ 
                      \partial F
                      \end{array}
                      \right)_{(i)}
                 =    e^{f_{ij}(z)} M_{ij} 
                      \left(
                      \begin{array}{c}
                      X \\
                      \partial F
                      \end{array}
                      \right)_{(j)},
                      \label{symplecticvector}
              \EE
            with $ f_{ij} $ holomorphic and $ M_{ij} \in Sp(2m + 2, \mathbb{R})$ ,
      \item The transition functions satisfy the cocycle conditions (\ref{cocyclecondition}) 
            on overlap regions of three charts.
    \end{enumerate}
  This definition of a special K\"ahler manifold clearly depends on the existence of the prepotential $ F(X) $.
  In reference of \cite{Van}, it has been proved that these definitions are equivalent each other.
  
\section{Quaternionic K\"ahler manifolds}
  Let us turn to the hypermultiplet sector.
  There are $4$ real scalar fields for each hypermultiplet and, at least locally, 
  they can be regarded as $4$ components of a quaternion.
  These scalar fields span a quaternionic manifold $\CH \CM$.
  If we have $k$ hypermultiplets, the manifold $\CH\CM$ has dimension $4k$.
  
  Let us consider a $4k$-dimensional real manifold with a metric $h$:
    \BE
      d s^2
       =     h_{uv}(b) db^u \otimes db^v
             ~~~;~~~
             u, v = 1,\ldots, 4m,
    \EE
  and three complex structures $J^x$ that satisfy the quaternionic algebra
    \BE
      J^xJ^y
       =  - \delta^{xy} \I + \epsilon^{xyz} J^z,
            \label{quaternionicalgebra}
    \EE
  and respect to which the metric is hermitian: for any tangent vector $X,Y$ on $\CH \CM$,
    \BE
      h(J^x X, J^y Y)
       =     h(X, Y).
             \label{JXJY}
    \EE
  From (\ref{JXJY}), it follows that one can introduce a triplet of 2-forms
    \BEA
      K^x
      &=&    K^x_{uv} db^u \wedge db^v,
             \nonumber \\
      K^x_{uv}
      &=&    h_{uw} (J^x)^w_v
             \label{hyperKahlerform}
    \EEA
  The triplet of 2-forms, $K^x$, is named the hyperK\"ahler form.
  It provides the generalization of the K\"ahler form introduced in the complex case.
  It is an $SU(2)$ Lie-algebra valued 2-form in the same way as the K\"ahler form is a $U(1)$ Lie-algebra valued 2-form.
  In the complex case, the definition of K\"ahler manifold involves the statement that the K\"ahler 2-form is closed 
  and, in Hodge-K\"ahler manifold, the K\"ahler 2-form can be identified with the curvature of a line bundle.
  Similar steps can be taken also here which lead to quaternionic manifolds.
  
  Consider a principal $SU(2)$ bundle $\CS\CU \rightarrow \CH\CM$
  that play for hypermultiplets the same role played by the line bundle $\CL \rightarrow \CS\CM$ 
  in the case of vector multiplets.
  Let $\omega^x$ denote a connection on such a bundle.
  To obtain a quaternionic manifold we must impose the condition that the hyperK\"ahler 2-form is covariantly closed 
  with respect to the connection $\omega^x$
    \BE
      \nabla K^x
       \equiv
             d K^x
           + \epsilon^{xyz} \omega^y \wedge K^z
       =     0.
    \EE
  Furthermore, we define the $\CS \CU$ curvature by
    \BE
      \Omega^x
       =     d \omega^x
           + \frac{1}{2} \epsilon^{xyz} \omega^y \wedge \omega^z.
    \EE
  \\
  \textit{Definition 2.4.}
  A quaternionic manifold $\CH \CM$ is a $4m$-dimensional manifold with the structure described above 
  and such that the curvature of the $\CS\CU$ bundle is proportional to the HyperK\"ahler 2-form as
    \BE
      \Omega^x
       =     \lambda K^x.
             \label{Omega}
    \EE
  where $\lambda$ is a non-vanishing real number.
\subsubsection{Holonomy}
  As a consequence of the above structure, 
  the holonomy group of the manifold $\CH\CM$ is a subgroup of $Sp(2) \times Sp(2k, \mathbb{R})$.
  If we introduce flat indices $A,B=1,2;~\alpha,\beta=1,\ldots,2k$ that run  
  in the fundamental representations of $SU(2)$ and $Sp(2k, \mathbb{R})$ respectively,
  we can find vielbein 1-form
    \BE
      \CU^{A \alpha}
       =     \CU^{A \alpha}_u db^u,
    \EE
  such that
    \BE
      h_{uv}
       =     \CU^{A \alpha}_u \CU^{B \beta}_v \mathbb{C}_{\alpha \beta} \e_{AB},
             \label{huv}
    \EE
  where $\mathbb{C}_{\alpha \beta} = - \mathbb{C}_{\beta \alpha}$ and $\e_{AB} = - \e_{BA}$ are, respectively, 
  the flat $Sp(2m)$ and $Sp(2) \sim SU(2)$ invariant metrics
    \footnote[2]{Notations are fixed in the appendix A}.
  
  The vielbein $\CU^{A \alpha}$ is covariantly closed with respect to the $SU(2)$ connection $w^x$ 
  and to some $Sp(2k, \mathbb{R})$ Lie algebra valued connection $\Delta^{\alpha \beta}=\Delta^{\beta \alpha}$:
    \BE
      \nabla \CU^{A \alpha}
       \equiv     
             d \CU^{A \alpha} + \frac{i}{2} \omega^x (\e \sigma_x \e^{-1})^A_{~B} \wedge \CU^{B \alpha}
           + \Delta^{\alpha \beta} \wedge \CU^{A \gamma} \mathbb{C}_{\beta \gamma}
       =     0,
    \EE
  where $(\sigma^x)_A^{~B}$ are the standard Pauli matrices
    \footnote[8]{Also, see appendix A}
  and $Sp(2k, \mathbb{R})$ Lie algebra valued connection $\Delta^{\a \b}$ satisfies:
    \BEA
      \Delta_\alpha^{~\beta}
      &\equiv&
             \Delta^{\gamma \beta} \mathbb{C}_{\gamma \alpha},
             \NN \\
      \Delta^\alpha_{~\beta}
      &\equiv&
             \mathbb{C}_{\beta \gamma} \Delta^{\alpha \gamma}.
             \label{Spconnection}
    \EEA
  Furthermore, $\CU^{A \alpha}$ satisfies the reality condition:
    \BE
      \CU_{A \alpha}
       \equiv
             (\CU^{A \alpha})^*
       =     \e_{AB} \mathbb{C}_{\alpha \beta} \CU^{B \beta}.
             \label{vielbeinU}
    \EE
  Eq.(\ref{vielbeinU}) defines the rule to lower the symplectic indices 
  by means of the flat symplectic metrics $\e_{AB}$ and $ \mathbb{C}_{\alpha \beta}$.
  More specifically we can write a stronger version of (\ref{huv}) \cite{BaggerWitten}:
    \BEA
      (\CU^{A \alpha}_u \CU^{B \beta}_v + \CU^{A \alpha}_v \CU^{B \beta}_u) \mathbb{C}_{\alpha \beta}
      &=&    h_{uv} \e^{AB},
             \\
      (\CU^{A \alpha}_u \CU^{B \beta}_v + \CU^{A \alpha}_v \CU^{B \beta}_u) \e_{AB}
      &=&    h_{uv} \frac{1}{m} \mathbb{C}^{\alpha \beta}.
    \EEA
  We have also the inverse vierbein $\CU^u_{A \alpha}$ defined by the following equation:
    \BE
      \CU^u_{A \alpha} \CU_v^{A \alpha} 
       =     \delta^u_v,
    \EE
  Flatting a pair of indices of the Riemann tensor $\CR^{uv}_{~~ts}$ we obtain
    \BE
      \CR^{uv}_{~~ts} \CU^{A \alpha}_u \CU^{B \beta}_v
       =   - \frac{i}{2} \e^{AC} R_{C|ts}^{~B} \mathbb{C}^{\alpha \beta}
           + \mathbb{R}^{\alpha \beta}_{ts} \e^{AB},
             \label{CR}
    \EE
  where $R_{A|uv}^{~B}$ and $\mathbb{R}^{\alpha \beta}_{uv}$ are the curvature of the $SU(2)$ connection
  and the $Sp(2m)$ connection respectively:
    \BEA
      R_A^{~B}
      &=&    R_{A | uv}^{~B} db^u \wedge db^v,
             \NN \\
      &\equiv&
             \Omega^x_{uv} (\sigma_x)_A^{~B},
             \label{SU(2)curvature}
             \\
      \mathbb{R}^{\alpha \beta}
      &=&    \mathbb{R}^{\alpha \beta}_{ts} db^t \wedge db^s
             \nonumber \\
      &\equiv&
             d \Delta^{\alpha \beta} 
           + \Delta^{\alpha \gamma} \wedge \Delta^{\delta \beta} \mathbb{C}_{\gamma \delta}.
             \label{Spcurvature}
    \EEA
  Eq.(\ref{CR}) is the explicit statement 
  that the Levi-Civita connection associated with the metric $h$ has a holonomy group contained in $SU(2)\otimes Sp(2m)$.
  
  Let us consider (\ref{quaternionicalgebra}), (\ref{hyperKahlerform}) and (\ref{Omega}), 
  so we easily obtain the following relation:
    \BE
      h^{st} K^x_{us} K^y_{tw}
       =   - \delta^{xy} h_{uw}
           + \e^{xyz} K^z_{uw}.
             \label{K1}
    \EE
  By using (\ref{Omega}), (\ref{K1}) can be rewritten as follows:
    \BE
      h^{st} \Omega^x_{us} \Omega^y_{tw}
       =   - \lambda^2 \delta^{xy} h_{uw}
           + \lambda \e^{xyz} \Omega^z_{uw}.
             \label{K2}
    \EE
\chapter{The gauging}
  In this chapter, we discuss the gauging procedure.
  The gauge group is identified with a subgroup of isometries of the product manifold $\CS \CM \times \CH \CM$.
  The isometries are generated by Killing vectors.
  Also, the Killing vectors are written by the Killing potential which is, in geometrical point of view, 
  identical with the momentum map providing the Poisson realization of Lie algebra on the manifold.
  
  Firstly, we review about the Killing vectors and the Killing potential.
  We also see the notion of the momentum map.
  Furthermore, the construction of the momentum map on special K\"ahler manifolds or quaternionic K\"ahler manifolds 
  is considered respectively.
  Finally, we gauge all the connections which have been given in the last chapter.
  These are based on the works of \cite{Galicki, D'Auria, Andrianopoli}.
\section{The momentum map}
\subsection{The Killing vectors}
  Consider a K\"ahler manifold $\CM$ of complex dimension $m$.
  A vector field $X$ is said a Killing vector field if an infinitesimal displacement $\e X$ 
  generates an isometry, that is, under the displacement the K\"ahler metric is invariant:
    \BE
      (\CL_X g)_{\mu \nu}
       =     0,
             \label{Killing1}
    \EE
  where $\CL_X$ is the Lie derivative along the vector field $X$.
  Let $k^a_\Lambda$ ($a = 1 \dots m$) be a component of holomorphic Killing vectors, that is, 
    \BEA
      X
      &=&    a^\Lambda k_\Lambda,
             \NN \\
      k_\Lambda
      &=&    k^a_\Lambda \frac{\partial}{\partial z^a} + k^{a^*}_\Lambda \frac{\partial}{\partial z^{a^*}},
    \EEA
  where $k_\Lambda$ is a basis of the Killing vectors.
  The above statement can be rewritten such that, under the infinitesimal holomorphic coordinate transformations
    \BE
      \delta z^a
       =     \e^\Lambda k_\Lambda^a (z),
    \EE
  the K\"ahler metric is invariant.
  
  Eq.(\ref{Killing1}) implies the following Killing equations:
    \BEA
      \nabla_a k_{b \Lambda} + \nabla_b k_{a \Lambda}
      &=&    0,
             \nonumber \\
      \nabla_{a^*} k_{b \Lambda} + \nabla_b k_{a^* \Lambda}
      &=&    0,
             \label{Killing5}
    \EEA
  where we have defined as $k_b = g_{ba^*} k^{a^*}$, $k_{a^*}= g_{a^* b} k^b$.
  Holomorphicity of the Killing vectors means the following differential constraints:
    \BEA
      \partial_{b^*} k^a_\Lambda (z)
      &=&    0,
             \NN \\
      \partial_b k^{a^*}_\Lambda (z)
      &=&    0.
    \EEA
  
  Let us consider a compact Lie group $G$ acting on $\CM$ by means of Killing vector fields $X$.
  From (\ref{Killing1}), we obtain for the K\"ahler 2-form,
    \BE
      0
       =     \CL_X K
       =     i_X d K + d(i_X K)
       =     d(i_X K),
             \label{Killing2}
    \EE
  where $i_X$ denotes the contraction with $X$.
  In the third equality, we have used the fact that the K\"ahler 2-form is closed.
  In general, if $\CM$ is simply connected, first de Rham cohomology group is trivial
    \footnote[2]{see, for example, chapter 6 of \cite{Nakahara}}.
  Thus, the closed one form is also exact.
  In this case, the one form $i_X K$ is exact, then we can write
    \BE
      - \frac{1}{2\pi} d \CP_{X}
       =     i_{X} K,
             \label{Killing3}
    \EE
  where a function $\CP_{X}$ is defined up to a constant.
  If we expand $X = a^\Lambda k_\Lambda$ in a basis of Killing vectors $k_\Lambda$, 
  we can also expand $\CP_X$ as,
    \BE
      \CP_X
       =     a^\Lambda \CP_\Lambda.
    \EE
  Using this basis, we can rewrite (\ref{Killing3}) in components
    \BEA
      k_\Lambda^a
      &=&    i g^{ab^*} \partial_{b^*} \CP_\Lambda,
             \NN \\
      k_\Lambda^{a^*}
      &=&  - i g^{a^* b} \partial_{b} \CP_\Lambda.
             \label{Killing4}
    \EEA
  The function $\CP_\Lambda$ is called Killing potential.
  
  Indeed the Killing vectors which are written as (\ref{Killing4}) satisfy the Killing equations (\ref{Killing5})
  The first equation of (\ref{Killing5}) is automatically satisfied, because
    \BEA
      \nabla_a k_b
      &=&    \partial_a k_b + \Gamma^c_{ab} k_c
             \NN \\
      &=&    (\partial_a g_{bd^*} + \Gamma^c_{ab} g_{cd^*} + \Gamma^{c^*}_{ad^*} g_{bc^*}) k^{d^*}
             \NN \\
      &=&    (\nabla_a g_{bd^*}) k^{d^*}
             \NN \\
      &=&    0.
    \EEA
  where we have used $\Gamma^{c^*}_{ad^*}=0$.
  Also, since (\ref{Killing4}) implies
    \BEA
      \nabla_{a^*} k_{b \Lambda}
      &=&  - i \partial_{a^*} \partial_b \CP_\Lambda,
             \NN \\
      \nabla_b k_{a^* \Lambda}
      &=&    i \partial_b \partial_{a^*} \CP_\Lambda,
    \EEA
  thus the second equation of (\ref{Killing5}) is satisfied.
  In other words if we can find a function $\CP_\Lambda$ 
  such that the expression $i g^{ab^*} \partial_{b^*} \CP_\Lambda$ is holomorphic,
  then (\ref{Killing4}) defines a Killing vector.
  
  If we substitute $g_{ab^*}=\partial_a \partial_{b^*} \CK$ into (\ref{Killing4}),
  we obtain an expression for the Killing potential in terms of the K\"ahler potential,
    \BE
      i \CP_\Lambda 
       =     \frac{1}{2} (k_\Lambda^a \partial_a \CK - k_\Lambda^{a^*} \partial_{a^*} \CK)
       =     k_\Lambda^a \partial_a \CK
       =   - k_\Lambda^{a^*} \partial_{a^*} \CK.
             \label{momentmap1}
    \EE
  Eq. (\ref{momentmap1}) is true 
  if the K\"ahler potential is exactly invariant under the transformations of the isometry group $G$ 
  and not only up to a K\"ahler transformation as defined in (\ref{Kahlertr}).
  In other words (\ref{momentmap1}) is true if
    \BE
      0 
       =     \CL_\Lambda \CK
       =     k_\Lambda^a \partial_a \CK + k_\Lambda^{a^*} \partial_{a^*} \CK.
             \label{momentmap2}
    \EE
  Note that not all the isometries of a general K\"ahler manifold have such a property.
\subsection{The momentum map}
  The construction of the Killing potential can be stated in a more precise geometrical formulation 
  which involves the notion of momentum map.
  Let us review this construction which reveals another deep connection between supersymmetry and geometry.
  
  Let us firstly prove the following relation:
    \BE
      X (\CP_{Y})
       =     \CP_{[X,Y]}.
             \label{equivariance}
    \EE
  We refer to this as equivariance relation.
  \vspace{11pt}
  \\
  \textit{Proof.} 
  Since the K\"ahler 2-form is closed, we get
    \BEA
      0
       =     d K (X, Y, Z)
      &=&    X (K(Y, Z)) - Y (K(X,Z)) + Z (K(X,Y))
             \NN \\
      &=&  - K([X, Y], Z) + K([X,Z],Y) -K([Y,Z], X).
    \EEA
  On the other hand, using (\ref{Killing2}), we obtain
    \BEA
      0
       =     d (i_X K) (Y, Z)
      &=&    Y(K(X,Z)) - Z(K(X,Y)) - K(X, [Y,Z]),
             \\
      0
       =     d (i_Y K) (X, Z)
      &=&    X(K(Y,Z)) - Z(K(Y,X)) - K(Y, [X,Z]).
    \EEA
  The above three equations imply
    \BE
      Z(K(Y,X))
       =     K([X,Y],Z),
    \EE
  which leads to:
    \BE
      d \circ i_X \circ i_Y (K)
       =     i_{[X,Y]} K,
             \label{Killing6}
    \EE
  where we have used that for any function $f$, $X(f) = i_X (df)$.
  However, the left hand side of (\ref{Killing6}) can be rewritten as
    \BE
      - 2 \pi d ( i_X \circ i_Y (K))
       =     d ( i_X (d \CP_Y))
       =     d (X(\CP_Y)),
    \EE
  in the first equality we have used (\ref{Killing3}).
  Therefore, we have proved (\ref{equivariance}) up to constant.
  
  \hspace{400pt}
   
  
  \vspace{11pt}
  
  Suppose that we expand $X$ in a basis $k_\Lambda$ such that 
    \BE
      [k_\Lambda,k_\Sigma]
       =     f_{\Lambda\Sigma}^{~~\Gamma} k_\Gamma.
    \EE
  In the following we will use the shorthand notation $\CL_\Lambda, i_\Lambda$ 
  for the Lie derivative and the contraction along the chosen basis of Killing vectors $k_\Lambda$.
  The left hand side of the equivariance relation (\ref{equivariance}) can be represented 
  in terms of the Killing vectors as follows,
    \BE
      X(\CP_Y)
       =     a^\Lambda k_\Lambda (a^\Sigma \CP_\Sigma)
       =     i a^\Lambda a^\Sigma g_{ab^*} (k_\Lambda^a k_\Sigma^{b^*} - k_\Sigma^a k_\Lambda^{b^*}),
    \EE
  where we have used (\ref{Killing4}).
  On the other hand, the right hand side of (\ref{equivariance}) is written as
    \BE
      \CP_{[X,Y]}
       =     a^\Lambda a^\Sigma f_{\Lambda \Sigma}^{~~\Gamma} \CP_\Gamma.
    \EE
  Therefore, the equivariance relation implies 
    \BE
      i g_{ab^*} (k_\Lambda^a k_\Sigma^{b^*} - k_\Sigma^a k_\Lambda^{b^*})
       =     f_{\Lambda \Sigma}^{~~\Gamma} \CP_\Gamma.
             \label{equivariance2}
    \EE
\subsubsection{Momentum map}
  There is another way of stating the equivariance relation.
  It is based on the notion of the momentum map.
  A momentum map is constituted by $\CP_X$.
  This can be regarded as a map,
    \BE
      \CP~:~\CM
       \rightarrow
             \mathbb{R} \otimes \mathfrak{g}^*,
    \EE
  where $\mathfrak{g}^*$ denotes the dual of the Lie algebra $\mathfrak{g}$ of the Lie group $G$.
  Indeed let $g \in \mathfrak{g}$ be the Lie algebra element corresponding to the Killing vector $X$;
  then, for a given $p \in \CM$,
    \BE
      \mu(m)~:~g
       \rightarrow
             \CP_X (p) \in \mathbb{R}
    \EE
  is a linear functional on $\mathbb{R}$.
  
  The momentum map is the Hamiltonian function 
  providing the Poissonian realization of the Lie algebra on the K\"ahler manifold.
  Indeed the very existence of the closed K\"ahler 2-form $K$ guarantees 
  that every K\"ahler space is a symplectic manifold and that we can define a Poisson bracket.
  
  Consider (\ref{Killing4}).
  To every generator of the abstract Lie algebra $\mathfrak{g}$ we have associated a function $\CP_\Lambda$ on $\CM$.
  The Poisson bracket of $\CP_\Lambda$ with $\CP_\Sigma$ is defined as follows,
    \BE
      \{ \CP_\Lambda, \CP_\Sigma \} 
       \equiv
             4 \pi K(\Lambda, \Sigma),
             \label{Poisson4}
    \EE
  where $K(\Lambda, \Sigma) = K(\vec{k}_\Lambda, \vec{k}_\Sigma)$ is the value of $K$ along the pair of Killing vectors.
  \vspace{11pt}
  \\
  \textit{Lemma 3.1.}
  The following identity is true,
    \BE
      \{ \CP_\Lambda, \CP_\Sigma \}
       =     f_{\Lambda \Sigma}^{~~\Gamma} \CP_\Gamma
           + C_{\Lambda \Sigma},
             \label{Poisson1}
    \EE
  where $C_{\Lambda \Sigma}$ is a constant tensor satisfying the cocycle condition
    \BE
      f_{\Lambda \Sigma}^{~~\Delta} C_{\Delta \Gamma}
           + f_{\Sigma \Gamma}^{~~\Delta} C_{\Delta \Lambda}
           + f_{\Gamma \Lambda}^{~~\Delta} C_{\Delta \Sigma}
       =     0.
             \label{Poisson3}
    \EE
  \\
  \textit{Proof.}
  Using (\ref{Killing3}), we have
    \BEA
      4 \pi K(\Lambda,\Sigma)
      &=&  - 4 \pi K(\Sigma,\Lambda)
       =     2 \pi i_\Sigma i_\Lambda K
       =   - 2 \pi i_\Lambda i_\Sigma K
       =   - i_\Sigma d \CP_\Lambda
       =     i_\Lambda d \CP_\Sigma
             \NN \\
      &=&    \frac{1}{2} (i_\Lambda d \CP_\Sigma - i_\Sigma d \CP_\Lambda)
             \NN \\
      &=&    \frac{1}{2} (\CL_\Lambda \CP_\Sigma - \CL_\Sigma \CP_\Lambda).
    \EEA
  Since the exterior derivative commutes with the Lie derivative, $[d , \CL_\Lambda] = 0$, we find
    \BEA
      4 \pi d K(\Lambda, \Sigma)
      &=&    \frac{1}{2} (\CL_\Lambda d \CP_\Sigma - \CL_\Sigma d \CP_\Lambda)
             \NN \\
      &=&  - \pi (\CL_\Lambda i_\Sigma K - \CL_\Sigma i_\Lambda K)
             \NN \\
      &=&  - 2 \pi i_{[\Lambda, \Sigma]} K
             \NN \\
      &=&    f_{\Lambda \Sigma}^{~~\Gamma} d \CP_\Gamma,
    \EEA
  in the third equality, we have used $[i_\Lambda, \CL_\Sigma]K=i_{[\Lambda, \Sigma]}K$ and $\CL_\Lambda \CK = 0$.
  Using (\ref{Poisson4}), we obtain
    \BE
      d ( \{ \CP_\Lambda,  \CP_\Sigma \} - f_{\Lambda \Sigma}^{~~\Gamma} \CP_\Gamma )
       =     0,
    \EE
  which proves (\ref{Poisson1}).
  The cocycle condition (\ref{Poisson3}) follows from the Jacobi identities satisfied by (\ref{Poisson4}).
  \hspace{330pt}
   

  \vspace{11pt}
  
  If the Lie algebra $\mathfrak{g}$ has a trivial second cohomology group $H^2(\mathfrak{g})=0$, 
  then the cocycle $C_{\Lambda \Sigma}$ is a coboundary, namely 
  $ C_{\Lambda \Sigma} = f_{\Lambda \Sigma}^{~~\Gamma} C_\Gamma$ and 
  $C^\Gamma$ are suitable constants.
  Therefore, if we assume $H^2(\mathfrak{g})=0$, we can absorb $C_\Lambda$ in the definition of $\CP_\Lambda$:
    \BE
      \CP_\Lambda
       \rightarrow
             \CP_\Lambda + C_\Lambda,
    \EE
  and we obtain the stronger equation
    \BE
      \{ \CP_\Lambda, \CP_\Sigma \}
       =     f_{\Lambda \Sigma}^{~~\Gamma} \CP_\Gamma.
             \label{Poisson2}
    \EE
  Note that $H^2(\mathfrak{g})=0$ is true for all semi-simple Lie algebras.
  Using (\ref{Poisson4}), (\ref{Poisson2}) can be rewritten in components as follows,
    \BE
      \frac{i}{2} g_{ab^*} (k_\Lambda^a k_\Sigma^{b^*} - k_\Sigma^a k_\Lambda^{b^*})
       =     \frac{1}{2} f_{\Lambda \Sigma}^\Gamma \CP_\Gamma.
             \label{equivariance3}
    \EE
  This equation is identical with the equivariance relation (\ref{equivariance2}).

\subsubsection{The momentum map on special K\"ahler manifolds}
  We consider the momentum map on special K\"ahler manifolds.
  In order to distinguish the holomorphic momentum map from the triholomorphic one $\CP_\Lambda^x$ 
  which is defined on quaternionic K\"ahler manifolds in next subsection, we adopt the notation $\CP_\Lambda^0$.
  The Lie derivative of the covariantly holomorphic section $V$ defined in (\ref{sectionV2}) is
    \BE
      \CL_\Lambda V
       =
             k_\Lambda^a \partial_a V + k_\Lambda^{a^*} \partial_{a^*} V
       =     T_\Lambda V + Vf_\Lambda(z),
             \label{SPM1}
    \EE
  where
    \BE
      T_\Lambda
       =     \left(
             \begin{array}{cc}
             a_\Lambda & b_\Lambda \\
             c_\Lambda & d_\Lambda
             \end{array}
             \right)
       \in   Sp(2n + 2, \mathbb{R})
    \EE
  is some element of the real symplectic Lie algebra and $f_\Lambda(z)$ 
  corresponds to an infinitesimal K\"ahler transformation.
  
  As we see in the next chapter, the Lagrangian of gauged $\CN=2$ supergravity is not necessarily 
  invariant under the K\"ahler transformation (\ref{Kahlertr}).
  In order for the Lagrangian to be invariant, we should impose the following restriction:
    \BE
      f_\Lambda(z)
       =     0.
             \label{SPM2}
    \EE
  Under the above restriction, recalling (\ref{CK}) and (\ref{sectionV}), 
  for the sections $V$ and $\Omega$ we have,
    \BEA
      \CL_\Lambda V
      &=&    (k_\Lambda^a \partial_a \CK + k_\Lambda^{a^*} \partial_{a^*} \CK) V
           + e^{\CK/2} \CL_\Lambda \Omega
             \NN \\
      &=&    T_\Lambda e^{\CK/2} \Omega,
             \NN \\
      \CL_\Lambda \Omega
      &=&    T_\Lambda \Omega.
    \EEA
  Thus, we obtain
    \BE
      \CL_\Lambda \CK
       =     k_\Lambda^a \partial_a \CK + k_\Lambda^{a^*} \partial_{a^*} \CK
       =     0
    \EE
  that is identical with (\ref{momentmap2}). 
  Therefore, as discussed in the last subsection, the following equation is true
    \BE
      i \CP_\Lambda^0
       =     k_\Lambda^a \partial_a \CK
       =   - k_\Lambda^{a^*} \partial_{a^*} \CK.
    \EE
  Utilizing the definition in (\ref{Ua}) we easily obtain,
    \BE
      k_\Lambda^a U_a
       =     T_\Lambda V + i \CP_\Lambda^0 V.
             \label{SPM3}
    \EE
  Taking the symplectic scalar product of (\ref{SPM3}) with $\bar{V}$ and recalling (\ref{sectionV2}) we finally get:
    \BEA
      \CP_\Lambda^0
      &=&    \langle \bar{V} | T_\Lambda V \rangle
       =     \langle V | T_\Lambda \bar{V} \rangle
             \NN \\
      &=&    e^{\CK} \langle \bar{\Omega} | T_\Lambda \Omega \rangle.
             \label{SPM4}
    \EEA
  In the gauging procedure, we are interested in the symplectic image of whose generators is block-diagonal 
  and coincides with adjoint representation in each block.
  Namely,
    \BE
      T_\Lambda
       =     \left(
             \begin{array}{cc}
             f^\Sigma_{~\Lambda \Delta} & \textrm{\boldmath $0$} \\
             \textrm{\boldmath $0$} & - f^\Sigma_{~\Lambda \Delta}
             \end{array}
             \right).
    \EE
  Then (\ref{SPM4}) becomes
    \BE
      \CP_\Lambda^0
       =     e^{\CK}
             \left(
             F_\Delta f^\Delta_{~\Lambda \Sigma} \bar{X}^\Sigma + \bar{F}_\Delta f^\Delta_{~\Lambda \Sigma} X^\Sigma
             \right).
    \EE
\subsection{The triholomorphic momentum map}
  Let us turn to a discussion of isometries of the quaternionic manifold $\CH\CM$ associated with hypermultiplets.
  The triholomorphicity of the momentum map $\CP^x$ ($x=1,2,3$) comes from 
  the quaternionic algebra (\ref{quaternionicalgebra}) of $\CH\CM$.
  We must assume that on $\CH\CM$ we have an action by triholomorphic isometries of the same Lie group $\CG$ 
  that acts on the special K\"ahler manifold $\CS\CM$.
  This is means that on product manifold $\CS\CM \otimes \CH\CM$ 
  we have Killing vectors
    \BE
      \hat{k}_\Lambda
       =     k_\Lambda^a \frac{\partial}{\partial z^a} + k_\Lambda^{a^*} \frac{\partial}{\partial z^{a^*}}
           + k_\Lambda^u \frac{\partial}{\partial b^u}
    \EE
  which generate the transformation keeping invariant the metric, 
    \BE
      \hat{g} 
       =     \left(
             \begin{array}{cc}
             g_{ab^*} & 0 \\
             0 & h_{uv}
             \end{array}
             \right),
    \EE
  and satisfy the same Lie algebra as the corresponding Killing vectors on special K\"ahler manifolds $\CS\CM$:
    \BE
      [\hat{k}_\Lambda, \hat{k}_\Sigma]
       =     f_{\Lambda \Sigma}^{~~\Gamma} \hat{k}_\Gamma.
             \label{commutation}
    \EE
  
  In the previous section, we have obtained the K\"ahler 2-form is invariant under the action of Lie derivative.
  Similarly, the Killing vector fields leave the hyperK\"ahler form invariant.
  The only difference is the freedom to do the $SU(2)$ rotations in the $SU(2)$ bundle $\CS\CU$, that is, 
    \BEA
      \CL_\Lambda K^x
      &=&    \e^{xyz} K^y W^z_\Lambda,
             \NN \\
      \CL_\Lambda \omega^x
      &=&    \nabla W^x_\Lambda,
             \label{triholo1}
    \EEA
  where $W^x_\Lambda$ is an $SU(2)$ compensator associated with the Killing vector $k^u_\Lambda$.
  This can be rewritten, by using the identification between hyperK\"ahler forms and $SU(2)$ curvatures (\ref{Omega}), as
    \BE
      \CL_\Lambda \Omega^x
       =     \e^{xyz} \Omega^y W^z_\Lambda.
             \label{triholo4}
    \EE
  The compensator $W^x_\Lambda$ necessarily fulfills the cocycle condition:
    \BE
      \CL_\Lambda W^x_\Sigma - \CL_\Sigma W^x_\Lambda + \e^{xyz} W^y_\Lambda W^z_\Sigma
       =     f_{\Lambda \Sigma}^{~\Gamma} W^x_\Gamma.
    \EE
  
  In full analogy with the case of K\"ahler manifolds, 
  to each Killing vector we can associate a triplet $\CP^x_\Lambda(b)$ of Killing potentials.
  Indeed,
    \BEA
      \nabla W_\Lambda^x
      &=&    \CL_\Lambda \omega^x
             \NN \\
      &=&    d (i_\Lambda \omega^x ) + i_\Lambda (d \omega^x)
             \NN \\
      &=&    \lambda (i_\Lambda K^x) + \nabla (i_\Lambda \omega^x),
    \EEA
  where $\nabla$ is defined such that, for $SU(2)$ vector $V^x$, $\nabla V^x = d V^x + \e^{xyz} \omega^y V^z$.
  Therefore, if we set
    \BE
      \CP^x_\Lambda
       \equiv
             \lambda^{-1} (i_\Lambda \omega^x - W^x_\Lambda),
    \EE
  we obtain
    \BE
      i_\Lambda K^x
       =   - \nabla \CP^x_\Lambda
       \equiv
           - (d \CP^x_\Lambda + \e^{xyz} \omega^y \CP^z_\Lambda).
             \label{triholo2}
    \EE
  If we expand $X = a^\Lambda \hat{k}_\Lambda$ on a basis of Killing vectors $k_\Lambda$ 
  satisfying the commutation relation (\ref{commutation}), we can set
    \BE
      \CP_X^x
       =     a^\Lambda \CP^x_\Lambda.
    \EE
  With this basis, we can also rewrite (\ref{triholo2}) as,
    \BE
      i_X K^x
       =   - \nabla \CP_X^x.
    \EE
  This is a generalization of the relation (\ref{Killing3}).
  
  Furthermore, we need a generalization of the equivariance relation obtained by (\ref{equivariance}).
  It should be written in terms of $\CP_X^x$ as follows:
    \BE
      X ( \CP_Y^x )
       =     \CP_{[X, Y]}^x,
             \label{triholo3}
    \EE
  where the left-hand side of (\ref{triholo3}) is interpreted as
    \BE
      X ( \CP_Y^x )
       =     i_X \nabla \CP_Y^x
       =     X^u \nabla_u \CP_Y^x.
    \EE
    
\subsubsection{The triholomorphic momentum map}
  As in the K\"ahler manifold case, (\ref{triholo2}) defines a momentum map:
    \BE
      \CP^x~:~\CM
       \rightarrow 
             \mathbb{R}^3 \otimes \mathfrak{g}^*,
    \EE
  where $\mathfrak{g}^*$ denotes the dual of the Lie algebra $\mathfrak{g}$ of the group $G$.
  Indeed let $g \in \mathfrak{g}$ be the Lie algebra element corresponding to the Killing vector $X$, 
  then, for a given $p \in \CM$
    \BE
      \mu (p)~:~g
       \rightarrow
            \CP_X^x (p) \in \mathbb{R}^3
    \EE
  is a linear functional on $\mathfrak{g}$.
  
  Correspondingly, the triholomorphic Poisson bracket is defined as follows:
    \BE
      \{ \CP_\Lambda, \CP_\Sigma \}^x
       \equiv
             2 K^x (\Lambda, \Sigma)
           - \lambda \e^{xyz} \CP^y_\Lambda \CP^z_\Sigma
             \label{Triholo6}
    \EE
  \\
  \textit{Lemma 3.2.}
  The following identity is true,
    \BE
      \{ \CP_\Lambda, \CP_\Sigma \}^x
       =     f_{\Lambda \Sigma}^{~~\Gamma} \CP_\Gamma^x
           + C_{\Lambda \Sigma}^x,
             \label{TriPoisson1}
    \EE
  where $C_{\Lambda \Sigma}^x$ is covariantly constant, namely, $\nabla C^x_{\Lambda \Sigma} = 0$ 
  and fulfills the cocycle condition
    \BE
      f_{\Lambda \Sigma}^{~~\Delta} C_{\Delta \Gamma}^x
           + f_{\Sigma \Gamma}^{~~\Delta} C_{\Delta \Lambda}^x
           + f_{\Gamma \Lambda}^{~~\Delta} C_{\Delta \Sigma}^x
       =     0.
             \label{TriPoisson3}
    \EE
  \\
  \textit{Proof.}
  It is analogous to the proof of lemma 3.1.
  The difference is that we have to introduce covariant exterior derivative and covariant Lie derivative 
  $\tilde{\CL}_\Lambda$ instead of the ordinary ones.
  $\tilde{\CL}_\Lambda$ is defined such as for any $SU(2)$ vector $V^x$,
    \BE
      \tilde{\CL}_\Lambda V^x
       =     \CL_\Lambda V^x + \e^{xyz} W^y V^z.
    \EE
  With this definition, (\ref{triholo4}) is rewritten as
    \BE
      \tilde{\CL}_\Lambda \Omega^x
       =     0.
    \EE
  Using (\ref{triholo2}), we have
    \BEA
      2 K^x(\Lambda,\Sigma)
      &=&    i_\Sigma i_\Lambda K^x
       =   - i_\Lambda i_\Sigma K^x
             \NN \\
      &=&    \frac{1}{2} (i_\Lambda \nabla \CP_\Sigma - i_\Sigma \nabla \CP_\Lambda)
             \NN \\
      &=&    \frac{1}{2} 
             (\CL_\Lambda \CP_\Sigma^x - \CL_\Sigma \CP_\Lambda^x
           + \e^{xyz} (i_\Lambda \omega)^y \CP_\Sigma^z - \e^{xyz} (i_\Sigma \omega)^y \CP_\Lambda^z).
    \EEA
  Thus, we obtain
    \BE
      2 K^x(\Lambda,\Sigma) - \lambda \e^{xyz} \CP_\Lambda^y \CP_\Sigma^z
       =     \frac{1}{2} (\tilde{\CL}_\Lambda \CP_\Sigma^x - \tilde{\CL}_\Sigma \CP_\Lambda^x).
    \EE
  
  Since the covariant exterior derivative commutes with the covariant Lie derivative for any $SU(2)$ vector $V^x$, 
  $[\nabla , \tilde{\CL}_\Lambda] V^x = 0$, we find
    \BEA
      \nabla \left( 2 K^x(\Lambda, \Sigma) - \lambda \e^{xyz} \CP_\Lambda^y \CP_\Sigma^z \right)
      &=&    \frac{1}{2} (\tilde{\CL}_\Lambda \nabla \CP_\Sigma^x - \tilde{\CL}_\Sigma \nabla \CP_\Lambda^x)
             \NN \\
      &=&  - \frac{1}{2} (\tilde{\CL}_\Lambda i_\Sigma K^x - \tilde{\CL}_\Sigma i_\Lambda K^x)
             \NN \\
      &=&  - i_{[\Lambda, \Sigma]} K^x
             \NN \\
      &=&    f_{\Lambda \Sigma}^{~~\Gamma} \nabla \CP_\Gamma^x,
    \EEA
  in the third equality, we have used the fact that $[i_\Lambda, \tilde{\CL}_\Sigma]K^x = i_{[\Lambda, \Sigma]}K^x$.
  Using (\ref{Triholo6}), we get
    \BE
      \nabla ( \{ \CP_\Lambda,  \CP_\Sigma \}^x - f_{\Lambda \Sigma}^{~~\Gamma} \CP_\Gamma^x )
       =     0.
    \EE
  Thus, the lemma 3.2 is proved.
  \hspace{265pt}
   
  \vspace{11pt}
  
  If we assume that the second cohomology group is trivial, then we have
    \BE
      C_{\Lambda \Sigma}^x
       =     f_{\Lambda \Sigma}^{~~\Gamma} C_\Gamma^x,
    \EE
  and the constant $C_\Lambda^x$ is absorbed by $\CP^x_\Lambda$.
  Therefore, we obtain the Poissonian realization of the Lie algebra
    \BE
      \{ \CP_\Lambda, \CP_\Sigma \}^x
       =     f_{\Lambda \Sigma}^{~\Gamma} \CP^x_\Gamma,
    \EE
  which, in components, leads to
    \BE
      K^x_{uv} k^u_\Lambda k^v_\Sigma - \frac{\lambda}{2} \e^{xyz} \CP^y_\Lambda \CP^z_\Sigma
       =     \frac{1}{2} f_{\Lambda \Sigma}^{~\Gamma} \CP^x_\Gamma.
             \label{triholo5}
    \EE
  Eq.(\ref{triholo5}), which is the most convenient way of expressing equivariance relation in a coordinate basis, 
  plays a fundamental role in the construction of the supersymmetric action, supersymmetry transformation rules 
  and of the superpotential for $\CN=2$ supergravity on a general quaternionic manifold.
\section{Gauging of the composite connections}
  On the special and quaternionic manifold, 
  we have introduced several connection 1-forms related with different bundles.
  Gauging the corresponding supergravity theory is done by gauging these composite connections 
  of the underlying $\sigma$-model.
  
  For the Levi-Civita connection the gauging is standard on K\"ahler manifold $\CM$.
  Let $k^a_\Lambda(z)$ be the Killing vectors which defined in the last section.
  The ordinary differential $d z^a$ is replaced by the covariant differential defined as
    \BEA
      d z^a
      &\rightarrow&
             \nabla z^a
       =     d z^a + g A^\Lambda k^a_\Lambda (z),
             \\
      d \bar{z}^{a^*}
      &\rightarrow&
             \nabla \bar{z}^{a^*}
       =     d \bar{z}^{a^*} + g A^\Lambda k^{a^*}_\Lambda (\bar{z}),
    \EEA
  where $g$ is the gauge coupling constant and $A^\Lambda$ is the gauge 1-form.
  The Levi-Civita connection $\Gamma^a_b = \Gamma^a_{bc} d z^c$ is replaced by its gauged counterpart as
    \BEA
      \Gamma^a_b 
      &\rightarrow&
             \hat{\Gamma}^a_b
       =     \Gamma^a_{bc} \nabla z^c + g A^\Lambda \partial_b k^a_\Lambda,
             \label{GaugeGamma}
             \\
      \Gamma^{a^*}_{b^*} 
      &\rightarrow&
             \hat{\Gamma}^{a^*}_{b^*}
       =     \Gamma^{a^*}_{b^* c^*} \nabla \bar{z}^{c^*} + g A^\Lambda \partial_{b^*} k^{a^*}_\Lambda.
    \EEA
  The gauged curvature 2-form is 
    \BEA
      \hat{R}^a_b
      &\equiv&
             d \hat{\Gamma}^a_b + \hat{\Gamma}^a_c \wedge \hat{\Gamma}^c_b
             \NN \\
      &=&    R^a_{~bc^*d} \nabla \bar{z}^{c^*} \wedge \nabla z^d + g F^\Lambda \partial_b k^a_\Lambda (z),
    \EEA
  where $F^\Lambda$ is the gauged field strength:
    \BE
      F^\Lambda
       =     d A^\Lambda + \frac{1}{2} g f_{\Sigma \Gamma}^{~~\Lambda} A^\Sigma \wedge A^\Gamma.
             \label{GaugeF}
    \EE
  In an analogous way, the gauging of the $Sp(2k)$ connection gives
    \BE
      \Delta^{\a \b}
       \rightarrow
             \hat{\Delta}^{\a \b}
       =     \Delta^{\a \b} + g A^\Lambda \partial_u k^v_\Lambda \CU^{u | A \a} \CU^\b_{A | v},
    \EE
  and the associated gauged curvature $\mathbb{R}$ becomes
    \BE
      \mathbb{R}^{\a \b}
       \rightarrow
             \hat{\mathbb{R}}^{\a \b}
       =     \mathbb{R}^{\a \b}_{uv} db^u \wedge db^v 
           + g F^\Lambda \partial_u k^v_\Lambda \CU^{u | A \a} \CU^\b_{A | v}.
             \label{GaugemathbbR}
    \EE
  
  The existence of the Killing potentials allow the following definitions 
  for the $U(1)$ connection and the $SU(2)$ connection:
    \BEA
      \CQ
      &\rightarrow&
             \hat{\CQ}
       =     \CQ + g A^\Lambda \CP^0_\Lambda,
             \\
      \omega^x
      &\rightarrow&
             \hat{\omega}^x
       =     \omega^x + g A^\Lambda \CP^x_\Lambda.
    \EEA
  By computing the associated gauged curvatures, one finds the gauge covariant expressions:
    \BEA
      \hat{K}
      &=&    d \hat{\CQ}
             \NN \\
      &=&    i g_{ab^*} \nabla z^a \wedge \nabla \bar{z}^{b^*} + g F^\Lambda \CP^0_\Lambda,
             \label{GaugeK}
             \\
      \hat{\Omega}^x
      &=&    d \hat{\omega}^x + \frac{1}{2} \e^{xyz} \hat{\omega}^y \wedge \hat{\omega}^z
             \NN \\
      &=&    \Omega^x_{uv} \nabla b^u \wedge \nabla b^v + g F^\Lambda \CP^x_\Lambda.
             \label{GaugeOmega}
    \EEA

\chapter{The Lagrangian of gauged $\mathcal{N}=2$ supergravity}
  In this chapter we introduce the gauged $\mathcal{N}=2$ supergravity Lagrangian
  and the supersymmetry transformation laws 
  in terms of the geometric quantities which have been given in the last two chapters.
  
  In \cite{deWitvanHolten, BaggerWitten, deWitVan, deWitLauwers, deWitLauwers2, Cremmer, itoyama}, 
  the Lagrangian of $\CN=2$ supergravity was constructed.
  In particular, the Lagrangian of gauged $\CN=2$ supergravity was introduced 
  in \cite{D'Auria, Andrianopoli2, Andrianopoli}.
\section{The Lagrangian of gauged $\mathcal{N}=2$ supergravity}
  The complete Lagrangian can be found in \cite{Andrianopoli}.
  It is constructed such that its equations of motion are consistent to the solutions of the Bianchi identities.
  We will discuss with the Bianchi identities and its solutions in the appendix B.
  The construction of the Lagrangian is very complicated and tedious.
  Here we do not see the derivation of the complete Lagrangian.
  The general derivation of the supergravity Lagrangian is given in \cite{SUGRA}.
  Also, \cite{Castellani, D'Auria} will help you deriving the complete Lagrangian.
  
  The gauged $\CN=2$ supergravity action is given by
    \BE
      S
       =     \int d^4 x \sqrt{- g} 
             \left(
             \CL_{\rm{kin}} + \CL_{\rm{Pauli}} 
           + \CL_{\rm{4Fermi}}^{\rm{non-inv}} 
           + \CL_{\rm{4Fermi}}^{\rm{inv}} 
           + \CL_{\rm{Yukawa}} - V(z,\bar{z},b)
             \right).
    \EE
  $\CL_{\rm{kin}}$ consists of the kinetic terms of the component fields and is written as,
    \BEA
      \CL_{\rm{kin}}
      &=&    R + g_{ab^*} \nabla_{\mu} z^a \nabla^\mu \bar{z}^{b^*}
           + h_{uv} \nabla_\mu b^u \nabla^\mu b^v 
           + \frac{\epsilon^{\mu \nu \lambda \sigma}}{\sqrt{-g}}
             (\bar{\psi}^A_\mu \gamma_\nu \nabla_\lambda \psi_{A \sigma}
           - \bar{\psi}_{A \mu} \gamma_\nu \nabla_\lambda \psi^A_\sigma)
             \nonumber \\
      & &  + \frac{1}{4} (\Im \mathcal{N})_{\Lambda \Sigma} F^\Lambda_{\mu \nu} F^{\Sigma \mu \nu}
           + \frac{1}{4} (\Re \mathcal{N})_{\Lambda \Sigma} F^\Lambda_{\mu \nu} \widetilde{F}^{\Sigma \mu \nu} 
             \nonumber \\
      & &  - i g_{ab^*} 
             \left(
             \bar{\lambda}^{aA} \gamma_\mu \nabla^\mu \lambda^{b^*}_A 
           + \bar{\lambda}^{b^*}_A \gamma_\mu \nabla^\mu \lambda^{aA}
             \right)
           - 2 i 
             \left(
             \bar{\zeta}^{\alpha} \gamma_\mu \nabla^\mu \zeta_\alpha
           + \bar{\zeta}_{\alpha} \gamma_\mu \nabla^\mu \zeta^\alpha
             \right)
             \nonumber \\
      & &  + [
             g_{ab^*} \nabla_\mu \bar{z}^{b^*} \bar{\lambda}^{aA} \gamma^{\mu \nu} \psi_{A \nu}
           - g_{ab^*} \nabla_\mu \bar{z}^{b^*} \bar{\psi}^\mu_A \lambda^{aA}
             \nonumber \\
      & &  + 2 \CU^{\alpha A}_u \nabla_\mu b^u \bar{\zeta}_{\alpha} \gamma^{\mu \nu} \psi_{A \nu}
           - 2 \CU^{\alpha A}_u \nabla_\mu b^u \bar{\psi}^\mu_A \zeta_{\alpha}
           + h.c.
             ].
             \label{Lkin}
             \\
      & &    \NN
    \EEA
  $z^a$ and $\lambda^{aA}$ are, respectively, scalar fields and $SU(2)$ doublet gaugini of the vector multiplets.
  On the other hand, $b^u$ and $\zeta^\a$ are scalar fields and $SU(2)$ doublet hyperini of the hypermultiplets. 
  The $SU(2)$ doublet gravitini are represented by $\psi^A_\mu$.
  The field strengths $F^\Lambda_{\mu \nu}$
    \footnote[2]{We have redefined the field strengths $F^\Lambda_{\mu \nu}$ in (\ref{appendixfieldstrength})
                 in the appendix B as $F^\Lambda_{\mu \nu} \rightarrow  \frac{1}{2} F^\Lambda_{\mu \nu}$.}
  of the gauge fields $A^a_\mu$ and the graviphoton field $A^0_\mu$ are
    \BE
      F^\Lambda_{\mu \nu}
       =     \partial_\mu A^\Lambda_\nu
           - \partial_\nu A^\Lambda_\mu 
           + g f^\Lambda_{~\Sigma \Gamma} A^\Sigma_\mu A^\Gamma_\nu,
    \EE
  and $ \tilde{F}^\Lambda $ are their Hodge duals:
    \BE
      \tilde{F}^\Lambda_{\mu \nu}
       =     \frac{1}{2} \e_{\mu \nu \lambda \sigma} F^{\Lambda \lambda \sigma}.
    \EE
  The normalization of the kinetic terms for the hypermultiplet scalar fields $b^u$ depend 
  on the constant $\lambda$ of the quaternionic K\"ahler manifold.
  We have chosen that $\lambda = 1$ because of the positivity of the kinetic terms.
  
  Although we need not the explicit forms of $\CL_{\rm{Pauli}}$, $\CL_{\rm{4Fermi}}^{\rm{inv}}$ 
  and $\CL_{\rm{4Fermi}}^{\rm{non-inv}}$, we give them for completeness:
    \BEA
      \CL_{\rm{Pauli}}
      &=&    F^{- \Lambda}_{\mu \nu} (\Im \CN)_{\Lambda \Sigma}
             [
             2 L^{\Sigma} \bar{\psi}^{A \mu} \psi^{B \nu} \epsilon_{AB}
           - 2 i \bar{f}^{\Sigma}_{a^*} \bar{\lambda}^{a^*}_A \gamma^\nu \psi^\mu_B \epsilon^{AB}
             \nonumber \\
      & &  + \frac{1}{4} \nabla_a f_b^{\Sigma} \bar{\lambda}^{aA} \gamma^{\mu \nu} \lambda^{bB} \epsilon_{AB}
           - \frac{1}{2} L^{\Sigma} \bar{\zeta}_\alpha \gamma^{\mu \nu} \zeta_\beta \mathbb{C}^{\alpha \beta}
             ]
           + h.c.,
             \\
      & &    \NN
             \\
      \CL_{\rm{4fermi}}^{\rm{inv}}
      &=&    \frac{i}{2}
             \left(
             g_{ab^*} \bar{\lambda}^{aA} \gamma_\sigma \lambda^{b^*}_B
           - 2 \delta^A_B \bar{\zeta}^\alpha \gamma_\sigma \zeta_\alpha
             \right)
             \bar{\psi}_{A \mu} \gamma_\lambda \psi^B_\nu \frac{\epsilon^{\mu \nu \lambda \sigma}}{\sqrt{-g}}
             \nonumber \\
      & &  - \frac{1}{6}
             \left(
             C_{abc} \bar{\lambda}^{aA} \gamma^\mu \psi_\mu^B \bar{\lambda}^{bC} \lambda^{cD} \e_{AC} \e_{BD} 
           + h.c.
             \right)
             \nonumber \\
      & &  - 2 \bar{\psi}^A_\mu \psi^B_\nu \bar{\psi}^\mu_A \psi_B^\nu 
           + 2 g_{ab^*} \bar{\lambda}^{aA} \gamma_\mu \psi^B_\nu \bar{\lambda}^{a^*}_A \gamma^\mu \psi^\nu_B
             \nonumber \\
      & &  + \frac{1}{4}
             \left(
             R_{ab^*cd^*} + g_{ad^*} g_{cb^*} - \frac{3}{2} g_{ab^*} g_{cd^*}
             \right)
             \bar{\lambda}^{aA} \lambda^{cB} \bar{\lambda}^{b^*}_A \lambda^{d^*}_B
             \nonumber \\
      & &  + \frac{1}{4} g_{ab^*} \bar{\zeta}^\alpha \gamma_\mu \zeta_\alpha \bar{\lambda}^{aA} \gamma^\mu \lambda^{b^*}_A
           + \frac{1}{2} \CR^\alpha_{\beta ts} \CU^t_{A \gamma} \CU^s_{B \delta} 
             \e^{AB} \mathbb{C}^{\delta \eta} \bar{\zeta}_\alpha \zeta_\eta \bar{\zeta}^\beta \zeta^\gamma
             \nonumber \\
      & &  - \frac{1}{12}
             \left(
             i \nabla_a C_{bcd} \bar{\lambda}^{bA} \lambda^{aB} \bar{\lambda}^{cC} \lambda^{dD} \e_{AC} \e_{BD} + h.c.
             \right)
             \nonumber \\
      & &  + g_{ab^*} \bar{\psi}^A_\mu \lambda^{b^*}_A \bar{\psi}^\mu_B \lambda^{aB} 
           + 2 \bar{\psi}^A_\mu \zeta^\alpha \bar{\psi}^\mu_A \zeta_\alpha
           + \left(
             \e_{AB} \mathbb{C}_{\alpha \beta} \bar{\psi}^A_\mu \zeta^\alpha \bar{\psi}^{B \mu} \zeta^\beta 
           + h.c.
             \right),
             \NN \\
      & &    \\
      \CL_{\rm{4Fermi}}^{\rm{non-inv}}
      &=&    (\Im \CN)_{\Lambda \Sigma}
             [
             2 L^\Lambda L^\Sigma (\bar{\psi}^A_\mu \psi^B_\nu)^- (\bar{\psi}^{C \mu} \psi^{D \nu})^- \e_{AB} \e_{CD}
             \nonumber \\
      & &  - 8 i L^\Lambda \bar{f}^\Sigma_{a^*} 
             (\bar{\psi}^A_\mu \psi^B_\nu)^- (\bar{\lambda}^{a^*}_A \gamma^\nu \psi^\nu_B)^-
             \nonumber \\
      & &  - 2 \bar{f}^\Lambda_{a^*} \bar{f}^\Sigma_{b^*} (\bar{\lambda}^{a^*}_A \gamma^\nu \psi^\mu_B)^- 
             (\bar{\lambda}^{b^*}_C \gamma_\nu \psi_{D \mu})^- \e^{AB} \e^{CD}
             \nonumber \\
      & &  + \frac{i}{2} L^\Lambda \bar{f}^\Sigma_{d^*} g^{cd^*} C_{abc} (\bar{\psi}^A_\mu \psi^B_\nu)^-
             \bar{\lambda}^{aC} \gamma^{\mu \nu} \lambda^{bD} \e_{AB} \e_{CD}
             \NN
    \EEA
    \BEA
      & &  + \bar{f}^\Lambda_{e^*} \bar{f}^\Sigma_{d^*} g^{cd^*} C_{abc} 
             (\bar{\lambda}^{e^*}_A \gamma_\nu \psi_{B \mu})^- 
             \bar{\lambda}^{aA} \gamma^{\mu \nu} \lambda^{bB}
             \NN \\
      & &  - L^\Lambda L^\Sigma (\bar{\psi}^A_\mu \psi^B_\nu)^- 
             \bar{\zeta}_\alpha \gamma^{\mu \nu} \zeta_\beta \e_{AB} \mathbb{C}^{\alpha \beta}
             \NN \\
      & &  + i L^\Lambda \bar{f}^\Sigma_{a^*} (\bar{\lambda}^{a^*}_A \gamma^\nu \psi^\mu_B)^-
             \bar{\zeta}_\alpha \gamma_{\mu \nu} \zeta_\beta \e^{AB} \mathbb{C}_{\alpha \beta}
             \NN \\
      & &  - \frac{1}{32} C_{abc} C_{def} g^{cg^*} g^{fh^*} \bar{f}^\Lambda_{g^*} \bar{f}^\Sigma_{h^*} 
             \bar{\lambda}^{aA} \gamma_{\mu \nu} \lambda^{bB} 
             \bar{\lambda}^{dC} \gamma^{\mu \nu} \lambda^{eD} \e_{AB} \e_{CD}
             \NN \\
      & &  - \frac{1}{8} L^\Lambda \nabla_a f^\Sigma_b 
             \bar{\zeta}_\alpha \gamma_{\mu \nu} \zeta_\beta
             \bar{\lambda}^{aA} \gamma^{\mu \nu} \lambda^{bB} \e_{AB} \mathbb{C}^{\alpha \beta}
             \NN \\
      & &  + \frac{1}{8} L^\Lambda L^\Sigma 
             \bar{\zeta}_\alpha \gamma_{\mu \nu} \zeta_\beta \bar{\zeta}_\gamma \gamma^{\mu \nu} \zeta_\delta
             \mathbb{C}^{\alpha \beta} \mathbb{C}^{\gamma \delta}
             ]
           + h.c.,
    \EEA
  where we have used, $F^{\pm \Lambda}_{\mu \nu} 
  = \frac{1}{2} (F^\Lambda_{\mu \nu} \pm \frac{i}{2} \e^{\mu \nu \lambda \sigma} F^\Lambda_{\lambda \sigma})$.
  Also, $(\ldots)^-$ denotes the self dual part of the fermion bilinears.
  
  Because of the gauging, we obtain the following Yukawa coupling terms which include the mass terms of the fermions
  and scalar potential terms:
    \BEA
      \CL_{\rm{Yukawa}} 
      &=&    g [
             2 S_{AB} \bar{\psi}^A_\mu \gamma^{\mu \nu} \psi^B_\nu 
           + i g_{ab^*} W^{aAB} \bar{\lambda}^{b^*}_A \gamma_\mu \psi^\mu_B
           + 2 i N^A_\alpha \bar{\zeta}^\alpha \gamma_\mu \psi^\mu_A
             \nonumber \\
      & &  + \mathcal{M}^{\alpha \beta} \bar{\zeta}_\alpha \zeta_\beta
           + \mathcal{M}^\alpha_{aB} \bar{\zeta}_\alpha \lambda^{aB}
           + \mathcal{M}_{aA \vert bB} \bar{\lambda}^{aA} \lambda^{bB}
             ]
           + h.c.,
             \label{Lmass}
             \\
      & &    \NN
             \\
      V(z,\bar{z},b)
      &=&   g^2
            [
            g_{ab^*} k^a_\Lambda k^{b^*}_\Sigma \bar{L}^\Lambda L^\Sigma
          + g^{ab^*} f^\Lambda_a \bar{f}^\Sigma_{b^*} \mathcal{P}^x_\Lambda \mathcal{P}^x_\Sigma
            \NN \\
      & & + 4~ h_{uv} k^u_\Lambda k^v_\Sigma \bar{L}^\Lambda L^\Sigma
          - 3~ \bar{L}^\Lambda L^\Sigma \mathcal{P}^x_\Lambda \mathcal{P}^x_\Sigma
            ].
            \label{V}
    \EEA
  The coupling constant $g$ in $\CL_{\rm{Yukawa}}$, $V$ is a symbolic notation to remind 
  that these terms are produced by the gauging.
  Therefore, there is no Yukawa terms and the scalar potential term in the ungauged theory ($g=0$).
  
  $\CL_{\rm{Yukawa}}$ are written by the following matrices,
    \begin{eqnarray}
      S_{AB}
      &=&    \frac{i}{2} (\sigma_x)_{AB} \mathcal{P}^x_\Lambda L^\Lambda,
             \label{S}
             \\
      W^{aAB}
      &=&   \epsilon^{AB} k^a_\Lambda \bar{L}^\Lambda
          + i (\sigma_x)^{AB} \mathcal{P}^x_\Lambda g^{ab^*} \bar{f}^\Lambda_{b^*},
            \\
      N^A_\alpha
      &=&   2~ \mathcal{U}^A_{\alpha u} k^u_\Lambda \bar{L}^\Lambda,
            \\
      \mathcal{M}^{\alpha \beta}
      &=& - \mathcal{U}^{A \alpha}_u \mathcal{U}^{B \beta}_v \epsilon_{AB} \nabla^{[u} k^{v]}_\Lambda L^\Lambda,
            \\
      \mathcal{M}^\alpha_{bB}
      &=& - 4~ \mathcal{U}^\alpha_{Bu} k^u_\Lambda f^\Lambda_b,
            \\
      \mathcal{M}_{aA \vert bB}
      &=&   \frac{1}{2} 
            \left( 
            \epsilon_{AB} g_{ac^*} k^{c^*}_\Lambda f^\Lambda_b
          + i (\sigma_x)_{AB} \mathcal{P}^x_\Lambda \nabla_b f^\Lambda_a
            \right).
    \end{eqnarray}
  In the subsequent chapter, 
  it is convenient to divide $ W^{aAB} $ and $ \mathcal{M}_{aA \vert bB} $ into two parts respectively 
  such that 
    \BEA
      W^{aAB}_1 
      &=&    \epsilon^{AB} k^a_\Lambda \bar{L}^\Lambda,
             \\
      W^{aAB}_2 
      &=&    i (\sigma_x)^{AB} \mathcal{P}^x_\Lambda g^{ab^*} \bar{f}^\Lambda_{b^*},
             \label{W2}
             \\
      \mathcal{M}_{1;aA \vert bB}
      &=&    \frac{1}{2} \epsilon_{AB} g_{ac^*} k^{c^*}_\Lambda f^\Lambda_b,
             \\
      \mathcal{M}_{2;aA \vert bB}
      &=&    \frac{i}{2} (\sigma_x)_{AB} \mathcal{P}^x_\Lambda \nabla_b f^\Lambda_a.
             \label{M2}
    \EEA
  In the following, we refer to these matrices as mass matrices.
  
  The covariant derivatives are defined as follows:
    \begin{subequations}
    \begin{align}
      \nabla_\mu \psi_{A \nu}
      &=     \partial_\mu \psi_{A \nu}
           - \frac{1}{4} \g_{ij} \omega^{ij}_\mu \psi_{A \nu}
           + \frac{i}{2} \hat{\CQ}_\mu \psi_{A \nu}
           + \hat{\omega}_{A \mu}^{~B} \psi_{B \nu},
             \\
      \nabla_\mu \psi^A_\nu
      &=     \partial_\mu \psi^A_\nu
           - \frac{1}{4} \g_{ij} \omega^{ij}_\mu \psi_{A \nu}
           - \frac{i}{2} \hat{\CQ}_\mu \psi^A_\nu
           + \hat{\omega}^A_{~B \mu} \psi^B_\nu,
             \\
      \nabla_\mu z^a
      &=     \partial_\mu z^a
           + g A^\Lambda_\mu k^a_\Lambda,
             \\
      \nabla_\mu \lambda^{aA}
      &=     \partial_\mu \lambda^{aA}
           - \frac{1}{4} \g_{ij} \omega^{ij}_\mu \lambda^{aA}
           - \frac{i}{2} \hat{\CQ}_\mu \lambda^{aA}
           + \hat{\Gamma}^a_{~b \mu} \lambda^{bA}
           + \hat{\omega}^A_{B \mu} \lambda^{aB},
             \\
      \nabla_\mu \lambda^{a^*}_A
      &=     \partial_\mu \lambda^{a^*}_A
           - \frac{1}{4} \g_{ij} \omega^{ij}_\mu \lambda^{a^*}_A
           + \frac{i}{2} \hat{\CQ}_\mu \lambda^{a^*}_A
           + \hat{\Gamma}^{a^*}_{~b^* \mu} \lambda^{b^*}_A
           + \hat{\omega}_{A \mu}^{~B} \lambda^{a^*}_B,
             \\
      \nabla_\mu b^u
      &=     \partial_\mu b^u
           + g A^\Lambda _\mu k^u_\Lambda,
             \\
      \nabla_\mu \zeta_\a
      &=     \partial_\mu \zeta_\a
           - \frac{1}{4} \g_{ij} \omega^{ij}_\mu \zeta_\a
           - \frac{i}{2} \hat{\CQ}_\mu \zeta_\a
           + \hat{\Delta}_{\a \mu}^\b \zeta_\b,
             \\
      \nabla_\mu \zeta^\a
      &=     \partial_\mu \zeta^\a
           - \frac{1}{4} \g_{ij} \omega^{ij}_\mu \zeta^\a
           + \frac{i}{2} \hat{\CQ}_\mu \zeta^\a
           + \hat{\Delta}^\a_{\b \mu} \zeta^\b.
    \end{align}
    \end{subequations}
  In the above equations, $\hat{\CQ}_\mu$, $\hat{\Gamma}^a_{~b \mu}$, $\hat{\omega}_{A \mu}^{~B}$ 
  and $\hat{\Delta}_{\a \mu}^{~\b}$ are the space-time components of the gauged connections 
  which have introduced at the end of the last chapter:
    \begin{subequations}
    \begin{align}
      \hat{\CQ}_\mu
      &=   - \frac{i}{2}
             (
             \partial_a \CK \partial_\mu z^a 
           - \partial_{a^*} \CK \partial_\mu \bar{z}^{a^*}
             )
           + g A^\Lambda_\mu \CP^0_\Lambda,
             \\
      \hat{\Gamma}^a_{~b \mu}
      &=     \Gamma^a_{bc} \partial_\mu z^c
           + g A^\Lambda_\mu \partial_b k^a_\Lambda,
             \\
      \hat{\omega}_{A \mu}^{~B}
      &=     \frac{i}{2} \hat{\omega}^x (\sigma_x)_A^{~B}
       =     \frac{i}{2} 
             \left(
             \omega^x_u \partial_\mu b^u
           + g A^\Lambda_\mu \CP^x_\Lambda 
             \right)
             (\sigma_x)_A^{~B},
             \\
      \hat{\Delta}_{\a \mu}^{~\b}
      &=     \Delta_{\a u}^{~\b} \partial_\mu b^u
           + g A^\Lambda_\mu \partial_u k^v_\Lambda \CU^{u | A \a} \CU^\b_{A | v}.
    \end{align}
    \end{subequations}
  where $\omega^{ij}_\mu$ is the spin-connection.
  Also, the coupling constant $g$ in the covariant derivatives is a symbolic notation which comes from the gauging.
  If we consider the ungauged theory, these terms are vanish 
  and all the gauged covariant derivatives reduce to ordinary ones.
  
\section{The supersymmetry transformation laws}
  In order to discuss whether supersymmetry are broken or not in the subsequent chapter, 
  we need the supersymmetry transformation laws of all the fermions.
  Let $\e^A$ be the infinitesimal parameter of the supersymmetry transformations.
  The supersymmetry transformation laws of the fermion fields are given by
    \BEA
      \delta \psi_{A \mu}
      &=&    \CD_\mu \e_A 
           - \frac{1}{4} 
             \left(
             \partial_a \CK \bar{\lambda}^{aB} \e_B - \partial_{a^*} \bar{\lambda}^{a^*}_B \e^B
             \right) \psi_{A \mu}
             \NN \\
      & &  - \omega_{Au}^{~B} \CU^u_{C \a} 
             (\e^{CD} \mathbb{C}^{\a \b} \bar{\zeta}_\b \e_D + \bar{\zeta}^\a \e^C) \psi_{B \mu}
             \NN \\
      & &  + (A_A^{~\nu B} \eta_{\mu \nu} + A_A^{'~\nu B} \g_{\mu \nu}) \e_B
             \NN \\
      & &  + \e_{AB} (T^-_{\mu \nu} + U^+_{\mu \nu}) \g^\nu \e^B
           + i g S_{AB} \eta_{\mu \nu} \g^\nu \e^B
             \label{SUSYtrpsi}
             \\
      \delta \lambda^{aA}
      &=&    \frac{1}{4} 
             (\partial_b \CK \bar{\lambda}^{bB} \e_B - \partial_{b^*} \CK \bar{\lambda}^{b^*}_B \e^B) \lambda^{aA}
             \NN \\
      & &  - \omega^{A}_{~Bu} \CU^u_{C\a} 
             (\e^{CD} \mathbb{C}^{\a \b} \bar{\zeta}_\b \e^D + \bar{\zeta}^\a \e^C) \lambda^{aB}
             \NN \\
      & &  - \Gamma^a_{bc} \bar{\lambda}^{cB} \e_B \lambda^{bA} 
           + i (\nabla_\mu z^a - \bar{\lambda}^{aA} \psi_{A \mu}) \g^\mu \e^A
             \NN \\
      & &  + G^{- a}_{\mu \nu} \g^{\mu \nu} \e_B \e^{AB} + Y^{aAB} \e_B
           + g W^{aAB} \epsilon_B
             \label{SUSYtrlambda}
             \\
      \delta \zeta_\alpha
      &=&  - \Delta_{\a u}^{~\b} \CU^u_{A \gamma} 
             \left(
             \e^{AB} \mathbb{C}^{\gamma \beta} \bar{\zeta}_\beta \e_B + \bar{\zeta}^\gamma \e^A
             \right)
             \zeta_\b
             \NN \\
      & &  + \frac{1}{4} 
             (\partial_{a} \CK \bar{\lambda}^{aB} \e_B - \partial_{a^*} \CK \bar{\lambda}^{a^*}_B \e^B) \zeta_\a
             \NN \\
      & &  + i 
             \left(
             \CU_u^{B \b} \nabla_\mu b^u 
           - \e^{BC} \mathbb{C}^{\b \gamma} \bar{\zeta}_\gamma \psi_C
           - \bar{\zeta}^\b \psi^B
             \right)
             \g^\mu \e^A \e_{AB} \mathbb{C}_{\a \b} 
             \NN \\
      & &  + g N^A_\alpha \epsilon_A.
             \label{SUSYtrzeta}
    \EEA
  The supersymmetry transformation laws of the bosons are
    \BEA
      \delta e_\mu^i
      &=&  - i \bar{\psi}_{A \mu} \g^i \e^A - i \bar{\psi}^A_\mu \g^i \e_A,
             \\
      \delta A^\Lambda_\mu
      &=&    2 \bar{L}^\Lambda \bar{\psi}_{A \mu} \e_B \e^{AB} 
           + 2 L^\Lambda \bar{\psi}^A_\mu \e^B \e_{AB}
             \NN \\
      & &  + i (f^\Lambda_a \bar{\lambda}^{aA} \g_\mu \e^B \e_{AB} 
           + \bar{f}^\Lambda_{a^*} \bar{\lambda}^{a^*}_A \g_\mu \e_B \e^{AB}),
             \\
      \delta z^a
      &=&    \bar{\lambda}^{aA} \e_A,
             \\
      \delta b^u
      &=&    \CU^u_{A \a} (\bar{\zeta}^\a \e^A + \mathbb{C}^{\a \b} \e^{AB} \bar{\zeta}_\b \e_B),
    \EEA
  where 
    \begin{subequations}
    \begin{align}
      A_A^{~\mu B}
      &=   - \frac{i}{4} g_{a^* b}
             \left(
             \bar{\lambda}^{a^*}_A \g^\mu \lambda^{bB} 
           - \delta_A^B \bar{\lambda}^{a^*}_C \g^\mu \lambda^{b C}
             \right),
             \\
      A_A^{'\mu B}
      &=     \frac{i}{4} g_{a^* b}
             \left(
             \bar{\lambda}^{a^*}_A \g^\mu \lambda^{bB} 
           - \frac{1}{2} \delta_A^B \bar{\lambda}^{a^*}_C \g^\mu \lambda^{b C}
             \right)
           - \frac{i}{4} \delta_A^B \bar{\zeta}_\a \g^\mu \zeta^\a,
    \end{align}
    \end{subequations}
    \begin{subequations}
    \begin{align}
      T^-_{\mu \nu}
      &=     i (\Im \CN)_{\Lambda \Sigma} L^\Lambda
             \left(
             \tilde{F}^{\Sigma -}_{\mu \nu} 
           + \frac{1}{4} \nabla_a f^\Sigma_b \bar{\lambda}^{aA} \g_{\mu \nu} \lambda^{bB} \e_{AB}
           - \frac{1}{2} \mathbb{C}^{\a \b} \bar{\zeta}_\a \g_{\mu \nu} \zeta_\b L^\Sigma
             \right),
             \\
      T^+_{\mu \nu}
      &=     i (\Im \CN)_{\Lambda \Sigma} \bar{L}^\Lambda
             \left(
             \tilde{F}^{\Sigma +}_{\mu \nu} 
           + \frac{1}{4} \nabla_{a^*} \bar{f}^\Sigma_{b^*} \bar{\lambda}^{a^*}_A \g_{\mu \nu} \lambda^{b^*}_B \e^{AB}
           - \frac{1}{2} \mathbb{C}_{\a \b} \bar{\zeta}^\a \g_{\mu \nu} \zeta^\b \bar{L}^\Sigma
             \right),
    \end{align}
    \end{subequations}
    \begin{subequations}
    \begin{align}
      U^-_{\mu \nu}
      &=   - \frac{i}{4} \mathbb{C}^{\a \b} \bar{\zeta}_\a \g_{\mu \nu} \zeta_\b,
             \\
      U^+_{\mu \nu}
      &=   - \frac{i}{4} \mathbb{C}_{\a \b} \bar{\zeta}^\a \g_{\mu \nu} \zeta^\b,
    \end{align}
    \end{subequations}
    \begin{subequations}
    \begin{align}
      G^{a-}_{\mu \nu}
      &=   - \frac{1}{2} g^{ab^*} \bar{f}^\Sigma_{b^*} (\Im \CN)_{\Sigma \Lambda}
             \left(
             \tilde{F}^{\Lambda -}_{\mu \nu} 
           + \frac{1}{4} \nabla_a f^\Lambda_b \bar{\lambda}^{aA} \g_{\mu \nu} \lambda^{bB} \e_{AB}
           - \frac{1}{2} \mathbb{C}^{\a \b} \bar{\zeta}_\a \g_{\mu \nu} \zeta_\b L^\Lambda
             \right),
             \\
      G^{a^* +}_{\mu \nu}
      &=   - \frac{1}{2} g^{a^* b} f^\Sigma_b (\Im \CN)_{\Sigma \Lambda}
             \left(
             \tilde{F}^{\Lambda +}_{\mu \nu} 
           + \frac{1}{4} \nabla_{c^*} \bar{f}^\Lambda_{d^*} \bar{\lambda}^{c^*}_A \g_{\mu \nu} \lambda^{d^*}_B \e^{AB}
           - \frac{1}{2} \mathbb{C}_{\a \b} \bar{\zeta}^\a \g_{\mu \nu} \zeta^\b \bar{L}^\Lambda
             \right),
    \end{align}
    \end{subequations}
    \BE
      Y^{aAB}
       =     \frac{i}{2} g^{ab^*} C_{b^* c^* d^*} \bar{\lambda}^{c^*}_C \lambda^{d^*}_D \e^{AC} \e^{BD}.
    \EE
  In the above equations, $\tilde{F}^\Lambda_{\mu \nu}$ are the supercovariant field strengths defined by:
    \BEA
      \tilde{F}^\Lambda_{\mu \nu}
      &=&    F^\Lambda_{\mu \nu} + \bar{L}^\Lambda \bar{\psi}^A_\mu \psi^B_\nu \e_{AB}
           + \bar{L}^\Lambda \bar{\psi}_{A \mu} \psi_{B \nu} \e^{AB}
             \NN \\
      & &  - i f_a^\Lambda \bar{\lambda}^{aA} \g_{[\nu} \psi^B_{\mu]} \e_{AB}
           - i \bar{f}_{a^*}^\Lambda \bar{\lambda}^{a^*}_A \g_{[\nu} \psi_{B\mu]} \e^{AB}.
    \EEA
\chapter{Partial supersymmetry breaking in the $\CN=2$ $U(1)$ gauged model}
  The simplest realization of the partial breaking of $\CN=2$ local supersymmetry has been discussed 
  in reference \cite{Ferrara1}.
  The model includes a $U(1)$ vector multiplet and a hypermultiplet.
  In this chapter, we review this model and it will be generalized to $U(N)$ gauged model in next chapter.
  
  First of all, the parametrizations of the vector multiplet and the hypermultiplet are given in section 5.1.
  In section 5.2, we see that the $\CN=2$ local supersymmetry is broken to $\CN=1$.
  This is confirmed by the appearance of a Nambu-Goldstone fermion and
  the mass spectrum which is computed in section 5.3.
  Finally, in section 5.4, we summarize the results and discuss its applications.
\section{$\CN=2$ $U(1)$ gauged supergravity model}
\subsection{Vector multiplet}
  The $U(1)$
  \footnote[2]{To be precise, the gauge symmetry is $U(1) \times U(1)_{graviphoton}$.} 
  vector multiplet contains a complex scalar field $z$ 
  which spans the special K\"ahler manifold of complex dimension $1$.
  We start from the case in which the holomorphic prepotential $F(X^0, X^1)$ exists.
  Since the prepotential $F(X)$ is a homogeneous of second degree of the coordinates $X^\Lambda(z)$ ($\Lambda=0,1$),
  we can write $F$ as follows,
    \BE
      F(X^0, X^1)
       =     (X^0)^2 \CF(X^1/X^0).
    \EE
  We can evaluate $F_\Lambda$ by taking the derivative of $F$ with respect to $X^\Lambda$,
    \BEA
      F_0
      &=&    2 X^0 \CF(X^1/X^0) - X^1 \frac{\partial}{\partial (X^1/X^0)} \CF(X^1/X^0),
             \NN \\
      F_1
      &=&    X^0 \frac{\partial}{\partial (X^1/X^0)} \CF(X^1/X^0).
             \label{U100}
    \EEA
  It is natural to choose the coordinate $X^\Lambda(z)$ linearly independent: 
    \BE
      X^0 (z)
       =     \frac{1}{\sqrt{2}},
             ~~~
      X^1 (z)
       =     \frac{1}{\sqrt{2}} z.
             \label{U101}
    \EE
  Substituting (\ref{U101}) into (\ref{U100}), we have
    \BEA
      F_0(z)
      &=&   \frac{1}{\sqrt{2}} \left( 2 \CF(z) - z \frac{\partial \CF(z)}{\partial z} \right),
            \NN \\  
      F_1(z)
      &=&   \frac{1}{\sqrt{2}} \frac{\partial \CF(z)}{\partial z}.
    \EEA
  To specify the model, we have to choose a particular form of the holomorphic function $\CF(z)$.
  Our choice, here, is
    \BE
      \CF (z)
       =  - i z.
    \EE
  This is the simplest choice of the function $\CF(z)$: it corresponds to consider the microscopic (or bare) theory.
  The holomorphic section $\Omega(z)$ can be written as,
    \BE
      \Omega(z)
       =     \left(
             \begin{array}{c}
             X^\Lambda \\
             F_\Lambda
             \end{array}
             \right)
       =     \frac{1}{\sqrt{2}}
             \left(
             \begin{array}{c}
             1 \\
             z \\
             - i z \\
             - i
             \end{array}
             \right).
    \EE
  We can easily compute the K\"ahler potential (\ref{CK}) and its derivative:
    \BEA
      \CK
      &=&  - \log (z + \bar{z})
             \label{U11},
             \\
      \partial_{z} \CK
      &=&  - (z + \bar{z})^{-1}
             \label{U12}.
    \EEA
  By using above equations, the K\"ahler metric and the Levi-Civita connection (\ref{LeviCivita}) are
    \BEA
      g_{z \bar{z}}
      &=&    \partial_z \partial_{\bar{z}} \CK
       =     (z + \bar{z})^{-2},
             \label{U13}
             \\
      \Gamma^z_{zz}
      &=&  - g^{zz^*} \partial_z g_{zz^*}
       =     2 (z + \bar{z})^{-1},
             \label{U14}
    \EEA
  where we have used the shorthand notation: 
  $\partial_z = \partial/\partial z$ and $\partial_{z^*} = \partial/\partial \bar{z}$.
  
  Now perform the symplectic transformation, 
    \BE
      \Omega
       \rightarrow
            \tilde{\Omega}
       =    \CS \Omega
       =    \left(
            \begin{array}{cccc}
            1 & 0 & 0 & 0 \\
            0 & 0 & 0 & -1 \\
            0 & 0 & 1 & 0 \\
            0 & 1 & 0 & 0
            \end{array}
            \right)
            \Omega
       =    \frac{1}{\sqrt{2}}
            \left(
            \begin{array}{c}
            1 \\
            i \\
            -iz \\
            z 
            \end{array}
            \right),
            \label{U15}
    \EE
  where $\CS \in Sp(4, \mathbb{R})$.
  After this mapping, the transformed section $\tilde{\Omega}$ can no longer be written 
  in the standard form with the prepotential $\tilde{F}$, 
  because the last two components clearly can not be written as functions of the first two.
  Thus no prepotential $\tilde{F}(\tilde{X}^0, \tilde{X}^1)$ exists.
  We still have the same K\"ahler manifold (\ref{U11})-(\ref{U14}) 
    \footnote[8]{The K\"ahler potential $\CK$ has been defined by the symplectic invariant way.
                 Therefore, the K\"ahler metric and the Levi-Civita connection are same.}, 
  but with different couplings of the scalar fields to the vectors and to the fermions.
  
  In the following, we use the transformed section (\ref{U15}) and omit tilde $\tilde{}$:
    \BE
      X^\Lambda
       =     \frac{1}{\sqrt{2}}
             \left(
             \begin{array}{c}
             1 \\
             i 
             \end{array}
             \right),
             ~~~~~
      F_\Lambda
       =     \frac{1}{\sqrt{2}}
             \left(
             \begin{array}{c}
             -iz \\
             z 
             \end{array}
             \right).
    \EE
  
  For future reference, let us evaluate $f^\Lambda_z$ and $h_{\Lambda | z}$.
  Substituting (\ref{U11})-(\ref{U15}) into (\ref{Ua}), we obtain
    \BEA
      f^\Lambda_z
      &=&    e^{\CK/2} (\partial_z + \partial_z \CK) X^\Lambda
             \NN \\
      &=&  - \frac{1}{\sqrt{2}} (z + \bar{z})^{-3/2} 
             \left(
             \begin{array}{c}
             1 \\
             i
             \end{array}
             \right),
             \\
      h_{\Lambda | z}
      &=&    e^{\CK/2} (\partial_z + \partial_z \CK) F_\Lambda
             \NN \\
      &=&    \frac{1}{\sqrt{2}} \bar{z} (z + \bar{z})^{-3/2}
             \left(
             \begin{array}{c}
             -i \\
             1
             \end{array}
             \right).
    \EEA
  Furthermore, the covariant derivative of $f^\Lambda_z$ can be obtained 
  by substituting (\ref{U11})-(\ref{U15}) into (\ref{nablaU}), 
    \BEA
      \nabla_z f^\Lambda_z
      &=&    \partial_z f^\Lambda_z + \frac{1}{2} \partial_z \CK f^\Lambda_z + \Gamma^z_{zz} f^\Lambda_z
             \NN \\
      &=&    0.
             \label{U1nablaf}
    \EEA
  If we compare (\ref{U1nablaf}) with (\ref{Cabc}), we obtain
    \BE
      C_{zzz}
       =     0.
    \EE
  This is because $\CF$ has been chosen as the simplest function.
  
\subsection{Hypermultiplet}
  Before going to details, we must explain why one hypermultiplet is needed to break half supersymmetry.
  This discussion is same for the $U(N)$ gauged case in the next chapter.
  If $\mathcal{N}=2$ local supersymmetry is broken to $\mathcal{N}=1$ spontaneously,
  one gravitino remains massless but the other becomes massive by super-Higgs mechanism.
  So, it is important to note that in addition to super-Higgs mechanism Higgs mechanism must occur.
  Since $\mathcal{N}=1$ supersymmetry is manifest, 
  the massive gravitino forms $\mathcal{N}=1$ massive multiplet of spin $ (3/2,1,1,1/2) $.
  Thus two gauge bosons, that is $U(1)$ gauge boson and graviphoton, must become massive by absorbing the scalar fields,
  in other words, Higgs mechanism must happen so as to keep $\mathcal{N}=1$ supersymmetry.
  Therefore we need at least one hypermultiplet 
  and at least two $U(1)$ translational isometries which provide Higgs mechanism.
  
  In this chapter (and in the next chapter), 
  we take the same parametrizations as those of \cite{Ferrara1, Ferrara2, David, IM}.
  The $SU(2)$ connection $\omega^x$ and the $\CS\CU$ curvature (\ref{Omega}) are parametrized as,
    \begin{equation}
      \omega^x_u 
       =    \frac{1}{b^0} \delta^x_u,~~~~
      \Omega^x_{0u}
       =  - \frac{1}{2(b^0)^2} \delta^x_u,~~~~
      \Omega^x_{yz}
       =    \frac{1}{2(b^0)^2} \epsilon^{xyz}. 
    \end{equation}
  Furthermore, by using (\ref{K2}), the metric $ h_{uv} $ of this manifold is given by
    \begin{equation}
      h_{uv}
       =    \frac{1}{2(b^0)^2} \delta_{uv}.
            \label{U16}
    \end{equation}
  In order to write down the fermion mass matrices and the supersymmetry transformation laws,
  we also need the symplectic vielbein $ \mathcal{U}^{\alpha A}_u db^u $ ($ \alpha, A = 1,2$).
  It leads:
    \BE
      \mathcal{U}^{\alpha A}
       =    \frac{1}{2b^0} \epsilon^{\alpha \beta} (db^0 - i \sigma^x db^x)_\beta^{~A},
    \EE
  where $ \sigma^x $ are the standard Pauli matrices
   \footnote[7]{The notations are explained in appendix A.}.
  
  The metric (\ref{U16}) is invariant under arbitrary constant translations of the coordinates $b^1, b^2, b^3$ 
  because it does not depend on $b^1, b^2, b^3$.
  As we have seen above, in order for the $\CN=2$ supersymmetry to be broken partially, 
  we have to gauge two of the isometries.
  Here we choose the following translational isometries of two coordinates $b^2, b^3$ as the gauged isometries:
    \BEA
      b^2
      &\rightarrow&
             b^2 + \e^2 (g_2 + g_3),
             \\
      b^3
      &\rightarrow&
             b^3 + \e^3 g_1,
    \EEA
  where $ g_1, g_2, g_3, \in \mathbb{R}$.
  The Killing vectors $ k^u_\Lambda $ which generate these isometries can be written as follows:
    \BEA
      k^u_0 
      &=&    g_1 \delta^{u3} + g_2 \delta^{u2},
             \label{U17}
             \\
      k^u_1
      &=&    g_3 \delta^{u2},
    \EEA
  and the associated Killing potentials $ \mathcal{P}^x_\Lambda $ for these vectors are
    \BEA
      \mathcal{P}^x_0 
      &=&    \frac{1}{b^0} (g_1 \delta^{x3} + g_2 \delta^{x2}),
             \\
      \mathcal{P}^x_1
      &=&    \frac{1}{b^0} g_3 \delta^{x2}.
             \label{U18}
    \EEA
\section{Partial supersymmetry breaking}
  In this section, we see that the local $\CN=2$ supersymmetry is broken to $\CN=1$ in the simple model
  which has been given in the last section.
  This will be confirmed by the appearance of a Nambu-Goldstone fermion: 
  the fermion whose supersymmetry transformation on the vacuum is non-zero.
  Since the supersymmetry transformation laws of the fermions are mainly written by the mass matrices, 
  let us evaluate it here.
  By substituting the parametrizations in the last section into (\ref{S})-(\ref{M2}), 
  we have
    \BEA
      S_{AB}
      &=& - \frac{i e^{\CK/2}}{2 \sqrt{2} b^0} 
            \left(
            \begin{array}{cc}
            i g_2 + g_3 & g_1 \\
            g_1 & i g_2 + g_3
            \end{array}
            \right),
            \label{U1S} \\
      W^{z AB}_1 
      &=&   0,
            \\
      W^{z AB}_2
      &=&  - \frac{e^{\CK/2}}{\sqrt{2} b^0} (z + \bar{z})
             \left(
             \begin{array}{cc}
             g_2 + i g_3 & i g_1
             \\
             i g_1 & g_2 + i g_3
             \end{array}
             \right),
    \EEA
    \BEA
      N^A_\alpha
      &=&   \frac{i e^{\CK/2}}{\sqrt{2} b^0} 
            \left(
            \begin{array}{cc}
            g_1 & - i g_2 + g_3 \\
            i g_2 - g_3 & - g_1
            \end{array}
            \right),
            \label{U1N} \\
      \mathcal{M}^{\alpha \beta}
      &=&   \frac{i e^{\CK/2}}{\sqrt{2} b^0}
            \left(
            \begin{array}{cc}
            - i g_2 - g_3 & g_1 \\
            g_1 & - i g_2 - g_3
            \end{array}
            \right),
            \\
      \mathcal{M}^\alpha_{z B}
      &=&   \frac{\sqrt{2} i e^{\CK/2}}{b^0} (z + \bar{z})^{-1}
            \left(
            \begin{array}{cc}
            g_1 & i g_2 +  g_3 
            \\
            - i g_2 - g_3 & - g_1
            \end{array}
            \right),
            \\
      \mathcal{M}_{1;z A \vert z B}
      &=&   0,
            \\
      \mathcal{M}_{2;z A \vert z B}
      &=&   0.
            \label{U1M2}
    \EEA
  where (\ref{U1M2}) comes from (\ref{U1nablaf})
\subsubsection{The scalar potential}
  The scalar potential $V(z, \bar{z}, b)$ is given by (\ref{V}).
  If we substitute the parametrizations in the last section into $V$, we have
    \BEA
      V(z, \bar{z}, b)
      &=&    g^{zz^*} \CP^x_\Lambda \CP^x_\Sigma f^\Lambda_z \bar{f}^\Sigma_{\bar{z}}
           + 4 h_{uv} k^u_\Lambda k^v_\Sigma \bar{L}^\Lambda L^\Sigma
           - 3 \CP^x_\Lambda \CP^x_\Sigma \bar{L}^\Lambda L^\Sigma
             \NN \\
      &=&    0,
    \EEA
  identically, for any value of $g_1, g_2, g_3$ and also of $z$ and $b^u$.
  In the following, suppose that we choose one of the vacua, that is, we know the value of $z$ at the vacuum 
  and write the vacuum expectation value of $\dots$ as $\expect{\dots}$.
  In this model, we cannot determine the vacuum expectation value of $z$.
  In the next chapter, the scalar potential takes the non-trivial form 
  and we can determine the value of $z$ at the vacuum.
\subsubsection{Supersymmetry transformations of the fermions}
  Using (\ref{U1S})-(\ref{U1N}), the vacuum expectation values of the supersymmetry transformation laws of 
  the fermions (\ref{SUSYtrpsi})-(\ref{SUSYtrzeta}) are
    \BEA
      \expect{\delta \psi_{A \mu}}
      &=&    i \expect{S_{AB}} \g_\mu \e^B
             \NN \\
      &=&    \expect{\frac{e^{\CK/2}}{2\sqrt{2} b^0}} 
             \left(
             \begin{array}{cc}
             i g_2 + g_3 & g_1 \\
             g_1 & i g_2 + g_3
             \end{array}
             \right)
             \g_\mu
             \left(
             \begin{array}{c}
             \e^1 \\
             \e^2
             \end{array}
             \right),
             \label{U1S2}
             \\
      \expect{\delta \lambda^{z A}}
      &=&    W^{z AB} \e_B
             \NN \\
      &=&  - \expect{\frac{e^{\CK/2} (z + \bar{z})}{\sqrt{2}b^0}}
             \left(
             \begin{array}{cc}
             g_2 + i g_3 & i g_1
             \\
             i g_1 & g_2 + i g_3
             \end{array}
             \right)
             \left(
             \begin{array}{c}
             \e_1 \\
             \e_2
             \end{array}
             \right),
             \\
      \expect{\delta \zeta_\a}
      &=&    N_\a^{A} \e_A
             \NN \\
      &=&    \expect{\frac{e^{i \CK/2}}{\sqrt{2}b^0}}
             \left(
             \begin{array}{cc}
             g_1 & - i g_2 + g_3 \\
             i g_2 - g_3 & - g_1
             \end{array}
             \right)
             \left(
             \begin{array}{c}
             \e_1 \\
             \e_2
             \end{array}
             \right).
             \label{U1N2}
    \EEA
  If each matrix has one zero-eigenvalue, the supersymmetry transformation corresponding to one direction is zero 
  and that corresponding to another direction is non-zero, 
  so we can see the $\CN=2$ local supersymmetry is broken to $\CN=1$.
  In order for each matrix to have one zero-eigenvalue, 
  we have to impose the following conditions on the coupling constants:
    \BE
      g_2 
       =     0,
             ~~~
      g_1
       =     \pm g_3.
             \label{U1vacuumcondition1}
    \EE
  Notice that $g_1, g_2, g_3 \in \mathbb{R}$.
  In other words, when the conditions (\ref{U1vacuumcondition1}) are satisfied, 
  $\CN=2$ local supersymmetry is broken to $\CN=1$.
  
  In the following, we choose the coupling constants as follows, 
    \BE
      g_2 
       =     0,
             ~~~
      g_1
       =     g_3.
             \label{U1vacuumcondition2}
    \EE
  Substituting (\ref{U1vacuumcondition2}) into (\ref{U1S2})-(\ref{U1N2}),
    \BEA
      \expect{\delta \psi_{A \mu}}
      &=&    \expect{\frac{e^{\CK/2}}{2\sqrt{2} b^0}} g_1
             \left(
             \begin{array}{cc}
             1 & 1 \\
             1 & 1
             \end{array}
             \right)
             \g_\mu
             \left(
             \begin{array}{c}
             \e^1 \\
             \e^2
             \end{array}
             \right),
             \\
      \expect{\delta \lambda^{z A}}
      &=&  - \expect{\frac{i e^{\CK/2} (z + \bar{z})}{\sqrt{2}b^0}} g_1
             \left(
             \begin{array}{cc}
             1 & 1
             \\
             1 & 1
             \end{array}
             \right)
             \left(
             \begin{array}{c}
             \e_1 \\
             \e_2
             \end{array}
             \right),
             \\
      \expect{\delta \zeta_\a}
      &=&    \expect{\frac{i e^{\CK/2}}{\sqrt{2}b^0}} g_1
             \left(
             \begin{array}{cc}
             1 & 1 \\
             - 1 & - 1
             \end{array}
             \right)
             \left(
             \begin{array}{c}
             \e_1 \\
             \e_2
             \end{array}
             \right).
    \EEA
  If we define $ \phi_{\pm} = \frac{1}{\sqrt{2}} (\phi_1 \pm \phi_2) $ where $ \phi \in {\psi, \zeta, \lambda} $ 
  (the left or right chirality are denoted by the upper or lower position of the index $ \pm $), 
  their supersymmetry transformations on the vacuum are
    \BEA
      \expect{\delta \psi_{+ \mu}}
      &=&    \expect{\frac{e^{\mathcal{K}/2}}{\sqrt{2} b^0}} g_1
             \gamma_\mu (\epsilon_1 + \epsilon_2),
             \nonumber \\
      \expect{\delta \lambda^{z +}}
      &=&  - \expect{\frac{\sqrt{2} i e^{\CK/2} (z + \bar{z})}{b^0}} g_1
             (\epsilon_1 + \epsilon_2),
             \nonumber \\
      \expect{\delta \zeta_{-}}
      &=&    \expect{\frac{\sqrt{2} i e^{\mathcal{K}/2}}{b^0}} g_1
             (\epsilon_1 + \epsilon_2),
             \nonumber \\
      \expect{\delta \psi_{- \mu}}
      &=&    \expect{\lambda^{z -}}
       =     \expect{\zeta_{+}}
       =     0.
    \EEA
  As we will see in the next section, the gravitino $\psi^+_\mu$ becomes massive by the super-Higgs mechanism, 
  while $\psi^-_\mu$ remains massless.
  We define linear combinations of the fermions $\lambda^{z +}$ and $\zeta_{-}$ such that
    \BEA
      \chi_{\bullet}
      &\equiv&    
           - \expect{(z + \bar{z})^{-1}} \lambda^{z +} + 2 \zeta_{-},
             \nonumber \\
      \eta_\bullet 
      &\equiv&
             \expect{(z + \bar{z})^{-1}} \lambda^{z +} + \zeta_{-},
    \EEA
  whose supersymmetry transformations on the vacuum are
    \BEA
      \expect{\delta \chi_\bullet}
      &=&    \expect{\frac{3 \sqrt{2} i e^{\mathcal{K}/2}}{b^0}} g_1
             (\epsilon_1 + \epsilon_2),
             \NN \\
      \expect{\delta \eta_\bullet}
      &=&    0,
    \EEA
  where the upper or lower position of dot represents left or right chirality respectively.
  Note that we cannot make the expectation values of the supersymmetry transformations 
  of all the ( spin-1/2 ) fermions zero 
  and only the supersymmetry transformations of the fermion $\chi_\bullet$ takes the non-zero value.
  Thus, $\chi_\bullet$ is the Nambu-Goldstone fermion and the $\CN=2$ supersymmetry is broken to $\CN=1$.
\section{The mass spectrum}
\subsection{Fermion mass}
  Let us discuss the occurrence of the super-Higgs mechanism.
  To see this, we have to evaluate the fermion mass terms of the Lagrangian $\CL_{\rm{Yukawa}}$ (\ref{Lmass}).
    \BEA
      \CL_{\rm{Yukawa}}
      &=&  - \expect{\frac{\sqrt{2} i g_1 e^{\CK/2}}{b^0}}
             \left(
             \bar{\psi}^{+}_\mu \gamma^{\mu \nu} \psi^{+}_\nu
           - i \bar{\chi}^\bullet \gamma_\mu \psi^\mu_{+}
           + \frac{1}{3} \bar{\chi}_\bullet \chi_\bullet
           - \frac{1}{3} \bar{\eta}_\bullet \eta_\bullet
             \right)
           + \ldots
           + h.c.
             \NN \\
      & &
    \EEA
  We can, also, confirm $\chi_\bullet$ is the Nambu-Goldstone fermion 
  from the fact that it couples to the gravitino $\psi^{+}_\mu$ in the second term.
  Such a field can be gauged away by a suitable gauge transformation of the gravitino $\psi^{+}_\mu$:
    \BE
      \psi^+_\mu
       \rightarrow \psi^+_\mu + \frac{i}{6} \gamma_{\mu} \chi_\bullet,
    \EE
  which results in
    \BE
      \CL_{\rm{Yukawa}}
       =   - i  \expect{\frac{\sqrt{2} g_1 e^{\mathcal{K}/2}}{b^0}}
             \left(
             \bar{\psi}^{+}_\mu \gamma^{\mu \nu} \psi^{+}_\nu
           - \frac{1}{3} \bar{\eta}_\bullet \eta_\bullet
             \right)
           + \ldots
           + h.c.
    \EE
  The super-Higgs mechanism has occurred, that is, by absorbing the Nambu-Goldstone fermion, 
  the gravitino $ \psi^+_\mu $ has acquired a mass.
  
  The kinetic terms of the massive fermions in the $\CL_{\rm{kin}}$ (\ref{Lkin}) are  
    \BE
      \CL_{\rm{kin}}
       =     \frac{\epsilon^{\mu \nu \lambda \sigma}}{\sqrt{-g}}
             \bar{\psi}^A_\mu \gamma_\nu \partial_\lambda \psi_{A \sigma}
           - \frac{i}{3} \bar{\eta}^\bullet \gamma_\mu \partial^\mu \eta_\bullet
           + \ldots
           + h.c.
    \EE
  It is easy to find the mass of the fermions by evaluating the equations of motion of them.
  As a result, the gravitino mass $m$ is
    \BE
      m
       =     | \expect{\frac{\sqrt{2} g_1 e^{\mathcal{K}/2}}{b^0}} |.
             \label{U1m}
    \EE
  Notice that the mass of the physical fermion $ \eta_\bullet $ is the same as the gravitino, that is, $ m $.
  Therefore, we can anticipate that $ \psi^+_\mu $ and $ \eta_\bullet $ form a $\CN=1$ massive multiplet of spin (3/2,1,1,1/2).
  On the other hand, $ \lambda^{z -} $, together with the scalar fields, 
  are expected to form $ \mathcal{N} =1 $ massless chiral multiplet.
  To make sure, we must consider the masses of the gauge bosons and the scalar fields.
\subsection{Boson mass}
  Let us consider the masses of the gauge bosons.
  They appear in the kinetic term of the hypermultiplet scalar:
    \BEA
      h_{uv} \nabla_\mu b^u \nabla^\mu b^v
      &=&    \frac{1}{2 (b^0)^2} \delta_{uv}
             (\partial_\mu b^u + A^\Lambda_\mu k_\Lambda^u)(\partial^\mu b^v + A^{\Lambda \mu} k_\Lambda^v)
             \NN \\
      &=&    \frac{1}{2 (b^0)^2} 
             \left(
             g_1^2 A^0_\mu A^{0 \mu} + g_1^2 A^1_\mu A^{1 \mu}
           + 2 g_1 (A_\mu^0 \partial^\mu b^3 + A_\mu^1 \partial^\mu b^2)
             \right)
           + \ldots,
             \label{U1gaugemass}
    \EEA
  Furthermore the massless scalar fields $b^2, b^3$ can be eliminated from (\ref{U1gaugemass}) 
  by employing the gauge transformations of $A_\mu^0, A_\mu^1$:
    \BE
      A_\mu^0
       \rightarrow
             A_\mu^0 - \frac{1}{g_1} \partial_\mu b^3,
             ~~~~~~~
      A_\mu^1
       \rightarrow
             A_\mu^1 - \frac{1}{g_1} \partial_\mu b^2.
    \EE
  This is the ordinary Higgs mechanism and the scalar fields $b^2, b^3$ are the Nambu-Goldstone bosons.
  
  The gauge kinetic terms are 
  $ \frac{1}{4} (\Im \mathcal{N})_{\Lambda \Sigma} F^\Lambda_{\mu \nu} F^{\Sigma \mu \nu} $, 
  so we must compute the generalized coupling matrix $ \mathcal{N} $ (\ref{CouplingMatrix}):
    \BE
      \bar{\CN}_{\Lambda \Sigma}
       =     h_{\Lambda | I} \circ (f^{-1})^I_{~\Sigma},
    \EE
  with
    \BEA
      h_{\Lambda | I}
      &=&    \left(
             \begin{array}{cc}
             \bar{M}_\Lambda & h_{\Lambda | z}
             \end{array}
             \right) 
       =     \frac{e^{\CK/2} \bar{z}}{\sqrt{2}}
             \left(
             \begin{array}{cc}
             i & - \frac{i}{z + \bar{z}} \\
             1 & \frac{1}{z + \bar{z}}
             \end{array}
             \right),
             \\
      f^\Lambda_I
      &=&    \left(
             \begin{array}{cc}
             \bar{L}^\Lambda & f^\Lambda_z
             \end{array}
             \right)
       =     \frac{e^{\CK/2}}{\sqrt{2}}
             \left(
             \begin{array}{cc}
             1 & - \frac{1}{z + \bar{z}} \\
             - i & - \frac{i}{z + \bar{z}}
             \end{array}
             \right).
    \EEA
  Therefore, the matrix $\CN_{\Lambda \Sigma}$ is
    \BE
      \CN_{\Lambda \Sigma}
       =   - i z
             \left(
             \begin{array}{cc}
             1 & 0 \\
             0 & 1
             \end{array}
             \right).
    \EE
  The kinetic terms of the gauge bosons are written as follows,
    \BE
      \frac{1}{4} \Im \expect{\mathcal{N}_{\Lambda \Sigma}} F^\Lambda_{\mu \nu} F^{\Sigma \mu \nu}
       =   - \frac{1}{4} \frac{\expect{e^{-\CK}}}{2}
             F^0_{\mu \nu} F^{0 \mu \nu}
           - \frac{1}{4} \frac{\expect{e^{-\CK}}}{2}
             F^1_{\mu \nu} F^{1 \mu \nu}.
             \label{U1gaugekin}
    \EE
  From (\ref{U1gaugemass}) and (\ref{U1gaugekin}), we can read off the masses of the gauge bosons.
  As a result, both of them agree with (\ref{U1m}), that is,
    \BE
      m
       =     | \expect{\frac{\sqrt{2} g_1 e^{\mathcal{K}/2}}{b^0}} |.
    \EE
  The $U(1)$ gauge boson and the graviphoton have acquired the mass.
  
  Since the scalar potential $V$ is identically zero, there is no mass term of the vector multiplet scalar $z$.
  Thus, all the scalar fields are massless.
  
  We summarize the mass spectrum of our model in the table 5.1:
  \\
  $\CN=1$ gravity multiplet contains the vierbein $e^i_{\mu}$ and the gravitino $\psi^-_\mu$.
  On the other hand, the massive gravitino $\psi^+_\mu$, the $U(1)$ gauge boson $A^1_\mu$, the graviphoton $A^0_\mu$ 
  and the fermion $ \eta_\bullet $ form a massive spin-3/2 multiplet.
  The gaugino $\lambda^{z -}$ and the complex scalar field $z$ form the massless chiral multiplet.
  The hyperino $ \zeta_+ $ and the scalar $ b^0,b^1 $ form $ \mathcal{N} = 1 $ chiral multiplet.
  Note that the $U(1) \times U(1)_{graviphoton}$ gauge symmetry is completely broken
  and the vacuum lie in the Higgs phase.
    \begin{table}[h]
      \begin{center}
        \begin{tabular}{|c|c|c|}
          \hline
          $ \CN = 1 $ multiplet & field & mass \\ \hline \hline
          gravity multiplet & $ e^i_{\mu} $, $ \psi^{-}_{\mu} $ & 0 \\ \hline
          spin-3/2 multiplet & $ \psi^{+}_{\mu} $, $ A^0_\mu $, $ A^1 $, $\eta_\bullet$ & $m$ \\ \hline
          chiral multiplet & $ \lambda^{z -} $, $z$ & $0$ \\ \hline
          chiral multiplet & $ \zeta_+ $, $ b^0,b^1 $ & 0 \\
          \hline
        \end{tabular}
      \caption{the mass spectrum}
      \end{center}
    \end{table}
\section{Conclusion and discussion}
  We have seen that, in the $\CN=2$ $U(1)$ gauged supergravity model, 
  the $\CN=2$ supersymmetry has been broken to $\CN=1$ counterpart.
  The super-Higgs and the Higgs mechanisms are observed:
  The Nambu-Goldstone fermion and the Nambu-Goldstone bosons are absorbed 
  by the gravitino and the gauge bosons respectively.
  Also the gauge symmetry is broken by the Higgs mechanisms.
  
  To make more realistic model, we have to consider non-Abelian gauge group.
  But, if $\CN=2$ local supersymmetry is broken to $\CN=1$, by the Higgs mechanisms, 
  the gauge symmetry corresponding to the graviphoton 
  and to one of the gauge bosons of the vector multiplet sector have to be broken.
  Thus, it is desirable that the gauge symmetry of the vector multiplet sector is $U(1) \times$ (non-Abelian).
  Indeed it has been showed that, in the case of the $U(1) \times$(compact group), 
  partial supersymmetry breaking has occurred \cite{Fre}.

  The model which we have used in this section is the simplest one 
  because we have chosen the holomorphic function $\CF$ as the simple function.
  The resulting theory is the microscopic (or bare) theory.
  This leads to the following question;
  \textit{Does partial supersymmetry breaking occur in $\CN=2$ effective theory ?}
  We will answer the question in the next chapter:
  we will use $U(N)$ gauged effective model by keeping $\CF$ general holomorphic function.

\chapter{Partial supersymmetry breaking in the $\CN=2$ $U(N)$ gauged model}
  In this chapter, we will see that, in the $\CN=2$ supergravity model 
  which includes a $U(N)$
    \footnote[2]{To be precise, the gauge symmetry is $U(N) \times U(1)$.
                  The $U(1)$ gauge group comes from the graviphoton.} 
  vector multiplet and a hypermultiplet,
  the $\CN=2$ supersymmetry is broken to $\CN=1$ spontaneously with keeping the $SU(N)$ gauge symmetry manifest \cite{IM}.
  This is a generalization of the model which we have reviewed in the last chapter.
  We do not set the section as the simplest function here.
  It leads to the effective theory which contains higher order coupling terms of the scalar fields.
  In this sense, it is not simple generalization.
  
  The organization is parallel to that of the chapter 5.
  First of all, the parametrizations of the vector multiplet and hypermultiplet are given in section 6.1.
  In section 6.2, we observe that the $\CN=2$ local supersymmetry is broken to $\CN=1$ spontaneously.
  The mass spectrum is computed in section 6.3 and the resulting $\CN=1$ Lagrangian is obtained in section 6.4.
  Finally, in section 6.5, we summarize the results.
  
\section{$\CN=2$ $U(N)$ gauged supergravity model}
\subsection{$U(N)$ vector multiplet}
  The vector multiplets contain the complex scalar fields $z^a$.
  The index $a=1,\ldots,N^2$ label the $U(N)$ gauge group and $N^2$ refers to the overall $U(1)$.
  For simplicity, we write $N^2=n$ below.
  We start from the case where the holomorphic prepotential $F(X^0, X^a)$ exists.
  The prepotential $F(X^0, X^a)$ can be written,
    \BE
      F(X^0, X^a)
       =     (X^0)^2 \CF(X^a/X^0).
    \EE
  By taking the $X^\Lambda$ ($\Lambda = 0, 1 \ldots ,n$) derivative of $F$, we obtain
    \BEA
      F_0
      &=&    2 X^0 \CF(X^b/X^0) - X^a \frac{\partial}{\partial (X^a/X^0)} \CF(X^b/X^0),
             \NN \\
      F_a
      &=&    X^0 \frac{\partial}{\partial (X^a/X^0)} \CF(X^b/X^0).
             \label{UN00}
    \EEA
\newpage
  It is natural to choose the upper part of the holomorphic section, $X^\Lambda(z)$, as 
    \BE
      X^0 (z)
       =     \frac{1}{\sqrt{2}},
             ~~~
      X^a (z)
       =     \frac{1}{\sqrt{2}} z^a,
             \label{UN01}
    \EE
  which leads to
    \BEA
      F_0(z)
      &=&   \frac{1}{\sqrt{2}} \left( 2 \CF(z) - z^a \frac{\partial \CF(z)}{\partial z^a} \right),
            \NN \\  
      F_a(z)
      &=&   \frac{1}{\sqrt{2}} \frac{\partial \CF(z)}{\partial z^a}.
    \EEA
  where we have defined $\partial_a=\partial/\partial z^a$.
  We keep $\CF(z)$ general function for a moment.
  
  Let us perform the symplectic rotation which is similar to that in the last chapter.
  We rotate the overall $U(1)$ part, that is, $X^n \rightarrow - F_n$ and $F_n \rightarrow X^n$.
  As a result, the holomorphic sections are
    \begin{eqnarray}
      X^0(z)
       =    \frac{1}{\sqrt{2}},~~~~
      F_0(z)
      &=&   \frac{1}{\sqrt{2}} \left( 2 \CF(z) - z^a \frac{\partial \CF(z)}{\partial z^a} \right),
            \nonumber \\  
      X^{\hat{a}}(z)
       =    \frac{1}{\sqrt{2}} z^{\hat{a}},~~~~
      F_{\hat{a}}(z)
      &=&   \frac{1}{\sqrt{2}} \frac{\partial \CF(z)}{\partial z^{\hat{a}}},
            \\ \label{prepotential}
      X^n(z)
       =    \frac{1}{\sqrt{2}} \frac{\partial \CF(z)}{\partial z^n},~~~~
      F_n(z)
      &=& - \frac{1}{\sqrt{2}} z^n,
            \nonumber 
            \label{UNsection}
    \end{eqnarray}
  where the index $ \hat{a} = 1,\ldots, n - 1, $ label the $ SU(N) $ subgroup.
  Our choice of sections is such that no holomorphic prepotential exists.
  
  With this choice, K\"ahler potential and its derivatives are
    \BEA
      \mathcal{K}
      &=&  - \log \mathcal{K}_0,
             \\  
      \mathcal{K}_0 
      &=&    i 
             \left(
             \CF - \bar{\CF} - \frac{1}{2} ( z^a - \bar{z}^a ) (\CF_a + \bar{\CF}_a) 
             \right)
             \\
      \partial_a \mathcal{K}
      &=&  - \frac{i}{2 \CK_0} 
             \left(
             \CF_a - \bar{\CF}_a - (z^c - \bar{z}^c) \CF_{ac} 
             \right),
             \label{UNpK}
    \EEA
  where $ \CF_a = \partial \CF/ \partial z^a $.
  The K\"ahler metric and the Levi-Civita connection are
    \BEA
      g_{ab^*}
      &=&    \partial_a \partial_{b^*} \mathcal{K}
             \nonumber \\
      &=&    \partial_a \CK \partial_{b^*} \CK - \frac{i}{2 \CK_0} (\CF_{ab} - \bar{\CF}_{ab}),
             \label{UNg}
             \\
      \Gamma^{a}_{bc}
      &=&  - g^{ad^*} \partial_c g_{bd^*}
             \NN \\
      &=&  - \delta^a_b \partial_c \CK - \delta^a_c \partial_b \CK
           + \frac{1}{\CK_0} g^{ad^*}
             (\partial_b \partial_c \CK_0 \partial_{d^*} \CK + \partial_b \partial_c \partial_{d^*} \CK_0).
             \label{UNLevi}
    \end{eqnarray}
  Furthermore, by using (\ref{Cabc}) and (\ref{UNLevi}), $\nabla_a f^0_b $ and $\nabla_a f^n_b$ are evaluated as follows,
    \begin{eqnarray}
      \nabla_a f^0_b
      &\equiv&    
             \frac{i e^{\mathcal{K}/2}}{\sqrt{2}} C_{abc} g^{cd^*} \partial_{d^*} \mathcal{K}
             \NN \\
      &\equiv& 
             \partial_a f^0_b 
           + \Gamma^c_{ab} f^0_c
           + \frac{1}{2} \partial_a \mathcal{K} f^0_b
             \\
      &=&    \frac{e^{\mathcal{K}/2}}{\sqrt{2}} 
             \left(
             \partial_a \partial_b \mathcal{K}
           - \partial_a \CK \partial_b \mathcal{K}
           + \frac{1}{\mathcal{K}_0} g^{cd^*} (\partial_a \partial_b \mathcal{K}_0 \partial_{d^*} \mathcal{K}
           + \partial_a \partial_b \partial_{d^*} \mathcal{K}_0) \partial_c \mathcal{K}
             \right)
             \label{cdf0},
             \NN \\
      \nabla_a f^n_b
      &\equiv&    
            \frac{i e^{\mathcal{K}/2}}{\sqrt{2}} C_{abc} g^{cd^*} 
            (\bar{\CF}_{nd} + \partial_{d^*} \mathcal{K} \bar{\CF}_n)
            \nonumber \\
      &\equiv& 
             \partial_a f^n_b 
           + \Gamma^c_{ab} f^n_c
           + \frac{1}{2} \partial_a \mathcal{K} f^n_b
             \nonumber \\
      &=&   \frac{e^{\mathcal{K}/2}}{\sqrt{2}} 
            \left(
            \CF_{nab}
          - \partial_a \mathcal{K} \partial_b \mathcal{K} \CF_n
          + \partial_a \partial_b \mathcal{K} \CF_n
            \right)
            \nonumber \\
      & & + \frac{e^{\mathcal{K}/2}}{\sqrt{2} \mathcal{K}_0} g^{cd^*} 
            \left(
            \partial_a \partial_b \mathcal{K}_0 \partial_{d^*} \mathcal{K}
          + \partial_a \partial_b \partial_{d^*} \mathcal{K}_0
            \right) 
            (\CF_{nc} + \partial_c \mathcal{K} \CF_n).
            \label{cdfn}
    \end{eqnarray}
  These equations will be used in the analysis of the scalar potential.
\subsubsection{$U(N)$ gauging}
  In order to gauge the vector multiplets,
  we define the Killing vectors as follows,
    \begin{eqnarray}
      k^c_a \partial_c 
       =   f^c_{ab} z^b \partial_c,~~~~
      k^{c^*}_a \bar{\partial}_{c^*}
       =   f^c_{ab} \bar{z}^{b^*} \bar{\partial}_{c^*}, 
           \label{UNgauging}
    \end{eqnarray}
  where $ f^a_{bc} $ is the structure constant of the $ U(N) $ gauge group satisfying
    \begin{equation}
      [t_a, t_b]
        =   i f^c_{ab} t_c.
    \end{equation}
  In this way, for example, the covariant derivative of the scalar fields can take the standard form;
    \begin{eqnarray}
      \nabla_\mu z^a
      &=&   \partial_\mu z^a + A^\Lambda_\mu k^a_\Lambda
            \nonumber \\
      &=&   \partial_\mu z^a + f^a_{bc} A^b_\mu z^c.
    \end{eqnarray}
  Note that these Killing vectors satisfy 
    \BE
      0
       =    \mathcal{L}_{\Lambda} \CK
       =    k^b_{\Lambda} \partial_b \CK + k^{b^*}_{\Lambda} \partial_{b^*} \CK.
            \label{killing}
    \EE
\subsection{Hypermultiplet}
  We take the same parametrizations as those of the section 5.1.2.
  Since in order for $\CN=2$ supersymmetry to be broken partially, as we have seen in the last chapter, 
  two gauge boson have to be massive by the Higgs mechanism, 
  let the overall $U(1)$ gauge boson and the graviphoton be the gauge bosons which become massive.
  For this purpose, we generalize (\ref{U17})-(\ref{U18}) as follows,
    \begin{eqnarray}
      k^u_0 
      &=&    g_1 \delta^{u3} + g_2 \delta^{u2},~~~~
      k^u_{\hat{a}}
       =     0,~~~~
      k^u_n
       =     g_3 \delta^{u2},
             \nonumber \\
      \mathcal{P}^x_0 
      &=&    \frac{1}{b^0} (g_1 \delta^{x3} + g_2 \delta^{x2}),~~~~
      \mathcal{P}^x_{\hat{a}}
       =     0,~~~~
      \mathcal{P}^x_n
       =     \frac{1}{b^0} g_3 \delta^{x2},
             \label{hypermultipletparame}
    \end{eqnarray}
\section{Partial supersymmetry breaking}
  In order to write down the supersymmetry transformations of the fermions, 
  let us evaluate the mass matrices here.
  Substituting (\ref{UNsection})-(\ref{UNg}), (\ref{UNgauging}) and the hypermultiplet parametrizations 
  into (\ref{S})-(\ref{M2}), 
  we have
    \begin{subequations}
    \begin{align}
      S_{AB}
      &=  - \frac{i e^{\CK/2}}{2 \sqrt{2} b^0} 
            \left(
            \begin{array}{cc}
            i (g_2 +g_3 \CF_n) & g_1 \\
            g_1 & i (g_2 +g_3 \CF_n)
            \end{array}
            \right),
            \label{UNS} \\
      W^{aAB}_1 
      &=  - i e^{\CK/2} \CD^a
            \left(
            \begin{array}{cc}
            0 & 1 \\
            -1 & 0
            \end{array}
            \right),
            \\
      W^{aAB}_2
      &=    \frac{e^{\CK/2}}{\sqrt{2} b^0} g^{ab^*}
            \left(
            \begin{array}{cc}
            \displaystyle\scriptstyle{g_2 \partial_{b^*} \CK + g_3 (\bar{\CF}_{nb} + \partial_{b^*} \CK \bar{\CF}_n)} 
            & i g_1 \partial_{b^*} \CK
            \\
            i g_1 \partial_{b^*} \CK
            & 
            \displaystyle\scriptstyle{g_2 \partial_{b^*} \CK + g_3 (\bar{\CF}_{nb} + \partial_{b^*} \CK \bar{\CF}_n)}
            \end{array}
            \right),
            \\
      N^A_\alpha
      &=    \frac{i e^{\CK/2}}{\sqrt{2} b^0} 
            \left(
            \begin{array}{cc}
            g_1 & - i (g_2 +g_3 \bar{\CF}_n) \\
            i (g_2 +g_3 \bar{\CF}_n) & - g_1
            \end{array}
            \right),
            \label{UNN} \\
      \mathcal{M}^{\alpha \beta}
      &=    \frac{i e^{\CK/2}}{\sqrt{2} b^0}
            \left(
            \begin{array}{cc}
            - i (g_2 +g_3 \CF_n) & g_1 \\
            g_1 & - i (g_2 +g_3 \CF_n)
            \end{array}
            \right),
            \\
      \mathcal{M}^\alpha_{bB}
      &=  - \frac{\sqrt{2} i e^{\CK/2}}{b^0}
            \left(
            \begin{array}{cc}
            g_1 \partial_{a} \CK &
            \displaystyle\scriptstyle{i (g_2 \partial_{a} \CK + g_3 (\CF_{na} + \partial_{a} \CK \CF_n))} 
            \\
            \displaystyle\scriptstyle{- i (g_2 \partial_{a} \CK + g_3 (\CF_{nb} + \partial_a \CK \CF_n))}
            & - g_1 \partial_{a} \CK
            \end{array}
            \right),
            \\
      \mathcal{M}_{1;aA \vert bB}
      &=  - \frac{i e^{\CK/2}}{2} g_{ac^*} (\partial_b + \partial_b \CK) \CD^c
            \left(
            \begin{array}{cc}
            0 & 1 \\
            -1 & 0
            \end{array}
            \right),
            \\
      \mathcal{M}_{2;aA \vert bB}
      &=  - \frac{1}{2 b^0} 
            \left(
            \begin{array}{cc}
            \displaystyle\scriptstyle{g_2 \nabla_b f_a^0 + g_3 \nabla_b f_a^n} 
            & - i g_1 \nabla_b f_a^0
            \\
            - i g_1 \nabla_b f_a^0
            & 
            \displaystyle\scriptstyle{g_2 \nabla_b f_a^0 + g_3 \nabla_b f_a^n}
            \end{array}
            \right)
            \\
      &=    \frac{i e^{\CK/2}}{2 \sqrt{2}} C_{abc} g^{cd^*} 
            \left(
            \begin{array}{cc}
            \displaystyle\scriptstyle{g_2 \partial_{b^*} \CK + g_3 (\bar{\CF}_{nb} + \partial_{b^*} \CK \bar{\CF}_n)} 
            & - i g_1 \partial_{b^*} \CK
            \\
            - i g_1 \partial_{b^*} \CK
            & 
            \displaystyle\scriptstyle{g_2 \partial_{b^*} \CK + g_3 (\bar{\CF}_{nb} + \partial_{b^*} \CK \bar{\CF}_n)}
            \end{array}
            \right),
            \label{UNM2}
    \end{align}
    \end{subequations}
  where we have introduced, 
    \BEA
      \CD^a
      &=&   \frac{i}{\sqrt{2}} f^a_{~bc} \bar{z}^{b^*} z^c.
    \EEA
\subsection{The scalar potential}
  By the gauging of hypermultiplet, the scalar potential $V$ takes a nontrivial form:
    \begin{eqnarray}
      V(z,\bar{z},b)
      &=&    e^{\CK} g_{ab^*} \CD^a \CD^b
           + \frac{e^{\mathcal{K}}}{(b^0)^2} g^{ab^*} D^x_a \bar{D}^x_{b^*}
             \nonumber \\
      & &  - \frac{e^{\mathcal{K}}}{2 (b^0)^2} 
             (\mathcal{E}^x + \mathcal{M}^x \CF_n)(\mathcal{E}^x + \mathcal{M}^x \bar{\CF}_n),
             \label{UNV1}
    \end{eqnarray}
  with
    \begin{eqnarray}
      D^x_a
      &=&    \frac{1}{\sqrt{2}} 
             \left(
             \mathcal{E}^x \partial_a \CK + \mathcal{M}^x (\CF_{na} + \partial_a \CK \CF_n)
             \right),
             \\
      \mathcal{E}^x
      &=&    (0,\ g_2,\ g_1),
             \nonumber \\
      \mathcal{M}^x
      &=&    (0,\ g_3,\ 0).
             \nonumber
    \end{eqnarray}
  The first term comes from $ U(N) $ gauging of the vector multiplet, 
  while the second and the last terms correspond to gauging of the hypermultiplet.
  Note that this potential is quite different from what we have seen in the last chapter.
  If we choose $\CF$ as the simplest function of the holomorphic sections $X^\Lambda$, 
  we obtain flat potential and cannot determine the expectation values of the scalar fields.
  On the other hand, in our model, we can obtain it by determining the minimum of the potential.

  Let us find the conditions which minimize the potential.
  Firstly, let us consider the variation of $ V $ with respect to the scalar field $ z^a $.
  The derivative of the second and the last terms of $ V $ is
    \BEA 
      \lefteqn{\partial_a 
             \left(
             \frac{e^{\mathcal{K}}}{(b^0)^2} g^{ab^*} D^x_a \bar{D}^x_{b^*}
           - \frac{e^{\mathcal{K}}}{2 (b^0)^2} 
             (\mathcal{E}^x + \mathcal{M}^x \CF_n)(\mathcal{E}^x + \mathcal{M}^x \bar{\CF}_n)
             \right)
             }
             \NN \\
      &=&    \frac{e^{\mathcal{K}}}{(b^0)^2} 
             \left(
             (\partial_a \mathcal{K}) g^{bc^*} D^x_b \bar{D}^x_{c^*}
           + (\partial_a g^{bc^*}) D^x_b \bar{D}^x_{c^*}
           + g^{bc^*} (\partial_a D^x_b) \bar{D}^x_{c^*}
             \right)
             \nonumber \\
      &=&    \frac{e^{\mathcal{K}}}{(b^0)^2} g^{bc^*} \bar{D}^x_{c^*}
             \left(
             \partial_a D^x_b
           - (\partial_b \mathcal{K}) D^x_a
           + \frac{1}{\mathcal{K}_0} g^{ed^*} (\partial_a \partial_b \mathcal{K}_0 \partial_{d^*} \mathcal{K}
           + \partial_a \partial_b \partial_{d^*} \mathcal{K}_0) D^x_e
             \right)
             \nonumber \\
      &=&    \frac{i e^{\mathcal{K}}}{(b^0)^2} C_{abc} g^{bd^*} \bar{D}^x_{d^*} g^{ce^*} \bar{D}^x_{e^*}
             \label{derivativeV0},
    \end{eqnarray}
  where we have used (\ref{cdf0}) and (\ref{cdfn}) in the last equality.
  Thus, the condition which minimizes the potential is
    \begin{equation}
      0
       =     \langle \partial_c V \rangle
       =     \langle
             \partial_c 
             \left(
             e^{\CK} g_{ab^*} \CD^a \CD^b
             \right)
             \rangle
           + \langle \frac{e^{\mathcal{K}}}{(b^0)^2} i C_{acd} g^{ab^*} \bar{D}^x_{b*} g^{de^*} \bar{D}^x_{e*} \rangle,
             \label{VC1}
      \end{equation}
      
  Of course, we must consider the variation with respect to the hypermultiplet scalar fields $ b^u $.
  Since $ V $ contains only $ b^0 $, it is straightforward to compute the derivative in terms of $b^0$.
  The condition which determines the vacuum is
    \BE
      0
       =     \langle \frac{\partial V}{\partial b^u} \rangle
       =   - \frac{e^{\mathcal{K}} }{(b^0)^3}
             \langle
             2 g^{ab^*} D^x_a \bar{D}^x_{b^*}
           - (\mathcal{E}^x + \mathcal{M}^x \CF_n)(\mathcal{E}^x + \mathcal{M}^x \bar{\CF}_n)
             \rangle \delta_{u0}. 
             \label{VC2}
    \EE
\subsubsection{The holomorphic function $\CF$}
  Since we would like to keep $SU(N)$ gauge symmetry manifest, 
  we will work on the condition $ \langle z^a \rangle = \delta^{an} \lambda $.
  Then $ \langle \CD^a \rangle = \langle \frac{i}{\sqrt{2}} f^a_{~bc} \bar{z}^{b^*} z^c \rangle = 0 $ holds.
  Thus we have $ \langle \partial_c (e^{\CK} g_{ab^*} \CD^a \CD^b) \rangle = 0 $.
  Moreover, we assume a form of the gauge invariant function $\CF(z)$, 
  which parallels that of \cite{FIS2}, as
    \begin{eqnarray}
      \CF(z)
      &=& - \frac{i C}{2} (z^n)^2 + \mathcal{G}(z),
            \\
      \mathcal{G}(z)
      &=&   \sum_{l=0}^k \frac{C_l}{l!} tr (z^a t_a)^l,
    \end{eqnarray}
  where $ C \in \mathbb{R} $ is constant.
  We will see that $ C $ must be non-zero in order for the inverse of the K\"ahler metric to exist.
  
  Let us compute the expectation values of some geometrical quantities.
  Firstly, the expectation value of the derivative of $ \CF $ is
    \BEA
      \langle \CF_a \rangle
      &=&    \delta_{an} \langle \CF_n \rangle,
             \nonumber \\
      \langle \CF_{na} \rangle
      &=&    \delta_{an} \langle \CF_{nn} \rangle,
             \nonumber \\
      \langle \CF_{\hat{a} \hat{b}} \rangle
      &=&    \delta_{\hat{a} \hat{b}} \langle \CF_{nn} - i C \rangle,
             \nonumber \\
      \langle \CF_{nab} \rangle
     &=&     \delta_{ab} \langle \CF_{nnn} \rangle,
             \label{fderivative}
    \EEA
  where the explicit forms of $ \langle \CF_n \rangle $ and $ \langle \CF_{nn} \rangle $ are
    \BEA
      \langle \CF_n \rangle
      &=&    \sum_{l} \frac{C_l}{(l-1)!} \lambda^{l-1} + i C \lambda,
             \nonumber \\
      \langle \CF_{nn} \rangle
      &=&    \sum_{l} \frac{C_l}{(l-2)!} \lambda^{l-2}.
    \EEA
  It is easy to compute the expectation value of $ \partial_a \mathcal{K} $ by using (\ref{UNpK}):
    \BEA
      \langle \partial_a \mathcal{K} \rangle
      &=&  - \frac{i \langle e^{\mathcal{K}} \rangle}{2} 
             \langle 
             \CF_a - \bar{\CF}_{a} - (\lambda - \bar{\lambda}) \CF_{an} 
             \rangle
             \nonumber \\
      &=&    \delta_{an} \langle \partial_n \mathcal{K} \rangle.
             \label{partialK}
    \EEA
  Furthermore, the vacuum expectation value of the K\"ahler metric $ g_{ab} $ (\ref{UNg}) can be evaluated as follows:
    \BEA
      \langle g_{ab^*} \rangle
      &=&    \left(
             \begin{array}{ccccc}
             \langle g_{11^*} \rangle &&&&
             \\
             & \langle g_{11*} \rangle &&& 0
             \\
             &&\ddots&
             \\
             &&&\ddots&
             \\
             0 &&&& \langle g_{nn} \rangle
             \end{array}
             \right),
    \EEA
  with
    \BEA
      \langle g_{11*} \rangle
      &=&  - \frac{i \langle e^{\mathcal{K}} \rangle}{2} \langle \CF_{nn} - \bar{\CF}_{nn} - 2iC \rangle,
             \nonumber \\
      \langle g_{nn} \rangle
      &=&    |\langle \partial_n \mathcal{K} \rangle|^2
           - \frac{i \langle e^{\mathcal{K}} \rangle}{2} \langle \CF_{nn} - \bar{\CF}_{nn} \rangle.
             \label{gVE1}
    \EEA
  Note that only the diagonal components are non-zero 
  and these take the same value except for $ \langle g_{nn} \rangle $.
  Finally, we have to compute $ \langle D^x_a \rangle $ and $ \langle C_{abc} \rangle $ by using the above equations;
    \BEA
      \langle D^x_a \rangle
      &=&    \delta_{an}
             \frac{1}{\sqrt{2}} 
             \langle
             \mathcal{E}^x \partial_a \mathcal{K} + \mathcal{M}^x (\CF_{na} + \partial_a \mathcal{K} \CF_n)
             \rangle,
             \nonumber \\
      &=&    \delta_{an} \langle D^x_n \rangle
             \nonumber \\
      \langle C_{abc} \rangle
      &=&    \frac{\langle e^{\mathcal{K}} \rangle}{2} \langle \CF_{abc} \rangle.
             \label{UNCabc}
    \EEA
\subsection{Vacuum conditions}
  Now we are ready to analyze (\ref{VC1}) and (\ref{VC2}).
  Substituting (\ref{fderivative})-(\ref{UNCabc}) into (\ref{VC1}), we obtain
    \BE
      0=
      \langle \frac{i e^{2 \mathcal{K}}}{2 (b^0)^2} \CF_{nnn} g^{n n^*} \bar{D}^x_{n^*} g^{n n^*} \bar{D}^x_{n^*} \rangle.
    \EE
  The points $ \langle \CF_{nnn} \rangle = 0 $ are unstable vacua 
  because $ \langle \partial_{a} \partial_{b^*} V \rangle = 0 $.
  Also, the points which satisfy $ \langle g^{n n^*} \rangle = 0 $ 
  and $ \langle \partial_n \mathcal{K} \rangle = 0 $ are not stable.
  Thus, the vacuum condition reduces to
    \begin{equation}
      \langle \bar{D}^x_{n^*} \bar{D}^x_{n^*} \rangle
       =     0,
             \label{VC3}
    \end{equation}
  which implies
    \begin{equation}
      \LLangle \CF_n + \frac{\CF_{nn}}{\partial_n \mathcal{K}} \RRangle
       =   - \left(
             \frac{g_2}{g_3} \pm i \frac{g_1}{g_3}
             \right).
             \label{VC4}
    \end{equation}
  where we use $ \llangle ... \rrangle $ for those vacuum expectation value 
  which are determined as the solutions to (\ref{VC3}).
  We have also assumed $ g_3 \neq 0 $.
  Note that if $ g_3 = 0 $, (\ref{VC3}) leads to $ g_1 = g_2 = 0 $ and the supersymmetry is unbroken.
  We are not interested in such a case.
  
  By using (\ref{VC4}), the second condition (\ref{VC2}) leads to
    \BEA
      0
      &=&    \llangle
             (\mathcal{E}^x + \mathcal{M}^x \CF_n)(\mathcal{E}^x + \mathcal{M}^x \bar{\CF}_n)
           - 2 g^{ab^*} D^x_a \bar{D}^x_{b^*}
             \rrangle
             \NN \\
      &=&    \left|
             g_1 \mp i g_3 \LLangle \frac{\CF_{nn}}{\partial_n \mathcal{K}} \RRangle
             \right|^2 
           + g_1^2 - \llangle g^{nn^*} |\partial_n \mathcal{K}|^2 \rrangle 2 g_1^2.
             \label{VC5}
    \EEA
  It can be shown that this condition (\ref{VC5}) leads to $\llangle \CF_{nn} \rrangle = 0$:
  If $\llangle \CF_{nn} \rrangle = 0$, (\ref{VC5}) is automatically satisfied.
  Thus, let us consider the case $\llangle \CF_{nn} \rrangle \neq 0$ 
  and prove that it conflicts with the assumption, $g_3 \neq 0$.
  We write $\CF_{nn}$ as
    \BE
      \CF_{nn}
       =     F_1 + i F_2,
    \EE
  where $F_1, F_2 \in \mathbb{R}$.
  From (\ref{partialK}), (\ref{gVE1}) and (\ref{VC4}), by using $F_1$ and $F_2$, we obtain
    \BEA
      \llangle g_{nn^*} \rrangle
      &=&    \llangle |\partial_n \CK|^2 \rrangle
           + \frac{F_2}{\llangle \CK_0 \rrangle},
             \label{A5}
             \\
      \llangle \partial_n \CK + \partial_{n^*} \CK \rrangle
      &=&    \frac{i}{\llangle \CK_0 \rrangle}
             \left(
             \LLangle \frac{\CF_{nn}}{\partial_n \CK} - \frac{\bar{\CF}_{nn}}{\partial_{n^*} \CK} \RRangle
           \pm
             2i \frac{g_1}{g_3}
           + (\lambda - \bar{\lambda}) F_1
             \right),
             \label{A1}
             \\
      \llangle \partial_n \CK - \partial_{n^*} \CK \rrangle
      &=&    \frac{1}{\llangle \CK_0 \rrangle}
             (\lambda - \bar{\lambda}) F_2.
             \label{A2}
    \EEA
  Then the condition (\ref{VC5}) can be written as
    \BE
      0
       =     2 g_1^2
           - 2 g_1^2 \llangle g^{nn^*} |\partial_n \CK|^2 \rrangle
           \mp
             i Y g_1 g_3
           + g_3^2 \LLangle |\frac{\CF_{nn}}{\partial_n \CK}|^2 \RRangle,
             \label{A3}
    \EE
  where we have defined $Y$ as
    \BEA
      Y
      &\equiv&    
             \LLangle \frac{\CF_{nn}}{\partial_n \CK} - \frac{\bar{\CF}_{nn}}{\partial_{n^*} \CK} \RRangle
             \NN \\
      &=&    \frac{1}{\llangle |\partial_n \CK|^2 \rrangle}
             \left[
             F_1 \llangle \partial_n \CK - \partial_{n^*} \CK \rrangle
           - \frac{F_2}{\llangle \CK_0 \rrangle} 
             \left(
             Y \pm 2i \frac{g_1}{g_3} + (\lambda - \bar{\lambda}) F_1
             \right)
             \right].
    \EEA
  In the second equality, we have used (\ref{A1}).
  Using (\ref{A5}), we can solve the above equation for $Y$:
    \BE
      Y \llangle g_{nn^*} \rrangle
       =   \mp
             \frac{F_2}{\llangle \CK_0 \rrangle} 2i \frac{g_1}{g_3}.
             \label{A4}
    \EE
  Substituting (\ref{A4}) into (\ref{A3}), we get
    \BEA
      0
      &=&    2 g_1^2
           - 2 g_1^2 \llangle g^{nn^*} |\partial_n \CK|^2 \rrangle
           - 2 g_1^2 \LLangle g^{nn^*} \frac{F_2}{\llangle \CK_0 \rrangle} \RRangle
           + g_3^2 \LLangle |\frac{\CF_{nn}}{\partial_n \CK}|^2 \RRangle
             \NN \\
      &=&    g_3^2 \LLangle |\frac{\CF_{nn}}{\partial_n \CK}|^2 \RRangle,
    \EEA
  where we have used (\ref{A5}).
  Therefore, we conclude that when $\llangle \CF_{nn} \rrangle \neq 0$, the vacuum condition leads to $g_3=0$.
  This conflicts with the assumption which is written in below (\ref{VC4}).
  Thus, we can say that the second vacuum condition implies 
    \BE
      \llangle \CF_{nn} \rrangle = 0.
           \label{VC6}
    \EE
  In the rigid theory \cite{FIS1, FIS2}, the second vacuum condition is not needed, 
  because there is no hypermultiplet in the model.
  In fact, we will see later that $ \mathcal{N} = 2 $ local supersymmetry is not broken partially 
  without the second vacuum condition.
  
  In the following, we choose the vacuum condition as
    \BE
      \llangle \CF_n \rrangle
       =   - \left(
             \frac{g_2}{g_3} + i \frac{g_1}{g_3}
             \right).
             \label{VC7}
    \EE
  With this, we can evaluate the expectation value of $ \partial_a \mathcal{K} $ and $ g_{ab^*} $ such that
    \BEA
      \llangle \partial_a \mathcal{K} \rrangle
      &=&  - \delta_{an} \llangle e^{\mathcal{K}} \rrangle \frac{g_1}{g_3},
             \nonumber \\
      \llangle g_{11^*} \rrangle
      &=&  - \llangle e^K \rrangle C,
             \nonumber \\
      \llangle g_{nn^*} \rrangle
      &=&    |\llangle \partial_n \mathcal{K} \rrangle|^2
       =     \llangle e^{2 \mathcal{K}} \rrangle 
             \left(
             \frac{g_1}{g_3}
             \right)^2.
             \label{VC8}
    \EEA
  Note that $ C = 0 $ is necessary for the K\"ahler metric to be invertible.
\subsection{Supersymmetry transformations of the fermions}
  Let us compute the vacuum expectation values of the mass matrices.
  Substituting (\ref{VC7})-(\ref{VC8}) into the mass matrices (\ref{UNS})-(\ref{UNM2}),
    \BEA
      \llangle S_{AB} \rrangle
      &=&  - \LLangle
             \frac{i e^{\mathcal{K}/2}}{2 \sqrt{2} b^0} g_1 
             \RRangle
             \left(
             \begin{array}{cc}
             1 & 1 
             \\
             1 & 1
             \end{array}
             \right),
             \label{Svev}
             \\
      \llangle W_2^{aAB} \rrangle
      &=&    \delta^{an}
             \LLangle
             \frac{i e^{\mathcal{K}/2}}{\sqrt{2} b^0} (\partial_n \mathcal{K})^{-1} g_1\RRangle
             \left(
             \begin{array}{cc}
             1 & 1
             \\
             1 & 1
             \end{array}
             \right),
             \label{Wvev}
             \\
      \llangle N^{A}_{\alpha} \rrangle
      &=&    \LLangle
             \frac{i e^{\mathcal{K}/2}}{\sqrt{2} b^0} g_1 
             \RRangle
             \left(
             \begin{array}{cc}
             1 & 1
             \\
             -1 & -1
             \end{array}
             \right),
             \label{Nvev}
             \\
      \llangle \CM^{\a \b} \rrangle
      &=&    \LLangle
             \frac{i e^{\mathcal{K}/2}}{\sqrt{2} b^0} g_1 
             \RRangle
             \left(
             \begin{array}{cc}
             -1 & 1
             \\
             1 & -1
             \end{array}
             \right),
             \\
      \llangle \CM^{\a}_{a B} \rrangle
      &=&  - \LLangle
             \frac{\sqrt{2} i e^{\mathcal{K}/2}}{b^0} g_1 \partial_a \CK
             \RRangle
             \left(
             \begin{array}{cc}
             1 & 1
             \\
             -1 & -1
             \end{array}
             \right),
             \\
      \llangle \CM_{2;aA | bB} \rrangle
      &=&    \LLangle
             \frac{e^{\mathcal{K}/2}}{2\sqrt{2} b^0} g_1 C_{abc} g^{cd^*} \partial_{d^*} \CK
             \RRangle
             \left(
             \begin{array}{cc}
             -1 & 1
             \\
             1 & -1
             \end{array}
             \right).
             \label{Mvev}
    \EEA
  Notice that each matrix has a zero eigenvalue.
  The expectation values of the supersymmetry transformations of the fermions are
    \BEA
      \llangle \delta \psi_{+ \mu} \rrangle
      &=&    \LLangle
             \frac{i e^{\mathcal{K}/2}}{2 b^0} g_1 
             \RRangle
             \gamma_\mu (\epsilon_1 + \epsilon_2),
             \nonumber \\
      \llangle \delta \lambda^{a +} \rrangle
      &=&    \delta^{an}
             \LLangle
             \frac{i e^{\mathcal{K}/2}}{ b^0} g_1 (\partial_n \mathcal{K})^{-1}
             \RRangle
             (\epsilon_1 + \epsilon_2),
             \nonumber \\
      \llangle \delta \zeta_{-} \rrangle
      &=&    \LLangle
             \frac{i e^{\mathcal{K}/2}}{b^0} g_1 
             \RRangle
             (\epsilon_1 + \epsilon_2),
             \nonumber \\
      \llangle \delta \psi_{- \mu} \rrangle
      &=&    \llangle \lambda^{a -} \rrangle
       =     \llangle \zeta_{+} \rrangle
       =     0.
    \EEA
  Note that only the supersymmetry transformations of $\psi^+_\mu$, $\lambda^{n +}$ and $\zeta_-$ at the vacuum 
  are non-zero.
  Furthermore, quite similar to the last chapter, we define linear combination of the fermions, 
  $\lambda^{n +}$ and $\zeta_{-}$ as
    \BEA
      \chi_\bullet
      &=&    \llangle \partial_n \mathcal{K} \rrangle \lambda^{n +} + 2 \zeta_{-},
             \nonumber \\
      \eta_\bullet
      &=&  - \llangle \partial_n \mathcal{K} \rrangle \lambda^{n +} + \zeta_{-},
    \EEA
  whose supersymmetry transformations are
    \BEA
      \llangle \delta \chi_\bullet \rrangle
      &=&    \LLangle
             \frac{3 i e^{\mathcal{K}/2}}{b^0} g_1 
             \RRangle
             (\epsilon_1 + \epsilon_2),
             \nonumber \\
      \llangle \delta \eta_\bullet \rrangle
      &=&    0.
             \label{SUSYtr}
    \EEA
  Therefore, the fermion $\chi_\bullet$ is the Nambu-Goldstone fermion.
\section{The mass spectrum}
\subsection{Fermion mass}
  Let us consider the fermion mass spectrum.
  By substituting (\ref{Svev})-(\ref{Mvev}) into $ \CL_{\rm{Yukawa}} $, we obtain
    \BEA
      \CL_{\rm{Yukawa}}
      &=&  - i  \LLangle \frac{\sqrt{2}e^{\mathcal{K}/2}}{b^0} g_1 \RRangle
             \left(
             \bar{\psi}^{+}_\mu \gamma^{\mu \nu} \psi^{+}_\nu
           - i \bar{\chi}^{} \gamma_\mu \psi^\mu_{+}
           + \frac{1}{3} \bar{\chi}_\bullet \chi_\bullet
           - \frac{1}{3} \bar{\eta}_\bullet \eta_\bullet
             \right)
             \nonumber \\
      & &  + \frac{1}{2 \sqrt{2}} \LLangle \frac{e^{\mathcal{K}/2}}{b^0} g_3 \CF_{aan} \RRangle 
             \bar{\lambda}^{a -} \lambda^{a -}
           + \ldots + h.c.,
    \EEA
  The Nambu-Goldstone fermion $\chi_\bullet$ couples to the gravitino $ \psi^+_\mu $ in the second term 
  and we can remove it from the action by redefining the gravitino such that, 
    \BE
      \psi^+_\mu
       \rightarrow \psi^+_\mu + \frac{i}{6} \gamma_{\mu} \chi_\bullet,
    \EE
  which results in
    \BEA
      \CL_{\rm{Yukawa}}
      &=&  - i  \LLangle \frac{\sqrt{2}e^{\mathcal{K}/2}}{b^0} g_1 \RRangle
             \left(
             \bar{\psi}^{+}_\mu \gamma^{\mu \nu} \psi^{+}_\nu
           - \frac{1}{3} \bar{\eta}_\bullet \eta_\bullet
             \right)
           + \frac{1}{2 \sqrt{2}} \sum_{a=1}^n \LLangle \frac{e^{\mathcal{K}/2}}{b^0} g_3 \CF_{aan} \RRangle 
             \bar{\lambda}^{a -} \lambda^{a -}
             \NN \\
      & &    ~~~~~~~~~~~~~~~~~~~~~~~~~~~~~~~~~~~~~~~~~~~~~~~~~~~~
           + \ldots + h.c.~.
    \EEA
  The gravitino $ \psi_+ $ has acquired a mass by the super-Higgs mechanism.
  
  The kinetic terms of the massive fermions are
    \BE
      \CL_{\rm{kin}}
       =     \frac{\epsilon^{\mu \nu \lambda \sigma}}{\sqrt{-g}}
             \bar{\psi}^A_\mu \gamma_\nu \partial_\lambda \psi_{A \sigma}
           - \frac{i}{3} \bar{\eta}^\bullet \gamma_\mu \partial^\mu \eta_\bullet
           - i \sum_{a=1}^n \llangle g_{aa^*} \rrangle \bar{\lambda}^{a -} \gamma_{\mu} \partial^{\mu} \lambda^{a^*}_- 
           + \ldots
           + h.c.
    \EE
  It is easy to find the mass of the fermions by evaluating the equations of motion.
  The gravitino mass $ m $ and the masses of the gauginos $ m_{a} $ are
    \BEA
      m
      &=&    \left|
             \LLangle 
             \frac{\sqrt{2}e^{\mathcal{K}/2}}{b^0} g_1 
             \RRangle 
             \right|,
             \label{m}
             \nonumber \\
      m_a
      &=&    \left| 
             \LLangle 
             \frac{e^{\mathcal{K}/2}}{\sqrt{2 }b^0} g_3 \CF_{aan} g^{aa^*} 
             \RRangle 
             \right|.
             \label{ma}
    \EEA
  Notice that the mass of the physical fermion $ \eta_\bullet$ is the same as the gravitino, that is, $ m $.
  We can expect that $ \psi^+_\mu $ and $ \eta_\bullet $ form a $\CN=1$ massive multiplet of spin (3/2,1,1,1/2).
  On the other hand, $ \lambda^{a -} $, together with the scalar fields, 
  form $ \mathcal{N} =1 $ massive chiral multiplet.
  To make sure, we must consider the masses of the gauge bosons and the scalar fields.
\subsection{Boson mass}
  Let us compute the masses of the scalar fields.
  If we define the fluctuations 
  from the expectation value of the scalar fields as the new scalar fields $ \tilde{z}^a $, 
  that is, $\tilde{z}^a=z^a- \llangle z^a \rrangle$. 
  the scalar potential can be expanded as, 
    \BE
      V
       =    \llangle V \rrangle
          + \llangle \partial_a \partial_{b*} V \rrangle \tilde{z}^a \bar{\tilde{z}}^{b^*}
          + \frac{1}{2} \llangle \partial_a \partial_{b} V \rrangle \tilde{z}^a \tilde{z}^{b}
          + \frac{1}{2} \llangle \partial_{a^*} \partial_{b^*} V \rrangle \bar{\tilde{z}}^{a^*} \bar{\tilde{z}}^{b^*}
          + \ldots.
    \EE
  where we use the fact that first derivative of $ V $ is zero by the vacuum condition.
  The second derivative can be easily evaluated,
    \BEA
      \llangle \partial_a \partial_{b*} V \rrangle
      &=&   \LLangle
            \frac{2 i e^{\mathcal{K}}}{(b^0)^2} 
            C_{acd} g^{ce^*} (\partial_{b^*} \bar{D}^x_{e^*}) g^{df^*} \bar{D}^x_{f^*}
            \RRangle
            \nonumber \\
      &=&   \delta_{ab}
            \LLangle
            \frac{e^{\mathcal{K}}}{2 (b^0)^2} |g_3 \CF_{aan}|^2 g^{aa^*}
            \RRangle,
            \nonumber \\
      \llangle \partial_a \partial_b V \rrangle
      &=&   \llangle \partial_{a^*} \partial_{b^*} V \rrangle 
       =    0.
    \EEA
  Thus, the kinetic terms and the mass terms of the scalar fields $\tilde{z}^a$ are
    \BE
            \sum_a 
            \left(
            \llangle g_{aa^*} \rrangle \partial_\mu \tilde{z}^a \partial^\mu \bar{\tilde{z}}^{a^*}
          - \LLangle
            \frac{e^{\mathcal{K}}}{2 (b^0)^2} |g_3 \CF_{aan}|^2 g^{aa^*}
            \RRangle
            \tilde{z}^a \bar{\tilde{z}}^{a^*}
            \right).
    \EE
  We can see the mass of $ \tilde{z}^a $ is the same as (\ref{ma}), namely, the mass of gaugino $ \lambda^{a -} $.
  Therefore, as we have anticipated, they form $ N^2 $ massive chiral multiplets.

  The gauge boson masses appear in the kinetic terms of the hypermultiplet scalars:
    \BEA
      h_{uv} \nabla_\mu b^u \nabla^\mu b^v
      &=&    \frac{1}{2 (b^0)^2} \delta_{uv}
             (\partial_\mu b^u + A^\Lambda_\mu k_\Lambda^u)(\partial^\mu b^v + A^{\Lambda \mu} k_\Lambda^v)
             \NN \\
      &=&    \frac{1}{2 (b^0)^2} 
             (g_1^2 A^0_\mu A^{0 \mu} + g_3^2 A'^n_\mu A'^{n \mu})
           + \ldots,
             \label{Amass}
    \EEA
  where we have defined the gauge boson $A'^n_\mu$ as
    \BE
      A^{\prime n}_\mu
       =     A^n_\mu 
           + \left( \frac{g_2}{g_3} \right) A^0_\mu
             \label{A'}.
    \EE
  Since the kinetic terms of the gauge bosons are 
  $ \frac{1}{4} (\Im \mathcal{N})_{\Lambda \Sigma} F^\Lambda_{\mu \nu} F^{\Sigma \mu \nu} $, 
  we must compute the generalized coupling matrix $ \mathcal{N} $ (\ref{CouplingMatrix}) on the vacuum:
    \BE
      \llangle \mathcal{N}_{\Lambda \Sigma} \rrangle
       =    \begin{pmatrix}
            \llangle \mathcal{N}_{00} \rrangle & 0 & \cdots & \cdots & 0 & \llangle \mathcal{N}_{0n} \rrangle
            \\
            0 & \llangle \mathcal{G}_{11} \rrangle & 0 & \cdots & 0 & 0
            \\
            \vdots & 0 & \llangle \mathcal{G}_{22} \rrangle & & \vdots & \vdots & 
            \\
            \vdots & \vdots & & \ddots & 0 & \vdots
            \\
            0 & 0 & \cdots & 0 & \llangle \mathcal{G}_{n-1,n-1} \rrangle & 0
            \\
            \llangle \mathcal{N}_{n0} \rrangle & 0 & \cdots & \cdots & 0 & \llangle \mathcal{N}_{nn} \rrangle
            \\
            \end{pmatrix},
    \EE
  with
    \BEA
      \Im \llangle \mathcal{N}_{00} \rrangle
      &=&    \LLangle
             \frac{e^{-\mathcal{K}}}{2} 
             \RRangle \frac{g_1^2 + g_3^2}{g_1^2},
             \nonumber \\
      \Im \llangle \mathcal{N}_{0n} \rrangle
      &=&    \Im \llangle \mathcal{N}_{n0} \rrangle
       =     \LLangle
             \frac{e^{-\mathcal{K}}}{2} 
             \RRangle \frac{g_2 g_3}{g_1^2},
             \nonumber \\
      \Im \llangle \mathcal{N}_{nn} \rrangle
      &=&    \LLangle
             \frac{e^{-\mathcal{K}}}{2} 
             \RRangle 
             \left(
             \frac{g_3}{g_1}
             \right)^2.
    \EEA
  Therefore the gauge boson kinetic terms are
    \BEA
      \frac{1}{4} \Im \llangle \mathcal{N}_{\Lambda \Sigma} \rrangle F^\Lambda_{\mu \nu} F^{\Sigma \mu \nu}
      &=&  - \LLangle
             \frac{e^{-\mathcal{K}}}{8} 
             \RRangle
             F^0_{\mu \nu} F^{0 \mu \nu}
           - \LLangle
             \frac{e^{-\mathcal{K}}}{8} 
             \RRangle
             \left(\frac{g_3}{g_1}\right)^2 F'^n_{\mu \nu} F'^{n \mu \nu}  
             \nonumber \\
      & &  + \frac{1}{4} \sum_{\hat{a}} \Im \llangle \mathcal{G}_{\hat{a} \hat{a}} \rrangle F^{\hat{a}}_{\mu \nu} F^{\hat{a} \mu \nu}
             \label{Akin},
    \EEA
  where we have defined $ F'_{\mu \nu} = \partial_\mu A'_\nu - \partial_\nu A'_\mu $.
  We can read off the masses of gauge boson $ A^0_\mu $ and $ A'^n_\mu $ from (\ref{Amass}) and (\ref{Akin}).
  As a result, both of them agree with (\ref{m}).
  Thus the $U(N) \times U(1)_{graviphoton}$ gauge symmetry is broken to $SU(N)$
  and the vacuum lies in the Higgs phase.
  
  We summarize the mass spectrum of our model in the table 6.1:
  \\
  $\CN=1$ gravity multiplet contains the vierbein and the gravitino $ \psi^-_\mu $, 
  and the massive gravitino $ \psi^+_\mu $, $ U(1) $ gauge boson $ A'^n_\mu $, the graviphoton $ A^0_\mu $ 
  and the fermion $ \eta_\bullet $ form a massive spin-3/2 multiplet.
  The $\CN=2$ vector multiplets have been divided into $\CN=1$ vector multiplets and chiral multiplets.
  $ \mathcal{N} = 1 $ vector multiplets described by massless gauge bosons $ A^{\hat{a}}_\mu $ 
  and gauginos $ \lambda^{\hat{a} +} $ which become to $ SU(N) $ vector multiplet.
  On the other hand, the gaugino $ \lambda^{\hat{a} -} $ and the scalar field $ z^{\hat{a}} $ form chiral multiplets 
  which belong to the $ SU(N) $ adjoint representation.
  The gaugino $\lambda^{n -}$ and the complex scalar $z^n$ form a $\CN=1$ massive chiral multiplet.
  Also the hyperino $ \zeta_+ $ and the scalar $ b^0,b^1 $ form a chiral multiplet.
  Note that althouh this mass spectrum is analogous to the rigid counterpart \cite{FIS3}, 
  the phase of the theory is different: 
  our vacuum lies in the Higgs phase, while the rigid one in the Coulomb phase.
    \begin{table}[h]
      \begin{center}
        \begin{tabular}{|c|c|c|}
          \hline
          $ \CN = 1 $ multiplet & field & mass \\ \hline \hline
          gravity multiplet & $ e^i_{\mu} $, $ \psi^{-}_{\mu} $ & 0 \\ \hline
          spin-3/2 multiplet & $ \psi^{+}_{\mu} $, $ A^0_\mu $, $ A'^n_\mu $, $ \eta_\bullet $ & $ m $ \\ \hline
          $ SU(N) $ vector multiplet & $ A^{\hat{a}}_\mu $, $ \lambda^{\hat{a} +} $ & 0 \\ \hline
          $ SU(N) $ adjoint chiral multiplet & $ \lambda^{\hat{a} -} $, $ z^{\hat{a}} $ & $ m^{\hat{a}} $ \\ \hline
          chiral multiplet & $ \lambda^{n -} $, $ z^{n} $ & $ m^n $ \\ \hline
          chiral multiplet & $ \zeta_+ $, $ b^0,b^1 $ & 0 \\
          \hline
        \end{tabular}
        \caption{the mass spectrum}
      \end{center}
    \end{table}
\section{$\CN = 1$ Lagrangian}
  In the last section, we have considered the lowest order terms 
  with respect to the fermion fields and the shifted scalar fields $ \tilde{z}^a $.
  Now, we would like to know all of the terms in $\CL_{\rm{Yukawa}}$ and $V$.
  This can be done exactly and, at the end of this section, 
  we will see that $ \CN = 1 $ Lagrangian can be written by the superpotentials.
  
  In terms of $\tilde{z}^a$, the holomorphic function $\CF(z)$ is expanded as,
    \BEA
      \CF (z)
      &=&   \CF (\llangle z \rrangle + \tilde{z})
            \\ \nonumber
      &=&   \llangle \CF \rrangle + \tilde{\CF},
    \EEA
  where 
    \BE
      \tilde{\CF}
       =    \llangle \CF_a \rrangle \tilde{z}^a + \frac{1}{2!} \llangle \CF_{ab} \rrangle \tilde{z}^a \tilde{z}^b
          + \frac{1}{3!} \llangle \CF_{abc} \rrangle \tilde{z}^a  \tilde{z}^b \tilde{z}^c
          + \ldots.
    \EE
  Similarly, $ \CF_a $ and $ \CF_{ab} $ are
    \BEA
      \CF_a
      &=&   \llangle \CF_a \rrangle + \llangle \CF_{ab} \rrangle \tilde{z}^b
          + \frac{1}{2!} \llangle \CF_{abc} \rrangle \tilde{z}^b \tilde{z}^c
          + \ldots.
            \\ \nonumber
      &=&   \tilde{\CF_a},
            \\
      \CF_{ab}
      &=&   \llangle \CF_{ab} \rrangle + \llangle \CF_{abc} \rrangle \tilde{z}^c
          + \ldots.
            \\ \nonumber
      &=&   \tilde{\CF_{ab}}.
    \EEA
  We have redefined $ \tilde{\CF}_a $ and $ \tilde{\CF}_{ab} $ 
  as the derivative of $ \tilde{\CF} $ with respect to $ \tilde{z}^a $.
  The K\"ahler potential, its derivative and the K\"ahler metric are, respectively,
    \BEA
      \CK
      &=& - \log i 
            \left[
            \llangle \CF - \bar{\CF} \rrangle + \tilde{\CF} - \bar{\tilde{\CF}} 
          - \frac{1}{2} ( \llangle z^a - \bar{z}^a \rrangle + z^a - \bar{z}^a ) (\tilde{\CF_a} + \bar{\tilde{\CF_a}})
            \right],
            \\
      \partial_a \CK
      &=& - \frac{i}{2 \CK_0} (\tilde{\CF_a} - \bar{\tilde{\CF_a}} 
          - ( \llangle z^a - \bar{z}^a \rrangle + z^a - \bar{z}^a ) \tilde{\CF}_{ab}) 
            \\ \nonumber
      &=&   \tilde{\partial_a} \CK,
            \\
      g_{ab^*}
      &=&   \tilde{\partial_a} \CK \tilde{\partial}_{b^*} \CK 
          - \frac{i}{2 \CK_0} ( \tilde{\CF}_{ab} - \bar{\tilde{\CF}}_{ab})
            \nonumber \\
      &=&   \tilde{g}_{ab^*},
    \EEA
  where $ \tilde{\partial_a} = \partial/\partial \tilde{z}^a $.
  In this way, we can write down all the terms in the Lagrangian in terms of the new scalar field $\tilde{z}^a$.
  In the following, let us see several terms of the Lagrangian.
\subsubsection{The Yukawa interaction terms}
  The Yukawa interaction terms $\CL_{\rm{Yukawa}}$ can be easily rewritten with these notations.
  First of all, we define the following quantities,
    \BEA
      \CW (\tilde{z}, \bar{\tilde{z}})
      &\equiv&  
            e^{\CK/2} W(\tilde{z})
       \equiv
            2(S_{11} - S_{12})
       =    \frac{e^{\CK/2}}{\sqrt{2} b^0} g_3 (\tilde{\CF_n} - \llangle \CF_n \rrangle),
            \\
      \CS (\tilde{z}, \bar{\tilde{z}})
      &\equiv&  
            e^{\CK/2} S(\tilde{z})
       \equiv
            2(S_{11} + S_{12})
       =    \frac{e^{\CK/2}}{\sqrt{2} b^0} ( 2 g_2 + g_3 (\tilde{\CF_n} + \llangle \CF_n \rrangle) ),
    \EEA
  where $ S_{AB} $ is the gravitino mass matrix.
  Note that $ \CW $ and $ \CS $ are related as follows:
    \BE
      \CW
       =    \CS + i \frac{\sqrt{2} g_1 e^{\CK/2}}{b^0}.
            \label{superpotentialrelation}
    \EE
  Thus they are not independent.
  However, in the following, for convenience, we will write down the resulting Lagrangian, using both $\CW$, $\CS$.
  
  The first term of $\CL_{\rm{Yukawa}}$ is rewritten in terms of $\CW$ and $\CS$ as
    \BEA
      2 S_{AB} \bar{\psi}^A_\mu \gamma^{\mu \nu} \psi^B_\nu 
      &=&   \CW \bar{\psi}^{-}_{\mu} \gamma^{\mu \nu} \psi^{-}_{\nu}
          + \CS \bar{\psi}^{+}_{\mu} \gamma^{\mu \nu} \psi^{+}_{\nu}.
    \EEA
  Thus, we refer to $\CW$ and $\CS$ as superpotentials.
  Since the usual 
    \footnote[2]{For example, \cite{WB} or \cite{Andri1, Andri2}.}
  $\CN=1$ supergravity models coupled to chiral multiplets 
  contain one gravitino, there is only one superpotential.
  But, in this $\CN=1$ model obtained through the partial supersymmetry breaking, there are two gravitini. 
  Therefore there exist two superpotentials which are related by eq. (\ref{superpotentialrelation}).
  
  The covariant derivatives of $\CW$ and that of $\CS$ are respectively
    \BEA
      \tilde{\nabla}_a\CW
      &=&   \frac{e^{\CK/2}}{\sqrt{2} b^0} g_3 
            ( \tilde{\CF}_{na} + \tilde{\partial_a} \CK \tilde{\CF_n} - \tilde{\partial_a} \CK \llangle \CF_n \rrangle )
            \nonumber \\
      &=&   g_{ab^*} (\bar{W}^{b^*}_{2;11} - \bar{W}^{b^*}_{2;12}),
            \\
      \tilde{\nabla}_a \CS
      &=&   \frac{e^{\CK/2}}{\sqrt{2} b^0} 
            (2 g_2 \tilde{\partial_a} \CK + g_3 
            ( \tilde{\CF_{na}} + \tilde{\partial_a} \CK \tilde{\CF_n} + \tilde{\partial_a} \CK \llangle \CF_n \rrangle ))
            \NN \\
      &=&   g_{ab^*} (\bar{W}^{b^*}_{2;11} + \bar{W}^{b^*}_{2;12}),
    \EEA
  and the second derivatives of them are evaluated as,
    \BEA
      \tilde{\nabla}_a \tilde{\nabla}_b \CW
      &=&   \sqrt{2} (\mathcal{M}_{2;a1 \vert b1} - \mathcal{M}_{2;a1 \vert b2}),
            \\
      \tilde{\nabla}_a \tilde{\nabla}_b \CS
      &=&   \sqrt{2} (\mathcal{M}_{2;a1 \vert b1} + \mathcal{M}_{2;a1 \vert b2}),
    \EEA
  where $ \bar{W}^{b^*}_{AB} = (W^{bAB})^* $.
  This can be used to rewrite the second term and the last term of $ \CL_{\rm{Yukawa}} $:
    \BEA
      i g_{ab^*} W^{aAB} \bar{\lambda}^{b^*}_A \gamma_\mu \psi^\mu_B
      &=&   e^{\CK/2} \tilde{g}_{ab^*} \CD^a 
            ( \bar{\lambda}^{b^*}_{-} \gamma_\mu \psi^\mu_{+} - \bar{\lambda}^{b^*}_{+} \gamma_\mu \psi^\mu_{-} )
            \nonumber \\
      & & ~~~+ ~i \bar{\tilde{\nabla}}_{a^*} \bar{\CW} \bar{\lambda}^{b^*}_{-} \gamma_\mu \psi^\mu_{-}
          + i \bar{\tilde{\nabla}}_{a^*} \bar{\CS} \bar{\lambda}^{b^*}_{+} \gamma_\mu \psi^\mu_{+},
            \\
      \mathcal{M}_{aA \vert bB} \bar{\lambda}^{aA} \lambda^{bB}
      &=&   \mathcal{M}_{1;a1 \vert b2} (\bar{\lambda}^{a-} \lambda^{b+} - \bar{\lambda}^{a+} \lambda^{b-})
            \nonumber \\
      & & ~~~+ ~\frac{1}{\sqrt{2}} \tilde{\nabla}_a \tilde{\nabla}_b \CW \bar{\lambda}^{a-} \lambda^{b-}
          + \frac{1}{\sqrt{2}} \tilde{\nabla}_a \tilde{\nabla}_b \CS \bar{\lambda}^{a+} \lambda^{b+}.
    \EEA
  In this way, we can evaluate all the terms of $ \CL_{\rm{Yukawa}} $.
  As a result, we obtain
    \BEA
      \CL_{\rm{Yukawa}}
      &=&   \CW \bar{\psi}^{-}_{\mu} \gamma^{\mu \nu} \psi^{-}_{\nu}
          + i (\bar{\tilde{\nabla}}_{a^*} \bar{\CW} \bar{\lambda}^{a^*}_{-} 
          - 2 \bar{\CW} \bar{\zeta}^{+}) \gamma_\mu \psi^\mu_{-}
          - e^{\CK/2} \tilde{g}_{ab^*} \CD^a \bar{\lambda}^{b^*}_{+} \gamma_\mu \psi^\mu_{-}
            \nonumber \\
      & & + \CS \bar{\psi}^{+}_{\mu} \gamma^{\mu \nu} \psi^{+}_{\nu}
          + i (\bar{\tilde{\nabla}}_{a^*} \bar{\CS} \bar{\lambda}^{a^*}_{+} 
          + 2 \bar{\CS} \bar{\zeta}^{-}) \gamma_\mu \psi^\mu_{+}
          + e^{\CK/2} \tilde{g}_{ab^*} \CD^a \bar{\lambda}^{b^*}_{-} \gamma_\mu \psi^\mu_{+}
            \nonumber \\
      & & + \CW \bar{\zeta}_+ \zeta_+ 
          + \CS \bar{\zeta}_- \zeta_-
          - 2 \tilde{\nabla}_a \CW \bar{\zeta}_+ \lambda^{a-}
          + 2 \tilde{\nabla}_a \CS \bar{\zeta}_- \lambda^{a+}
            \\
      & & + \mathcal{M}_{1;a1 \vert b2} (\bar{\lambda}^{a-} \lambda^{b+} - \bar{\lambda}^{a+} \lambda^{b-})
          + \frac{1}{\sqrt{2}} \tilde{\nabla}_a \tilde{\nabla}_b \CW \bar{\lambda}^{a-} \lambda^{b-}
          + \frac{1}{\sqrt{2}} \tilde{\nabla}_a \tilde{\nabla}_b \CS \bar{\lambda}^{a+} \lambda^{b+}.
            \nonumber
    \EEA
  Note that if we substitute $\tilde{z}^a=0$ into $\CL_{\rm{Yukawa}}$, 
  it reduces to the fermion mass terms which have been obtained in the last section.
\subsubsection{The scalar potential}
  We turn to the scalar potential $V$.
  It is useful to rewrite it in terms of the mass matrices as
    \BE
      V
       =  - 12 \bar{S}^{1A} S_{A1} + \tilde{g}_{ab^*} \bar{W}^{b^*}_{1A} W^{a1A} + 2 \bar{N}_1^{\alpha} N_{\alpha}^1.
            \label{UNV}
    \EE
  which is also obtained from the supergravity Ward identities in reference \cite{Cecotti1}.
  Note that $ \bar{S}^{AB} = (S_{AB})^* $ and $ \bar{N}_A^\alpha = (N_\alpha^A)^* $.
  
  Let us rewrite (\ref{UNV}) in terms of the superpotentials.
  The first term is
    \BEA
      -12 \bar{S}^{1A} S_{A1}
      &=& - 6 ((S_{11} - S_{12})(\bar{S}^{11} - \bar{S}^{12}) + (S_{11} + S_{12})(\bar{S}^{11} + \bar{S}^{12}))
            \nonumber \\
      &=& - \frac{3}{2} (|\CW|^2 + |\CS|^2),
            \label{SS}
    \EEA
  and the second term is
    \BEA
      \tilde{g}_{ab^*} \bar{W}^{b^*}_{1A} W^{a1A}
      &=&   \tilde{g}_{ab^*} 
            \left(
            \bar{W}^{b^*}_{1;12} W^{a12}_1 + \bar{W}^{b^*}_{2;11} W^{a11}_2 + \bar{W}^{b^*}_{2;12} W^{a12}_2
            \right)
            \nonumber \\
      &=&   e^{\CK} \tilde{g}_{ab^*} \mathcal{D}^a \mathcal{D}^b
          + \frac{1}{2} \tilde{g}^{ab^*} \tilde{\nabla}_a \CW \bar{\tilde{\nabla}}_a \bar{\CW}
          + \frac{1}{2} \tilde{g}^{ab^*} \tilde{\nabla}_a S \bar{\tilde{\nabla}}_a \bar{\CS}.
    \EEA
  In the first equality, we have used (\ref{killing}).
  The last term is
    \BEA
      2 \bar{N}_1^{\alpha} N_{\alpha}^1
      &=&   |\CW|^2 + |\CS|^2
            \nonumber \\
      &=&   \frac{1}{2} h^{uv} \nabla_u \CW \nabla_v \bar{\CW} + \frac{1}{2} h^{uv} \nabla_u \CS \nabla_v \bar{\CS},
            \label{NN}
    \EEA
  where $ u, v = 0,1 $ and $ h_{uv} \equiv \delta_{uv}/2(b^0)^2 $.
  Note that $ \nabla_u \CW = \partial_u \CW $.
  Substituting (\ref{SS})-(\ref{NN}) into (\ref{UNV}), we have
    \BEA
      V
      &=&   e^{\CK/2} g_{ab^*} \mathcal{D}^a \mathcal{D}^b
          + \frac{1}{2} \tilde{g}^{ab^*} \tilde{\nabla}_a \CW \bar{\tilde{\nabla}}_a \bar{\CW}
          + \frac{1}{2} \tilde{g}^{ab^*} \tilde{\nabla}_a \CS \bar{\tilde{\nabla}}_a \bar{\CS}
            \nonumber \\
      & & - \frac{3}{2} |\CW|^2 - \frac{3}{2} |\CS|^2
          + \frac{1}{2} h^{uv} \nabla_u \CW \nabla_v \bar{\CW} 
          + \frac{1}{2} h^{uv} \nabla_u \CS \nabla_v \bar{\CS}.
    \EEA
  This is the final form of the scalar potential.
  We can see that $\CL_{\rm{Yukawa}}$ and the scalar potential $V$ takes essentially the same form 
  as the usual $\CN=1$ supergravity model (such as \cite{WB} or \cite{Andri1, Andri2}).
  Note that the superpotentials $\CW$ and $\CS$ are related by (\ref{superpotentialrelation}).

\section{Conclusion}
  We have seen that, in the $\CN=2$ $U(N)$ gauged supergravity model, 
  the $\CN=2$ supersymmetry has been broken to $\CN=1$ counterpart spontaneously.
  In particular, we have not chosen the symplectic section as the simplest function.
  This leads to the effective theory which contains higher order coupling terms of the scalar fields.
  As we have seen in the last section, the Nambu-Goldstone fermion and the Nambu-Goldstone bosons are absorbed 
  by the gravitino and the gauge bosons respectively, through the super-Higgs and the Higgs mechanisms.
  All the masses of the fermions and the bosons have been evaluated.
  The gauge symmetry $U(N) \times U(1)_{graviphoton}$ is broken to $SU(N)$ and the resulting model lies in the Higgs phase.
  Finally, we have considered the $\CN=1$ Lagrangian.
  The $\CN=1$ Yukawa interaction terms and the $\CN=1$ scalar potential can be written 
  in terms of the superpotentials.
  
  As is pointed out in \cite{Ferrara2}, if we force the gravity and the hypermultiplet to decouple, 
  the gravitino mass (\ref{m}) becomes zero.
  Thus, the gauge boson corresponding to the overall $U(1)$ and the graviphoton become massless in this limit.
  The Higgs phase of $U(1) \times U(1)_{graviphoton}$ approaches the Coulomb phase.
  


\section*{Acknowledgement}
The author would like to thank his supervisor Hiroshi Itoyama for helpful advices.
The author thanks Kazuhito Fujiwara for useful discussions and
thanks Hironobu Kihara, Yasunari Kurita, Makoto Sakaguchi and Yukinori Yasui for kindly helps.
The author is grateful to the colleagues in our fundamental physics group for various stimulating discussions.

\appendix
\chapter{Conventions and notations}
Minkowski metric is defined as $\eta_{ij} \equiv (1, -1, -1,-1)$.
The Riemann Tensor is
    \BE
      R^\mu_{~\nu}
       =     d \Gamma^\mu_{~\nu}
           + \Gamma^\mu_{~\rho} \wedge \Gamma^\rho_{~\nu}
       \equiv
           - \frac{1}{2} R^\mu_{~\nu \rho \sigma} dx^\rho \wedge dx^\sigma.
    \EE
Decomposition of tensors in self-dual and antiself-dual parts is
    \BE
      T^{\pm}_{\mu \nu}
       =     \frac{1}{2} 
             \left(
             T_{\mu \nu} \pm \frac{i}{2} \e_{\mu \nu \rho \sigma} T^{\rho \sigma}
             \right).
    \EE
\section{Notations in quaternionic K\"ahler manifolds}
\subsubsection{$SU(2)$ and $Sp(2k)$ metrics}
  The flat $SU(2)$ and $Sp(2k)$ metrics satisfy:
    \BEA
      \e^{AB} \e_{BC}
      &=&   - \delta^A_C,
              ~~~~~~
      e^{AB}
       =    - \e^{BA},
              ~~~~~~
      e^{12}
       =      e_{12}
       =    + 1,
              \\
      \mathbb{C}^{\a \b} \mathbb{C}_{\b \gamma}
      &=&   - \delta^{\a}_\gamma
              ~~~~~~
      \mathbb{C}^{\a \b}
       =    - \mathbb{C}^{\b \a},
              ~~~~~~
      \mathbb{C}^{12}
       =      \mathbb{C}_{12}
       =    + 1.
    \EEA
  For any $SU(2)$ vector $P_A$ we have:
    \BEA
      \e_{AB} P^B
      &=&     P_A,
              \\
      \e^{AB} P_B
      &=&   - P^A,
    \EEA
  and for any $Sp(2k)$ vectors $P_\a$ we have:
    \BEA
      \mathbb{C}_{\a \b} P^\b
      &=&    P_\a,
             \\
      \mathbb{C}^{\a \b} P_\b
      &=&  - P^\a.
    \EEA
\subsubsection{Pauli matrices}
  The standard Pauli matrices $(\sigma^x)_A^{~B}$ $(x=1,2,3)$ are
    \BE
      (\sigma^1)_A^{~B}
       =     \left(
             \begin{array}{cc}
             0 & 1 \\
             1 & 0
             \end{array}
             \right),
             ~~~
      (\sigma^2)_A^{~B}
       =     \left(
             \begin{array}{cc}
             0 & -i \\
             i & 0
             \end{array}
             \right),
             ~~~
      (\sigma^3)_A^{~B}
       =     \left(
             \begin{array}{cc}
             1 & 0 \\
             0 & -1
             \end{array}
             \right).
    \EE
  The Pauli matrices with two lower indices are defined as follows:
    \BEA
      (\sigma^x)_{AB}
      &\equiv&
             (\sigma^x)_A^{~C} \e_{BC},
             \\
      (\sigma^x)^{AB}
      &\equiv&
           - (\sigma^x)_C^{~B} \e^{AC}.
    \EEA
  The above equations can be written as
    \BEA
      (\sigma^1)_{AB}
      &=&    \left(
             \begin{array}{cc}
             1 & 0 \\
             0 & -1
             \end{array}
             \right),
      (\sigma^2)_{AB}
       =     \left(
             \begin{array}{cc}
             -i & 0 \\
             0 & -i
             \end{array}
             \right),
      (\sigma^3)_{AB}
       =     \left(
             \begin{array}{cc}
             0 & -1 \\
             -1 & 0
             \end{array}
             \right),
             \\
      (\sigma^1)^{AB}
      &=&    \left(
             \begin{array}{cc}
             -1 & 0 \\
             0 & 1
             \end{array}
             \right),
             ~
      (\sigma^2)^{AB}
       =     \left(
             \begin{array}{cc}
             -i & 0 \\
             0 & -i
             \end{array}
             \right),
             ~
      (\sigma^3)^{AB}
       =     \left(
             \begin{array}{cc}
             0 & 1 \\
             1 & 0
             \end{array}
             \right).
    \EEA
  These imply the following equation:
    \BE
      (\sigma^x_{AB})^*
       =   - \sigma^{xAB}.
    \EE
  If we define $(\sigma^x)^A_{~B}$ as
    \BE
      (\sigma^x)^A_{~B}
       =   - \e^{AC} (\sigma^x)_{CB},
    \EE
  we obtain
    \BE
      (\sigma^1)^A_{~B}
       =     \left(
             \begin{array}{cc}
             0 & 1 \\
             1 & 0
             \end{array}
             \right),
             ~~~
      (\sigma^2)^A_{~B}
       =     \left(
             \begin{array}{cc}
             0 & i \\
             - i & 0
             \end{array}
             \right),
             ~~~
      (\sigma^3)^A_{~B}
       =     \left(
             \begin{array}{cc}
             1 & 0 \\
             0 & -1
             \end{array}
             \right).
    \EE
\section{Spinor conventions}
\subsubsection{Clifford algebra}
  The gamma matrices in 4-dimensions are given as follows:
    \begin{subequations}
    \begin{align}
      \{ \gamma_i, \gamma_j \} 
      &=     2 \eta_{ij},
             \NN \\
      [\gamma_i, \gamma_j]
      &=     2 \gamma_{ij},
             \NN \\
      \gamma_5
      &\equiv 
           - i \gamma_0 \gamma_1 \gamma_2 \gamma_3
             ,~~~~
      \g_5^\dagger
       =     \g_5,
             ~~~~~
      \gamma_5^2
       =     1,
             \NN \\
      \{ \g_5, \g_i \}
      &=     0,
             \NN \\
      \gamma_0^\dagger
      &=     \gamma_0,
             \NN \\
      \e_{ijkl} \g^{kl}
      &      2 i \g_{ij} \g_5,
             ~~~~~
      \e^{ijkl} \g_{kl}
       =     2 i \g^{ij} \g_5.
             \NN
    \end{align}
    \end{subequations}
  
\subsubsection{Chirality}
  The upper or lower position of the indices of the spinors fix their chirality as follows:
    \BEA
      \g_5
      \left(
      \begin{array}{c}
      \psi_A \\
      \lambda^{aA} \\
      \zeta_\a \\
      \chi_\bullet \\
      \eta_\bullet
      \end{array}
      \right)
      &=&    \left(
             \begin{array}{c}
             \psi_A \\
             \lambda^{aA} \\
             \zeta_\a \\
             \chi_\bullet \\
             \eta_\bullet
             \end{array}
             \right)
             ~~~:\rm{right ~handed},
             \\
      \g_5
      \left(
      \begin{array}{c}
      \psi^A \\
      \lambda^a_A \\
      \zeta^\a \\
      \chi^\bullet \\
      \eta^\bullet
      \end{array}
      \right)
      &=&  - \left(
             \begin{array}{c}
             \psi^A \\
             \lambda^a_A \\
             \zeta^\a \\
             \chi^\bullet \\
             \eta^\bullet
             \end{array}
             \right)
             ~~~:\rm{left ~handed}.
    \EEA
  
\subsubsection{Majorana conditions}
  For any fermion $\phi$, the Majorana condition is
    \BE
      \bar{\phi}
       \equiv
             \phi^\dagger \g_0
       =     \phi^T C,
    \EE
  where the charge conjugation matrix has the following properties:
    \BEA
      C^2
      &=&  - 1,
             ~~~~~
      C^T
       =   - C,
             \NN \\
      (C \g^i)^T
      &=&    C \g^i,
             ~~~~~
      (C \g^{ij})^T
       =     C \g^{ij},
             \NN \\
      (C \g_5)^T
      &=&  - C \g_5,
             ~~~~~
      (C \g_5 \g^i)^T
       =   - C \g_5 \g^i,
             \NN \\
      (C \g_5 \g^i \g^j)^T
      &=&  - C \g_5 \g^j \g^i,
             ~~~~~
      (C \g_5 \g^{ij})^T
       =     C \g_5 \g^{ij}.
    \EEA
\subsubsection{Hermiticity of currents}
  For 0-form spinors:
    \BEA
      (\bar{\chi}_\bullet \eta_\bullet)^\dagger
      &=&    \bar{\eta}^\bullet \chi^\bullet
       =     \bar{\chi}^\bullet \eta^\bullet,
             \\
      (\bar{\chi}_\bullet \g^i \eta^\bullet)^\dagger
      &=&    \bar{\eta}_\bullet \g^i \chi^\bullet
       =   - \bar{\chi}^\bullet \g^i \eta_\bullet,
             \\
      (\bar{\chi}_\bullet \g^{ij} \eta_\bullet)^\dagger
      &=&  - \bar{\eta}^\bullet \g^{ij} \chi^\bullet
       =     \bar{\chi}^\bullet \g^{ij} \eta^\bullet.
    \EEA
  For 1-form spinors
    \BEA
      (\bar{\psi}_A \psi_B)^\dagger
      &=&  - \bar{\psi}^B \psi^A
       =     \bar{\psi}^A \psi^B,
             \\
      (\bar{\psi}^A \g^i \psi_B)^\dagger
      &=&  - \bar{\psi}^B \g^i \psi_A
       =   - \bar{\psi}_A \g^i \psi^B,
             \\
      (\bar{\psi}^A \g^{ij} \psi^B)^\dagger
      &=&    \bar{\psi}_B \g^{ij} \psi_A
       =     \bar{\psi}_A \g^{ij} \psi_B.
    \EEA
\chapter{The Bianchi identities and its solutions}
\section{Ungauged case}
  Let us start from ungauged case.
  We introduce the basic fields of $\CN=2$ supergravity 
  and their associated curvatures as differential forms in superspace.
  We take the one-forms $e^i$ and $\psi_A$, as a basis for $\CN=2$ superspace, 
  whose space-time components $e^i_\mu$ and $\psi_{A \mu}$ are the ordinary vierbein and gravitino fields.
  The curvatures of $\CN=2$ ungauged supergravity coupled to the vector multiplets and hypermultiplet are as follows:
  
  The superspace curvatures in gravitational sector are defined as,
    \begin{subequations}
    \begin{align}
      T^i
      &\equiv
             \CD e^i - i \bar{\psi}_A \wedge \gamma^i \psi^A,
             \label{Ucurvature1}
             \\
      \rho_A
       \equiv
             \nabla \psi_A
      &\equiv
             d \psi_A - \frac{1}{4} \gamma_{ij} \omega^{ij} \wedge \psi_A 
           + \frac{i}{2} \CQ \wedge \psi_A + \omega_A^{~B} \wedge \psi_B,
             \\
      \rho^A
       \equiv
             \nabla \psi^A
      &\equiv
             d \psi^A - \frac{1}{4} \gamma_{ij} \omega^{ij} \wedge \psi^A 
           - \frac{i}{2} \CQ \wedge \psi^A + \omega^A_{~B} \wedge \psi^B,
             \\
      R^{ij}
      &\equiv
             d \omega^{ij} - \omega^i_{~k} \wedge \omega^{kj},
    \end{align}
    \end{subequations}
  where $T^i$ is the torsion 2-form, $R^{ij}$ is the space-time Ricci 2-form, 
  $\omega^{ij}$ is the spin-connection 1-form, 
  $\CQ$ is $U(1)$ bundle connection which has been defined in (\ref{Q}),
  $\omega_A^{~B} \equiv \frac{i}{2} (\sigma_x)_A^{~B} \omega^x$ ($\omega^A_{~B} = \e^{AC} \omega_C^{~D} \e_{DB}$) 
  are $SU(2)$ bundle connection which has been defined in the section 2.3.
  
  The curvatures and covariant derivatives in the vector multiplet sector are
    \begin{subequations}
    \begin{align}
      \nabla z^a
      &=    d z^a,
             \\
      \nabla \bar{z}^{a^*} 
      &=    d \bar{z}^{a^*},
             \\
      \nabla \lambda^{aA}
      &\equiv
             d \lambda^{aA} - \frac{1}{4} \gamma_{ij} \omega^{ij} \lambda^{aA}
           - \frac{i}{2} \CQ \lambda^{aA}
           + \Gamma^a_{~b} \lambda^{bA}
           + \omega^A_{~B} \lambda^{aB},
             \\
      \nabla \lambda^{a^*}_A
      &\equiv
             d \lambda^{a^*}_A - \frac{1}{4} \gamma_{ij} \omega^{ij} \lambda^{a^*}_A
           + \frac{i}{2} \CQ \lambda^{a^*}_A
           + \Gamma^a_{~b} \lambda^{b^*}_A
           + \omega^A_{~B} \lambda^{a^*}_B,
             \\
      F^\Lambda
      &\equiv
             d A^\Lambda
           + \bar{L}^\Lambda \bar{\psi}_A \wedge \psi_B \e^{AB} 
           + L^\Lambda \bar{\psi}^A \wedge \psi^B \e_{AB},
             \label{Ucurvature3}
    \end{align}
    \end{subequations}
  where $A^\Lambda$ ($\Lambda = 0, \ldots, m$) is the gauge connection 1-form: 
  the value $\Lambda=0$ corresponds to the graviphoton 
  and $\Lambda=1,\ldots,m$ corresponds to the gauge bosons of $m$ vector multiplets.
  The quantities $L^\Lambda$ and $\bar{L}^\Lambda$ are arbitrary functions of $z^a$ and $\bar{z}^{a^*}$ 
  on K\"ahler manifold, 
  but, as we will see later, the Bianchi identities constrain in such a way 
  that they coincide with the objects defined in (\ref{sectionV}).
  Note that we do not assume that the vector multiplet scalar sector is described by the special K\"ahler geometry 
  rather we only assume it is described by Hodge-K\"ahler geometry.
  
  The covariant derivatives in the hypermultiplet sector are
    \begin{subequations}
    \begin{align}
      \nabla \zeta_\a
      &\equiv
             d \zeta_\a
           - \frac{1}{4} \gamma_{ij} \omega^{ij} \zeta_\a
           - \frac{i}{2} \CQ \zeta_\a
           + \Delta_\a^{~\b} \zeta_\b,
             \\
      \nabla \zeta^\a
      &\equiv
             d \zeta^\a
           - \frac{1}{4} \gamma_{ij} \omega^{ij} \zeta^\a
           + \frac{i}{2} \CQ \zeta^\a
           + \Delta^\a_{~\b} \zeta^\b,
             \label{Ucurvature2}
    \end{align}
    \end{subequations}
  where $\Delta_\alpha^{~\beta}$ is the gauged Levi-Civita connection on $\CH\CM$ defined in the section 2.3, 
  satisfying the conditions (\ref{Spconnection}).
  It is convenient to convert the world index of the curvature $d b^u$ into a flat index $A$, $\a$ 
  by means of the quaternionic vielbein such that,
    \BE
      \CU^{A \alpha}
       \equiv
             \CU^{A \alpha}_u d b^u.
    \EE
\subsubsection{The Bianchi identities in the ungauged case} 
  We can derive the following Bianchi identities:
  in the gravitational sector, 
    \begin{subequations}
    \begin{align}
      \CD T^i
      &+     R^{ij} \wedge e_j
           - i \bar{\psi}^A \wedge \g^i \rho_A
           + i \bar{\rho}^A \wedge \g^i \psi_A
       =     0,
             \label{UBI1}
             \\
      \nabla \rho_A
      &+     \frac{1}{4} \g_{ij} R^{ij} \wedge \psi_A
           - \frac{i}{2} K \wedge \psi_A
           - R_A^{~B} \wedge \psi_B
       =     0,
             \\
      \nabla \rho^A
      &+     \frac{1}{4} \g_{ij} R^{ij} \wedge \psi^A
           + \frac{i}{2} K \wedge \psi^A
           - R^A_{~B} \wedge \psi^B
       =     0,
             \\
      \CD R^{ij}
      &=     0,
    \end{align}
    \end{subequations}
  where $R_A^{~B}$ is the $SU(2)$ curvature defined in (\ref{SU(2)curvature}) 
  and $K$ is the K\"ahler 2-form, $K=d \CQ$.
  The covariant derivative $\CD$ have defined as, for vector $V^i$ and tensor $V^{ij}$, 
    \BEA
      \CD V^i
      &=&    d V^i - \omega^i_{~j} \wedge V^j,
             \\
      \CD V^{ij}
      &=&    d V^{ij} - \omega^i_{~k} \wedge V^{kj} + \omega^j_{~k} \wedge V^{ik}.
    \EEA
  
  In the vector multiplet sector, we obtain
    \begin{subequations}
    \begin{align}
      d^2 z^a
      &=     d^2 \bar{z}^{a^*} = 0,
             \\
      \nabla^2 \lambda^{aA}
      &+     \frac{1}{4} \g_{ij} R^{ij} \lambda^{aA}
           + \frac{i}{2} K \lambda^{aA}
           + R^a_{~b} \lambda^{bA} 
           - \frac{i}{2} R^A_{~B} \lambda^{aB}
       =     0,
             \\
      \nabla^2 \lambda^{a^*}_A
      &+     \frac{1}{4} \g_{ij} R^{ij} \lambda^{a^*}_A
           - \frac{i}{2} K \lambda^{a^*}_A
           + R^{a^*}_{~b^*} \lambda^{b^*}_A 
           - \frac{i}{2} R^B_{~A} \lambda^{a^*}_A
       =     0,
             \\
      d F^\Lambda
      &=     \e_{AB} \nabla L^\Lambda \wedge \bar{\psi}^A \wedge \psi^B 
           + 2 \e_{AB} L^\Lambda \bar{\psi}^A \wedge \rho^B
             \NN \\
      &-     \e^{AB} \nabla \bar{L}^\Lambda \wedge \bar{\psi}_A \wedge \psi_B 
           + 2 \e^{AB} \bar{L}^\Lambda \bar{\psi}_A \wedge \rho_B = 0,
    \end{align}
    \end{subequations}
  where $R^a_{~b}$ is the Ricci 2-form of the Hodge-K\"ahler manifold.
  
  In the hypermultiplet sector, we have
    \begin{subequations}
    \begin{align}
      \nabla^2 \zeta_\a
      &+     \frac{1}{4} \g_{ij} R^{ij} \zeta_\a
           + \mathbb{R}_\a^{~\b} \zeta_\b
           + \frac{i}{2} K \zeta_\a
       =     0,
             \\
      \nabla^2 \zeta^\a
      &+     \frac{1}{4} \g_{ij} R^{ij} \zeta^\a
           + \mathbb{R}^\a_{~\b} \zeta^\b
           - \frac{i}{2} K \zeta^\a
       =     0,
             \\
      \nabla \CU^{A \a}
      &=     0,
             \label{UBI3}
    \end{align}
    \end{subequations}
  where $\mathbb{R}^\a_{~\b}$ is the $Sp(2m)$ curvature which has been defined in (\ref{Spcurvature}).
\subsubsection{The solutions of the Bianchi identities in the ungauged case}
  The solutions of the Bianchi identities (\ref{UBI1})-(\ref{UBI3}) in the ungauged case are written as follows:
    \BEA
      T^i
      &=&    0,
             \\
      \rho_A
      &=&    \rho_{A|ij} e^i \wedge e^j 
           + (
             A^B_{~A|j} \eta^{ij}
           + A'^B_{~A|j} \g^{ij}
             ) \psi_B \wedge e^i
             \NN \\
      & &  + \e_{AB} (T^-_{ij} + U^+_{ij}) \g^j \psi^B \wedge e^i,
             \\
      \rho^A
      &=&    \rho^A_{ij} e^i \wedge e^j 
           + (
             A^{A| j}_{~B} \eta_{ij}
           + A^{'A| j}_{~B} \g_{ij}
             ) \psi^B \wedge e^i
             \NN \\
      & &  + \e^{AB} (T^+_{ij} + U^-_{ij}) \g^j \psi_B \wedge e^i,
             \\
      R^{ij}
      &=&    R^{ij}_{kl} e^k \wedge e^l
           - i (\bar{\psi}_A \theta^{A|ij}_k + \bar{\psi}^A \theta^{ij}_{A|k}) \wedge e^k
             \NN \\
      & &  + \e^{ijkl} \bar{\psi}_A \wedge \g_k \psi_B (A^{'B}_{~A|k} - \bar{A}^{'~B}_{A|k})
             \NN \\
      & &  + i \e^{AB} \bar{\psi}_A \wedge \psi_B (T^{+ij} + U^{-ij})
           - i \e_{AB} \bar{\psi}^A \wedge \psi^B (T^{-ij} + U^{+ij}),
             \\
      & &    \NN \\
      d z^a
      &=&    Z^a_i e^i + \bar{\psi}_A \lambda^{aA},
             \\
      d \bar{z}^{a^*}
      &=&    Z^{a^*}_i e^i + \bar{\psi}^A \lambda^{a^*}_A,
             \\
      \nabla \lambda^{aA}
      &=&    \nabla_i \lambda^{aA} e^i 
           + i Z^a_i \g^i \psi^A 
           + G^{-a}_{ij} \g^{ij} \psi_B \e^{AB} 
           + Y^{aAB} \psi_B,
             \\
      \nabla \lambda^{a^*}_A
      &=&    \nabla_i \lambda^{a^*}_A e^i 
           + i \bar{Z}^{a^*}_i \g^i \psi_A 
           + G^{+a^*}_{ij} \g^{ij} \psi^B \e_{AB} 
           + Y^{a^*}_{AB} \psi^B,
             \\
      F^\Lambda
      &=&    F^\Lambda_{ij} e^i \wedge e^j 
           + i (
             f^\Lambda_a \bar{\lambda}^{aA} \g_i \psi^B \e_{AB} 
           + \bar{f}^\Lambda_{a^*} \bar{\lambda}^{a^*}_A \g_i \psi_B \e^{AB}
             ) \wedge e^i,
             \\
      & &    \NN \\
      \nabla \zeta_\a
      &=&    \nabla_i \zeta_\a e^i 
           + i \CU^{B\b}_i \g^i \psi^A \e_{AB} \mathbb{C}_{\a \b},
             \\
      \nabla \zeta^\a
      &=&    \nabla_i \zeta^\a e^i
           + i \CU^{A\a}_i \g^i \psi_A,
             \\
      \CU^{A \a}
      &=&    \CU^{A\a}_i e^i
           + \e^{AB} \mathbb{C}^{\a \b} \bar{\psi}_B \zeta_\b
           + \bar{\psi}^A \zeta^\a,
    \EEA
  where
    \begin{subequations}
    \begin{align}
      A_A^{~iB}
      &=   - \frac{i}{4} g_{a^* b} (\bar{\lambda}^{a^*}_A \g^i \lambda^{bB} 
           - \delta^B_A \bar{\lambda}^{a^*}_C \g^i \lambda^{bC}),
             \label{UDterm1}
             \\
      A_A^{'~iB}
      &=     \frac{i}{4} g_{a^* b} (\bar{\lambda}^{a^*}_A \g^i \lambda^{bB} 
           - \frac{1}{2} \delta^B_A \bar{\lambda}^{a^*}_C \g^i \lambda^{bC})
           + \frac{i}{4} \lambda \delta^B_A \bar{\zeta}_\a \g^i \zeta^\a,
    \end{align}
    \end{subequations}
    \BE
      \theta^{ij;k}_A
       =     2 \g^{[i} \rho^{j]k}_A + \g^k \rho^{ij}_A,
             ~~~~~
      \theta^{ij;k|A}
       =     2 \g^{[i} \rho^{j]k|A} + \g^k \rho^{ij|A},
    \EE
    \begin{subequations}
    \begin{align}
      T^-_{ij}
      &=     2 i (\Im \CN)_{\Lambda \Sigma} L^\Sigma
             \left(
             F^{- \Lambda}_{ij}
           + \frac{1}{8} \nabla_a f_b^\Lambda \bar{\lambda}^{aA} \g_{ij} \lambda^{bB} \e_{AB}
           + \frac{\lambda}{4} \mathbb{C}^{\a \b} \bar{\zeta}_\a \g_{ij} \zeta_\b L^\Lambda
             \right),
             \\
      T^+_{ij}
      &=     2 i (\Im \CN)_{\Lambda \Sigma} \bar{L}^\Sigma
             \left(
             F^{- \Lambda}_{ij}
           + \frac{1}{8} \nabla_{a^*} \bar{f}_{b^*}^\Lambda \bar{\lambda}^{a^*}_A \g_{ij} \lambda^{b^*}_B \e^{AB}
           + \frac{\lambda}{4} \mathbb{C}_{\a \b} \bar{\zeta}^\a \g_{ij} \zeta^\b \bar{L}^\Lambda
             \right),
    \end{align}
    \end{subequations}
    \begin{subequations}
    \begin{align}
      U^-_{ij}
      &=     \frac{i}{4} \lambda \mathbb{C}^{\a \b} \bar{\zeta}_\a \g_{ij} \zeta_\b,
             \\
      U^+_{ij}
      &=     \frac{i}{4} \lambda \mathbb{C}_{\a \b} \bar{\zeta}^\a \g_{ij} \zeta^\b,
    \end{align}
    \end{subequations}
    \begin{subequations}
    \begin{align}
      G^{a-}_{ij}
      &=   - g^{ab^*} \bar{f}^\Sigma_{b^*} (\Im \CN)_{\Sigma \Lambda}
             \left(
             F^{\Lambda -}_{ij} 
           + \frac{1}{8} \nabla_a f^\Lambda_b \bar{\lambda}^{aA} \g_{ij} \lambda^{bB} \e_{AB}
           + \frac{\lambda}{4} \mathbb{C}^{\a \b} \bar{\zeta}_\a \g_{ij} \zeta_\b L^\Lambda
             \right),
             \\
       G^{a^* +}_{ij}
      &=   - g^{a^* b} f^\Sigma_b (\Im \CN)_{\Sigma \Lambda}
             \left(
             F^{\Lambda +}_{ij} 
           + \frac{1}{8} \nabla_{a^*} \bar{f}^\Lambda_{b^*} \bar{\lambda}^{a^*}_A \g_{ij} \lambda^{b^*}_B \e^{AB}
           + \frac{\lambda}{4} \mathbb{C}_{\a \b} \bar{\zeta}^\a \g_{ij} \zeta^\b \bar{L}^\Lambda
             \right),
    \end{align}
    \end{subequations}
    \begin{subequations}
    \begin{align}
      Y^{a AB}
      &=     \frac{i}{2} g^{ab^*} C_{b^*c^*d^*} \bar{\lambda}^{c^*}_C \lambda^{d^*}_D \e^{AC} \e^{BD},
             \\
      Y^{a^*}_{AB}
      &=   - \frac{i}{2} g^{a^* b} C_{bcd} \bar{\lambda}^{cC} \lambda^{d D} \e_{AC} \e_{BD}.
             \label{UDterm2}
    \end{align}
    \end{subequations}
  Furthermore, from the closure of the Bianchi identities, we obtain the constraints 
  on $L^\Lambda$, $\bar{L}^\Lambda$, $f^\Lambda_a$, $\bar{f}^\Lambda_{a^*}$ and $C_{abc}$ 
  which are the geometrical object of the Hodge-K\"ahler manifold.
  This constraints restrict the Hodge-K\"ahler manifold to be a special K\"ahler manifold.
  We will discuss this at the end of the next section.
\section{Gauged case}
  Let us consider the modifications to the previous results when the theory is gauged.
  The discussion in this section is parallel to that in the previous section.
  In this case, we have to redefine the curvature (\ref{Ucurvature1})-(\ref{Ucurvature2}) 
  according to the discussions in the section 3.2.
  In particular, for the scalar fields $z^a$, $\bar{z}^{a^*}$ and $b^u$, we replace as follows:
    \begin{subequations}
    \begin{align}
      d z^a 
      &\rightarrow
             \nabla z^a
       =     d z^a + g A^\Lambda k_\Lambda^a (z),
             \\
      d \bar{z}^{a^*} 
      &\rightarrow
             \nabla \bar{z}^{a^*}
       =     d \bar{z}^{a^*} + g A^\Lambda k_\Lambda^{a^*} (\bar{z}),
             \\
      d b^u
      &\rightarrow
             \nabla b^u
       =     d b^u + g A^\Lambda k_\Lambda^u (b).
             \label{Gcurvature1}
    \end{align}
    \end{subequations}
  Eq.(\ref{Gcurvature1}) implies that the gauged quaternionic vielbein $\hat{\CU}^{A \a}$ is given by
    \BE
      \hat{\CU}^{A \a}
       =     \CU^{A \a}_{uv} \nabla b^u \wedge \nabla b^v.
    \EE
  We also have to replace the curvatures of the connections with thier gauged expressions:
    \begin{subequations}
    \begin{align}
      R^a_b
      &\rightarrow
             \hat{R}^a_b,
             \label{Gcurvature4}
             \\
      K
      &\rightarrow
             \hat{K},
             \\
      \mathbb{R}^{\a \b}
      &\rightarrow
             \hat{\mathbb{R}}^{\a \b},
             \\
      R_A^{~B}
      &\rightarrow
             \hat{R}_A^{~B}
       \equiv
             \frac{i}{2} (\sigma_x)_A^{~B} \hat{\Omega}^x.
             \label{Gcurvature5}
    \end{align}
    \end{subequations}
  where $\hat{R}^a_b$, $\hat{K}$, $\hat{\mathbb{R}}^{\a \b}$ and $\hat{R}_A^{~B}$ are given 
  in (\ref{GaugeGamma}), (\ref{GaugeK}), (\ref{GaugemathbbR}) and (\ref{GaugeOmega}).
  Furthermore, we have to replace $d A^\Lambda$ in (\ref{Ucurvature3}) with the complete gauge curvature (\ref{GaugeF}):
    \BE
      d A^\Lambda 
       \rightarrow
             F^\Lambda
       \equiv
             d A^\Lambda + \frac{1}{2} g f_{\Sigma \Gamma}^{~~\Lambda} A^\Sigma \wedge A^\Gamma.
             \label{appendixfieldstrength}
    \EE
  
  For completeness, let us collect the curvatures of all the sectors.
  The curvatures in gravitational sector are
    \begin{subequations}
    \begin{align}
      T^i
      &\equiv
             \CD e^i - i \bar{\psi}_A \wedge \gamma_i \psi^A,
             \\
      \rho_A
       \equiv
      \nabla \psi_A
      &\equiv
             d \psi_A - \frac{1}{4} \gamma_{ij} \omega^{ij} \wedge \psi_A 
           + \frac{i}{2} \hat{\CQ} \wedge \psi_A + \hat{\omega}_A^{~B} \wedge \psi_B,
             \\
      \rho^A
       \equiv
      \nabla \psi^A
      &\equiv
             d \psi^A - \frac{1}{4} \gamma_{ij} \omega^{ij} \wedge \psi^A 
           - \frac{i}{2} \hat{\CQ} \wedge \psi^A + \hat{\omega}^A_{~B} \wedge \psi^B,
             \\
      R^{ij}
      &\equiv
             d \omega^{ij} - \omega^i_{~k} \wedge \omega^{kj}.
    \end{align}
    \end{subequations}
  The curvatures and covariant derivatives in the vector multiplet sector are
    \begin{subequations}
    \begin{align}
      \nabla z^a
      &=     d z^a + g A^\Lambda k_\Lambda^a(z),
             \\
      \nabla \bar{z}^{a^*} 
      &=     d \bar{z}^{a^*} + g A^\Lambda k_\Lambda^{a^*} (\bar{z})
             \\
      \nabla \lambda^{aA}
      &\equiv
             d \lambda^{aA} - \frac{1}{4} \gamma_{ij} \omega^{ij} \lambda^{aA}
           - \frac{i}{2} \hat{\CQ} \lambda^{aA} + \hat{\Gamma}^a_{~b} \lambda^{bA}
           + \hat{\omega}^A_{~B} \lambda^{aB},
             \\
      \nabla \lambda^{a^*}_A
      &\equiv
             d \lambda^{a^*}_A - \frac{1}{4} \gamma_{ij} \omega^{ij} \lambda^{a^*}_A
           - \frac{i}{2} \hat{\CQ} \lambda^{a^*}_A + \hat{\Gamma}^a_{~b} \lambda^{b^*}_A
           + \hat{\omega}^A_{~B} \lambda^{a^*}_B,
             \\
      \hat{F}^\Lambda
      &\equiv
             F^\Lambda
           + \bar{L}^\Lambda \bar{\psi}_A \wedge \psi_B \e^{AB} + L^\Lambda \bar{\psi}^A \wedge \psi^B \e_{AB}.
    \end{align}
    \end{subequations}
  We have used the notation $\hat{F}^\Lambda$ in order to distinguish from 
  $F^\Lambda = d A^\Lambda + \frac{1}{2} g f_{\Sigma \Gamma}^{~~\Lambda} A^\Sigma \wedge A^\Gamma$.
  In the hypermultiplet sector the covariant derivatives are
    \begin{subequations}
    \begin{align}
      \CU^{A \alpha}
      &\equiv
             \CU^{A \alpha}_u \nabla b^u
       \equiv
             \CU^{A \alpha}_u (d b^u + g A^\Lambda k^u_\Lambda(b)),
             \\
      \nabla \zeta_\alpha
      &\equiv
             d \zeta_\alpha - \frac{1}{4} \gamma_{ij} \omega^{ij} \zeta_\alpha 
           - \frac{i}{2} \hat{\CQ} \zeta_\alpha + \hat{\Delta}_\alpha^{~\beta} \zeta_\beta,
             \\
      \nabla \zeta^\alpha
      &\equiv
             d \zeta^\alpha - \frac{1}{4} \gamma_{ij} \omega^{ij} \zeta^\alpha 
           + \frac{i}{2} \hat{\CQ} \zeta^\alpha + \hat{\Delta}^\alpha_{~\beta} \zeta^\beta.
    \end{align}
    \end{subequations}
  
\subsubsection{The Bianchi identities in the gauge case}
  The Bianchi identities in the gauged case is the same as the ungauged case (\ref{UBI1})-(\ref{UBI3}) 
  except that the curvatures are replaced as (\ref{Gcurvature4})-(\ref{Gcurvature5}).
  In the gravitational sector, we obtain the following Bianchi identities:
    \begin{subequations}
    \begin{align}
      \CD T^i
      &+     R^{ij} \wedge e_j
           - i \bar{\psi}^A \wedge \g^i \rho_A
           + i \bar{\rho}^A \wedge \g^i \psi_A
       =     0,
             \label{BI1}
             \\
      \nabla \rho_A
      &+     \frac{1}{4} \g_{ij} R^{ij} \wedge \psi_A 
           - \frac{i}{2} K \wedge \psi_A
           - \hat{R}_A^{~B} \wedge \psi_B
       =     0,
             \\
      \nabla \rho^A
      &+     \frac{1}{4} \g_{ij} R^{ij} \wedge \psi^A 
           + \frac{i}{2} K \wedge \psi^A 
           - \hat{R}^A_{~B} \wedge \psi^B 
       =     0,
             \\
      \CD R^{ij}
      &=     0.
    \end{align}
    \end{subequations}
  In the vector multiplet sector, we get
    \begin{subequations}
    \begin{align}
      \nabla^2 z^a
      &-     g 
             (
             \hat{F}^\Lambda 
           - \bar{L}^\Lambda \bar{\psi}_A \wedge \psi_B \e^{AB} 
           - L^\Lambda \wedge \bar{\psi}^A \wedge \psi^B \e_{AB}
             ) k^a_\Lambda
       =     0,
             \\
      \nabla^2 \bar{z}^{a^*}
      &-     g 
             (
             \hat{F}^\Lambda 
           - \bar{L}^\Lambda \bar{\psi}_A \wedge \psi_B \e^{AB} 
           - L^\Lambda \wedge \bar{\psi}^A \wedge \psi^B \e_{AB}
             ) k^{a^*}_\Lambda
       =     0,
             \\
      \nabla^2 \lambda^{aA}
      &+     \frac{1}{4} \g_{ij} R^{ij} \lambda^{aA}
           + \frac{i}{2} \hat{K} \lambda^{aA}
           + \hat{R}^a_{~b} \lambda^{bA} 
           - \frac{i}{2} \hat{R}^A_{~B} \lambda^{aB}
       =     0,
             \\
      \nabla^2 \lambda^{a^*}_A
      &+     \frac{1}{4} \g_{ij} R^{ij} \lambda^{a^*}_A
           - \frac{i}{2} \hat{K} \lambda^{a^*}_A
           + \hat{R}^{a^*}_{~b^*} \lambda^{b^*}_A 
           - \frac{i}{2} \hat{R}^B_{~A} \lambda^{a^*}_A 
       =     0,
             \\
      \nabla \hat{F}^\Lambda
      &-     \e_{AB} \nabla L^\Lambda \wedge \bar{\psi}^A \wedge \psi^B 
           + 2 \e_{AB} L^\Lambda \bar{\psi}^A \wedge \hat{\rho}^B
             \NN \\
      &-     \e^{AB} \nabla \bar{L}^\Lambda \wedge \bar{\psi}_A \wedge \psi_B 
           + 2 \e^{AB} \bar{L}^\Lambda \bar{\psi}_A \wedge \hat{\rho}_B 
       =     0.
    \end{align}
    \end{subequations}
  In the hypermultiplet sector, we derive
    \begin{subequations}
    \begin{align}
      \nabla^2 \zeta_\a
      &+     \frac{1}{4} \g_{ij} R^{ij} \zeta_\a
           + \hat{\mathbb{R}}_\a^{~\b} \zeta_\b
           + \frac{i}{2} \hat{K} \zeta_\a
       =     0,
             \\
      \nabla^2 \zeta^\a
      &+     \frac{1}{4} \g_{ij} R^{ij} \zeta^\a
           + \hat{\mathbb{R}}^\a_{~\b} \zeta^\b
           - \frac{i}{2} \hat{K} \zeta^\a
       =     0,
             \\
      \nabla \CU^{A \a}
      &-     g
             (
             \hat{F}^\Lambda 
           - \bar{L}^\Lambda \bar{\psi}_A \wedge \psi_B \e^{AB} 
           - L^\Lambda \wedge \bar{\psi}^A \wedge \psi^B \e_{AB}
             ) k^u_\Lambda (b) \CU^{A \a}_u
       =     0.
             \label{BI3}
    \end{align}
    \end{subequations}
\subsubsection{The solutions of the Bianchi identities in the gauged case}
  The solutions of the Bianchi identities (\ref{BI1})-(\ref{BI3}) in the gauged case are written as follows:
    \BEA
      T^i
      &=&    0,
             \\
      \rho_A
      &=&    \rho_{A|ij} e^i \wedge e^j 
           + ( A^B_{~A|j} \eta^{ij} + A'^B_{~A|j} \g^{ij} ) \psi_B \wedge e^i
             \NN \\
      & &  + \e_{AB} (T^-_{ij} + U^+_{ij}) \g^j \psi^B \wedge e^i
           + i g S_{AB} \eta_{ij} \psi^B \wedge e^i,
             \\
      \rho^A
      &=&    \rho^A_{ij} e^i \wedge e^j 
           + ( A^{A| j}_{~B} \eta_{ij} + A^{'A| j}_{~B} \g_{ij} ) \psi^B \wedge e^i
             \NN \\
      & &  + \e^{AB} (T^+_{ij} + U^-_{ij}) \g^j \psi_B \wedge e^i
           + i g \bar{S}^{AB} \eta_{ij} \g^j \psi_B \wedge e^i,
             \\
      R^{ij}
      &=&    R^{ij}_{kl} e^k \wedge e^l
           - i (\bar{\psi}_A \theta^{A|ij}_k + \bar{\psi}^A \theta^{ij}_{A|k}) \wedge \e^k
             \NN \\
      & &  + \e^{ijkl} \bar{\psi}_A \wedge \g_k \psi_B (A^{'B}_{~A|k} - \bar{A}^{'~B}_{A|k})
             \NN \\
      & &  + i \e^{AB} \bar{\psi}_A \wedge \psi_B (T^{+ij} + U^{-ij})
           - i \e_{AB} \bar{\psi}^A \wedge \psi^B (T^{-ij} + U^{+ij})
             \NN \\
      & &  - g S_{AB} \bar{\psi}^A \wedge \g^{ij} \psi^B - g \bar{S}^{AB} \bar{\psi}_A \wedge \g^{ij} \psi_B,
             \\
      & &    \NN \\
      \hat{F}^\Lambda
      &=&    F^\Lambda_{ij} e^i \wedge e^j 
           + i (
             f^\Lambda_a \bar{\lambda}^{aA} \g_i \psi^B \e_{AB} 
           + \bar{f}^\Lambda_{a^*} \bar{\lambda}^{a^*}_A \g_i \psi_B \e^{AB}
             ) \wedge e^i,
             \\
      \nabla \lambda^{aA}
      &=&    \nabla_i \lambda^{aA} e^i 
           + i Z^a_i \g^i \psi^A 
           + G^{-a}_{ij} \g^{ij} \psi_B \e^{AB} 
           + ( Y^{aAB} + g W^{aAB}) \psi_B,
             \\
      \nabla \lambda^{a^*}_A
      &=&    \nabla_i \lambda^{a^*}_A e^i 
           + i \bar{Z}^{a^*}_i \g^i \psi_A 
           + G^{+a^*}_{ij} \g^{ij} \psi^B \e_{AB} 
           + ( Y^{a^*}_{AB} + g W^{a^*}_{AB} ) \psi^B,
             \\
      \nabla z^a
      &=&    Z^a_i e^i + \bar{\psi}_A \lambda^{aA},
             \\
      \nabla \bar{z}^{a^*}
      &=&    Z^{a^*}_i e^i + \bar{\psi}^A \lambda^{a^*}_A,
             \\
      & &    \NN \\
      \nabla \zeta_\a
      &=&    \nabla_i \zeta_\a e^i 
           + i \CU^{B\b}_i \g^i \psi^A \e_{AB} \mathbb{C}_{\a \b}
           + g N_\a^A \psi_A,
             \\
      \nabla \zeta^\a
      &=&    \nabla_i \zeta^\a e^i
           + i \CU^{A\a}_i \g^i \psi_A
           + g N_A^\a \psi^A,
             \\
      \CU^{A \a}
      &=&    \CU^{A\a}_i e^i
           + \e^{AB} \mathbb{C}^{\a \b} \bar{\psi}_B \zeta_\b
           + \bar{\psi}^A \zeta^\a,
    \EEA
  where $A_A^{|iB}$, $A_A^{'|iB}$, $\theta^{ij;k}_A$, $T_{ij}$, $U_{ij}$, $G^a_{ij}$ and $Y^{aAB}$ 
  are identical with the ungauged case (\ref{UDterm1})-(\ref{UDterm2}).
  The constant $g$ is just a symbolic notation to remind that these terms are produced by the gauging 
  and the new terms $S_{AB}$, $W^{aAB}$ and $N_\a^A$ are
    \BEA
      S_{AB}
      &=&    \frac{i}{2} (\sigma_x)_A^{~C} \e_{BC} \CP^x_\Lambda L^\Lambda,
             \NN \\
      \bar{S}^{AB}
      &=&    \frac{i}{2} (\sigma_x)_C^{~B} \e^{CA} \CP^x_\Lambda \bar{L}^\Lambda,
    \EEA
    \BEA
      W^{aAB}
      &=&    \e^{AB} k^a_\Lambda \bar{L}^\Lambda
           + i (\sigma_x)_C^{~B} \e^{CA} \CP^x_\Lambda g^{ab^*} \bar{f}^\Lambda_{b^*},
             \NN \\
      W^{a^*}_{AB}
      &=&    \e_{AB} k^{a^*}_\Lambda L^\Lambda
           + i (\sigma_x)_A^{~C} \e_{BC} \CP^x_\Lambda g^{a^* b} f^\Lambda_b,
    \EEA
    \BEA
      N_\a^A
      &=&    2 \CU^A_{\a | u} k^u_\Lambda \bar{L}^\Lambda,
             \NN \\
      N_A^\a
      &=&  - 2 \CU^\a_{A | u} k^u_\Lambda L^\Lambda.
    \EEA
  
  The constraints following from the closure of the Bianchi identities are a set of differential constraints 
  on the upper parts $L^\Lambda$, $\bar{L}^\Lambda$, $f^\Lambda_a$ and $\bar{f}^\Lambda_{a^*}$ 
  of the symplectic sections $V$ and $U_a$ and on $C_{abc}$:
    \BEA
      \nabla_{a^*} L^\Lambda
      &=&    \nabla_a \bar{L}^\Lambda
       =     0,
             \label{constraint1}
             \\
      f_a^\Lambda
      &=&    \nabla_a L^\Lambda,
             \\
      \bar{f}_{a^*}^\Lambda
      &=&    \nabla_{a^*} \bar{L}^\Lambda,
             \\
      \nabla_{d^*} C_{abc}
      &=&    \nabla_d C_{a^*b^*c^*}
       =     0,
             \\
      \nabla_a f_b^\Lambda
      &=&    i g^{cd^*} \bar{f}_{d^*}^\Lambda C_{abc}.
             \label{constraint2}
    \EEA
  These equations are the same for that of the section 2.2.
  Therefore the constraints (\ref{constraint1})-(\ref{constraint2}) restrict the Hodge-K\"ahler manifold 
  which we start from to be a special K\"ahler manifold.

\end{document}